\documentclass[a4paper,11pt]{article}
\pdfoutput=1
\usepackage{graphicx}

\usepackage{jheppub} 

\usepackage[T1]{fontenc} 

\usepackage{xcolor}
\usepackage{arydshln}
\usepackage[font=footnotesize]{caption}
\usepackage{float}
\usepackage{braket}
\usepackage{extarrows}
\usepackage{bbm}
\usepackage{nicefrac}
 \usepackage{import}
 \usepackage{bbold}
\usepackage{mathrsfs}
\usepackage{pgf}
\usepackage[fleqn]{mathtools}
\usepackage{subcaption}
\usepackage{dsfont}
\usepackage{enumerate}
\usepackage{enumitem}
\usepackage{nccmath}
\usepackage{booktabs}
\usepackage{marvosym}

\usepackage{hyperref}
\usepackage{graphics}
\usepackage{bm}
\usepackage{tikz}
\usetikzlibrary{decorations.pathmorphing}
\usetikzlibrary{calc,fit}
\usetikzlibrary{cd}
\usetikzlibrary{shapes,arrows,positioning,chains,matrix}
\usepackage[framemethod=tikz]{mdframed}
\restylefloat{table}
\usepackage{stackrel}
\usepackage{mathdots}
\usepackage{blkarray}
\usepackage{clipboard}

\definecolor{oxfordblue}{rgb}{0.0, 0.13, 0.28}
\definecolor{forestgreen}{rgb}{0.0, 0.27, 0.13}
\definecolor{deepjunglegreen}{rgb}{0.0, 0.29, 0.29}
\definecolor{pinegreen}{rgb}{0.0, 0.47, 0.44}
\definecolor{turquoise}{rgb}{0.19, 0.84, 0.78}
\definecolor{skobeloff}{rgb}{0.0, 0.48, 0.45}
\definecolor{teal}{rgb}{0.0, 0.5, 0.5}
\definecolor{limegreen}{rgb}{0.2, 0.8, 0.2}
\definecolor{lincolngreen}{rgb}{0.11, 0.35, 0.02}
\definecolor{palatinatepurple}{rgb}{0.41, 0.16, 0.38}
\definecolor{orange-red}{rgb}{1.0, 0.27, 0.0}
\definecolor{outrageousorange}{rgb}{1.0, 0.43, 0.29}
\definecolor{orange(colorwheel)}{rgb}{1.0, 0.5, 0.0}
\definecolor{harvardcrimson}{rgb}{0.79, 0.0, 0.09}
\definecolor{princetonorange}{rgb}{1.0, 0.56, 0.0}
\definecolor{safetyorange}{rgb}{1.0, 0.3, 0.0}
\definecolor{portlandorange}{rgb}{1.0, 0.35, 0.21}
\definecolor{outrageousorange}{rgb}{1.0, 0.43, 0.29}
\definecolor{scarlet}{rgb}{1.0, 0.13, 0.0}
\definecolor{indigo(web)}{rgb}{0.29, 0.0, 0.51}
\definecolor{lasallegreen}{rgb}{0.03, 0.47, 0.19}
\definecolor{palatinatepurple}{rgb}{0.41, 0.16, 0.38}
\definecolor{patriarch}{rgb}{0.5, 0.0, 0.5}
\definecolor{jazzberryjam}{rgb}{0.65, 0.04, 0.37}
\definecolor{darkscarlet}{rgb}{0.34, 0.01, 0.1}
\definecolor{yellow(ncs)}{rgb}{1.0, 0.83, 0.0}
\definecolor{mikadoyellow}{rgb}{1.0, 0.77, 0.05}

\title{\boldmath State counting on fibered CY-3 folds and the non-Abelian Weak Gravity Conjecture}

\author[a,1]{Cesar Fierro Cota\note{Corresponding author.}}
\author[a,b]{Albrecht Klemm}
\author[c]{Thorsten Schimannek}

\affiliation[a]{Bethe Center for Theoretical Physics, Universit\"at Bonn, D-53115, Germany}
\affiliation[b]{Hausdorff Center for Mathematics, Universit\"at Bonn, D-53115, Germany}
\affiliation[c]{Faculty of Physics, University of Vienna, Boltzmanngasse 5, A-1090 Vienna, Austria}

\emailAdd{fierro@th.physik.uni-bonn.de} 
\emailAdd{aklemm@th.physik.uni-bonn.de}
\emailAdd{thorsten.schimannek@univie.ac.at}

\abstract{
We extend the dictionary between the BPS spectrum of Heterotic strings and the one of F-/M-theory compactifications on $K3$ fibered Calabi-Yau 3-folds to 
cases with higher rank non-Abelian gauge groups and in particular to dual pairs between Heterotic CHL orbifolds and compactifications on Calabi-Yau 3-folds with a compatible genus one fibration.
We show how to obtain the new supersymmetric index purely from the Calabi-Yau geometry by reconstructing the Noether-Lefschetz generators, which are vector-valued modular forms.
There is an isomorphism between the latter objects and vector-valued lattice Jacobi forms, which relates them to the elliptic genera and twisted-twined elliptic genera of six- and five-dimensional Heterotic strings.
The meromorphic Jacobi forms generate the dimensions of the refined cohomology of the Hilbert schemes of symmetric products of the fiber and allow us to refine the BPS indices in the 
fiber and therefore to obtain, conjecturally, actual state counts.  
Using the properties of the vector-valued lattice  Jacobi forms we also provide a mathematical proof of the 
non-Abelian weak gravity conjecture for F-/M-theory compactified on this general class of fibered Calabi-Yau 3-folds.}

\begin{document}
\rightline{
BONN-TH-2020-12,$\quad$UWThPh 2020-30
}
\maketitle

\flushbottom


\section{Introduction}
\label{sec:intro}
Analyzing supergravity theories~\cite{Morrison:1996pp,Bershadsky:1996nh} and superconformal quantum field theories (SCFT) without gravity~\cite{Klemm_1997} in six 
dimensions geometrically, using F-theory on elliptic -- or more general genus one fiberd Calabi-Yau 3-folds ${\cal E} \hookrightarrow M\rightarrow B$,  has been very successful.
While supergravity theories are defined on compact Calabi-Yau 3-folds  $M$, the SCFTs are  geometrically engineered on non-compact local  models, that are often realizable as limits of compact $M$, consisting  of elliptic fiberd 
surfaces $S$  over curves  $C_\beta\in B$, $\beta \in H^{\rm comp}_2(B,\mathbb{Z})$ with negative self intersection $\beta^2<0$.
In particular, in~\cite{Klemm_1997} it was proposed how to use topological string theory on $M$ with a skrinking dP$_9$ surface  $S$ and its mirror $W$ to extract information about the BPS spectrum of a prototypical $(0,1)$-SCFT, called  the $E$-string, after compactifying it on a circle to five dimensions.
This idea has been pushed much further for the non-gravitational theories by considering all possible fugacities~\cite{Huang:2013yta} and  more general 
configurations of contractable curves in the base~\cite{Heckman:2015bfa}. 
In~\cite{Haghighat:2013gba,Kim:2014dza,Haghighat:2014vxa} the elliptic genus, a BPS 
index exhibiting by construction the modular transformation and the quasi-periodicity of Jacobi forms, 
has been calculated using localization of an auxiliary two-dimensional non-abelian gauged linear sigma 
model describing the world sheet theory of the exceptional string in their infrared limit.

In~\cite{Huang:2015ada} it was realized that modular covariance and the pole structure of the  BPS index expansion of the topological 
string  partition function $Z(\bm{t},\lambda)$, depending in the A-model on the topological string coupling $\lambda$ and all 
complexified K\"ahler parameters $\bm{t}$, suggest for general, and in particular compact elliptic Calabi-Yau manifolds,  
a base degree $[\beta] \in H_2(B,\mathbb{Z})$ expansion of $Z(\bm{t},\lambda)$ whose coefficients $Z_\beta(\tau, \bm{z})$ are meromorphic Jacobi forms of a very specific structure.
Firstly, after an easily calculable shift~\cite{Huang:2015ada}, the K\"ahler parameter of the elliptic fiber becomes their {\sl modular parameter} $\tau$, while  $\lambda$, or in case of refined BPS counts  the refinement parameters
$\epsilon_1,\epsilon_2$  ($\lambda^2=-\epsilon_1,\epsilon_2$), as well as the Cartan K\"ahler parameters of all enhanced gauge groups, which become together with the flavor fugacities in the non-compact cases 
the {\sl elliptic parameters} $ \bm{z}$.   The denominators of  the $Z_\beta(\tau, \bm{z})$  are fixed by the pole structure  while the numerators  are  elements in  the finitely generated ring of weak Jacobi forms whose weight and index~\footnote{The  
contribution of the Cartan gauge elliptic parameters  to index  can be also understood by their chiral anomaly~\cite{Gu:2017ccq}.} is fixed geometrically  in terms  of the class of the base curve class $\beta $.
This finiteness allows to reconstruct the $Z_\beta(\tau, \bm{z})$ for compact Calabi-Yau  manifolds to a large extend and for certain local limits completely~\cite{Huang:2015ada}, even in the refined case~\cite{Gu:2017ccq,DelZotto:2017mee}.
In the general cases with enhanced gauge symmetries the ring of modular objects are Weyl invariant Jacobi forms~\cite{DelZotto:2017mee}.
The approach~\cite{Huang:2015ada} opened a successful strategy to obtain the automorphic transformation properties that restrict $Z_\beta(\tau, \bm{z})$ entirely 
geometrically on the Type II side~\footnote{A very systematic confirmation for the Jacobi form structure in the local cases are the elliptic blow up equations~\cite{Gu:2018gmy,Gu:2019dan,Gu:2019pqj,Gu:2020fem} 
derived from Nakajima and Yoshoka's, five dimensional blow up equations for $M$-theory on $M$, which also allow in 
most cases to fix all the boundary conditions for the refined partition function.} by constructing the monodromies of the periods on the mirror $W$ or, more 
efficiently, the  auto-equivalences of the derived category of coherent sheaves on $M$~\cite{Schimannek:2019ijf,Cota:2019cjx}.
In particular, in~\cite{Cota:2019cjx} it was shown that for genus one fibrations that do not have a section but only $N$-sections the relevant Jacobi forms exhibit modular transformation only under the finite index 
subgroups $\Gamma_1(N)$ of SL$(2,\mathbb{Z})$.

If the F-theory compactification on $M$ has a six-dimensional perturbative Heterotic limit, which requires that $M$ exhibits a (compatible) $K3$ fibration $\text{$K3$}\xhookrightarrow{} M\rightarrow \mathbb{P}^1_{\mathrm{b}}$, 
 the subset of the five-dimensional BPS spectrum with charges corresponding to curves in the $K3$ fiber can be checked against worldsheet supersymmetric indices of the Heterotic string in six or four dimensions.
In six dimensions, the corresponding invariants are encoded in the elliptic genus of Heterotic strings on $K3$, which is a lattice Jacobi form, while in four dimensions~\cite{Harvey:1995fq} it can be related to the {\sl new supersymmetric index} (NSI) of~\cite{Cecotti:1992qh}. 
The latter is essentially equivalent to the Noether-Leftshetz generator of the $K3$ fibration~\cite{Klemm:2004km,MR3114953,Enoki:2019deb}, which is a vector-valued modular form, and is in turn encoded in the genus zero topological string invariants that correspond to curves in the $K3$ fiber~\cite{Klemm:2004km}.  
The fact that this information can be extracted by mirror symmetry, as proven by Givental and Lian-Yau, led to a proof of the Yau-Zaslow formula~\cite{MR1398633} in terms of symmetric products of $K3$ for all classes~\cite{Klemm_2010}. 
While the elliptic genus and the new supersymmetric index contain equivalent information, they are very different automorphic objects.
One of the main results of this paper is, that they can in fact be directly translated into each other via a certain isomorphism between vector-valued modular forms and lattice Jacobi forms that has been first introduced in~\cite{MR2512363}.
This is explained in Section~\ref{sec:NLLJ} and illustrated at the hand of several examples in Sections~\ref{sec:examples} and~\ref{sec:genusone}.
It becomes particularly powerful in cases with a higher rank gauge group and Heterotic Wilson lines or when the gauge group is not realized perturbatively in the Heterotic duality frame.

If $M$ exhibits an elliptic and not only a genus one fibration, the base degree one topological string partition function $Z_{\beta=F}(\tau,\bm{z})$, with $F$ being the base of the $K3$ fiber and a curve of 
self-intersection $F^2=0$ in the  base $B$ for the genus one fibration, is identical to the elliptic genus~\footnote{Strictly speaking, it corresponds to a refinement of the elliptic genus with the topological string coupling as an additional elliptic parameter.}.
However, if $M$ is genus one fibered without a section but with $N$-sections, the Type IIA theory on $M$ is dual to Heterotic strings on $(K3\times T^2)/\mathbb{Z}_N$ and the corresponding new supersymmetric index
does not only contain information about the elliptic genus but also the twisted twined elliptic genera associated to the $K3$.
Similar to the topological string partition functions $Z_{\beta}$, the twisted twined elliptic genera are lattice Jacobi forms for $\Gamma_1(N)$.
However, as we discuss in Section~\ref{sec:genusone}, $Z_{\beta=F}$ is not identical to the untwisted untwined elliptic genus.
In particular, to reconstruct the BPS spectrum of Heterotic strings on $(K3\times S^1)/\mathbb{Z}_N$, the higher base degree invariants with $\beta=n\cdot F$ for $n=1,\dots,N$ have to be taken into account and encode different twisted sectors.
Nevertheless, the same isomorphism between vector-valued modular forms and lattice Jacobi forms can be applied to relate the twisted twined elliptic genera and the new supersymmetric index.
The only difference is, that the $\Gamma_1(N)$ lattice Jacobi forms are vector-valued lattice Jacobi forms under $\text{SL}(2,\mathbb{Z})$ transformations.

In cases where the $K3$ has itself an elliptic of genus one  fiberation, it might be viewed as comparable to the surface 
$S$ that arises for the non-critical strings.
However, there are two important differences.
The $K3$ fiber is not rigid, as $F^2=0$, and hence not contractable without simultaneously shrinking $M$.
This amounts to the fact that the tension of the Heterotic string can not be sent to zero without at the same time lowering the Planck scale such that in the tensionless limit the theory becomes trivial.
Moreover, the higher base degree partitions functions  $Z_{n\beta_0}(\tau,\bm{z})$ can be  obtained  from the primitive base degree one class $\beta_0$  contribution $Z_{\beta_0}(\tau,\bm{z})$, 
which is equivalent to the Heterotic elliptic genus, by applying Hecke transforms $T_n$ to $Z_{\beta_0}(\tau)$~\cite{Minahan:1998vr}.  
This is related to the fact that the $n^{(g)}_{\varphi,0}$, where $\varphi$ denotes the class of a curve in the $K3$ fiber, depend only on the self-intersection $\varphi^2$ of the class $\varphi$ in the corresponding Picard lattice.
The latter behaviour is captured by Noether-Lefschetz theory which we review in Section~\ref{sec:NLLJ}.
Physically, the properties just discussed can also be related to the fact that the Heterotic string has, as opposed to the non-critical strings considered in~\cite{Haghighat:2013gba,Kim:2014dza,Haghighat:2014vxa}, no non-trivial 
bound states.
As we already pointed out, a new observation in this paper is, that in the case of genus one fibered Calabi-Yau spaces with $N$-sections 
one needs, for a primitive class $\beta_0$ in the $K3$ fiber, information of  $Z_{k\beta_0}(\tau,\bm{z})$, $k=1,\ldots, N$ 
to reconstruct the NSI or equivalently the  Noether-Lefschetz generators, from which 
follow the topological string partition function $Z_{n\beta_0}(\tau,\bm{z})$ for all higher $n$.     

As further pointed out in~\cite{Kawai:2000px,Klemm:2004km} the higher genus informations can  be obtained using the Hilbert Scheme  formula for symmetric products of $K3$, an approach 
which also has been refined~\cite{Katz:2014uaa}.
There is considerable interest in the refinement  of the BPS invariants from $n^{(0)}_{\kappa}\in \mathbb{Z}$ to  
$N^{j_l,j_R}_\kappa\in \mathbb{N}$.
First of all, because it is the actual state count that should play a role in all of the weak gravity conjectures mentioned below, but also because the $N^{j_l,j_R}_\kappa$ decompose and therefore detect representations  of  symmetry groups  acting on the moduli 
space of the BPS states $\widehat{\cal M}_\kappa$, as for example the one of the  Matthieu group $M_{24}$ in~\cite{Katz:2014uaa}.
However, as reviewed in~\cite{Huang:2020dbh}, mathematicians currently struggle  with the canonical definition of the  orientation data  necessary to define the $N^{j_l,j_R}_\kappa\in \mathbb{N}$ in the compact case. 
Likewise a physical understanding in the supergravity context is, apart from the attempts in~\cite{Antoniadis_2014,Antoniadis_2015,Huang:2020dbh} to be developed.
This is in contrast to the local or gauge theory case, because here one has  a global $U(1)_R$ symmetry in the five dimensional Nekrasov background that allows the definition of a refined 
index~\cite{Nekrasov:2011bc} and a canonical choice of the  orientation data~\cite{Nekrasov:2014nea}, 
i.e. the  choice of the square root $(K_{Y,s})^\frac{1}{2}$ of the canonical bundle of the d-critical 
locus $(Y,s)$ in the moduli space of the BPS states $\widehat {\cal M}_\kappa$. The latter makes a mathematical definition 
of the $N_\kappa^{j_R,j_L}$ in the local case possible, as for example for the toric Calabi-Yau cases, where can be  
captured by  the Bialinicki-Birula decomposition of $\widehat {\cal M}_\kappa$  in the localisation formulas~\cite{Choi:2012jz}. 
In this work we extend the refinement conjecture of~\cite{Katz:2014uaa}, reviewed in Section \ref{sec:refined}, to 
many more examples of  elliptic $K3$ fibrations in Section \ref{sec:examples} as well as to the genus one fibrations  
in Section \ref{sec:genusone}.
As the $K3$ fiber can move, our new refinements results are somewhat in between the local cases and the generic compact cases.
Together with the techniques explained in the previous paragraph this allows us to reconstruct the NSI, the higher genus invariants as well as their refinements in many new cases and 
independent of  the question if there is a perturbative Heterotic dual.
                   
Another main purpose of this paper is to extend and use this good understanding of an important  sector of the spectrum of F- and M-theory compactifications on elliptic or genus one fiberd Calabi-Yau 3 folds\footnote{On 
higher dimensional elliptically fiberd Calabi-Yau manifolds, especially 4-folds the situation  for the monodromies and the
topological string amplitudes at genus zero and one  bears many similarities, which  is especially obvious if the latter  
exhibit a Calabi-Yau 3-fold fibration with a compatible genus one fibration structure of the above type.}, to prove 
the non-Abelian weak gravity conjecture that has been proposed in~\cite{Heidenreich:2015nta}.  

The weak gravity conjecture~\cite{ArkaniHamed:2006dz} (WGC) is based on the (argued) inconsistency of strongly gravitational quantum physics processes, which lead to a huge number of 
black hole remnants~\cite{Susskind:1995da}.
To avoid these problems and allow for possible decay channels, one requires that there must exist a species of superextremal states, i.e. their mass fulfills $m_{\text{el}}\lesssim  \sqrt{2}q\, g_{\text{el}}\, m_{\text{Pl}}$, where $q$ is the charge and $g_{\text{el}}$ is the gauge coupling. 
Together with electro-magnetic duality and with dynamics modelled on the  t'Hooft Polyakov monopole it also follows that the theory has as a cutoff scale $\Lambda\lesssim g_{\text{el}} m_{\text{Pl}}$.
This corroborates\footnote{The arguments are of course not independent as~\cite{Banks:2010zn} also ague with the unwanted 
stability  of remnants in the presence of global symmetries.} with the requirement that gravity cannot 
respect global symmetries, for which the arguments are more rigorous~\cite{Banks:2010zn,Harlow:2018tng}. 
In particular, when the limit $g_{\text{el}} \rightarrow 0$ is attempted  in an effective description 
of a  gauge theory embedded in a quantum theory of gravity, this should be at infinite distance in the moduli space and lead to one or both of the following scenarios:
\begin{enumerate}
	\item Gravity decouples, i.e. $m_{\text{Pl}}\rightarrow \infty$.
	\item A tower of light states appears, invalidating the effective  description of the theory. 
\end{enumerate}
In particular, the states in the second scenario are conjectured to always be excitation of a string that becomes weakly coupled in the limit~\cite{Lee:2019wij}.

In string theory, the latter breakdowns are argued to be inevitable~\cite{Ooguri:2006in} as the natural metric given by the kinetic terms of 
the moduli field in ${\cal M}$ is not geodesically closed and starting with a string  effective action at a point $p_{0}$ 
in the bulk of the moduli space ${\cal M}$ and moving by a distance $d(p,p_0)$ in the natural  
metric, the masses of infinite towers of  states  evolve in typical~\cite{Ooguri:2006in}, and potentially 
all situations like $m_p=m_{p_0} e^{-\alpha d(p,p_0)}$.
Here  $\alpha$ has been argued to be of order one in natural units in the refined swampland distance conjecture~\cite{Klaewer:2016kiy,Blumenhagen:2018nts}. 
The breakdown of the effective description due to a tower of states also suggests a refinement of the WGC to the so-called sublattice weak gravity conjecture (sLWGC) proposed in~\cite{Heidenreich:2015nta,Heidenreich:2016aqi}.
Based on a derivation in perturbative string theory and stability under Kaluza-Klein reduction, the sLWGC states that there exists a full rank sublattice $\Gamma_{\text{ext}}\subset\Gamma$ of the charge lattice $\Gamma$,
such that for every $\vec{q}\in\Gamma_{\text{ext}}$ there is a (not necessarily stable) particle of charge $\vec{q}$.

Compactifications on Calabi-Yau manifolds with $N=2$ supersymmetry have proven to be a rich testing ground for the swampland distance conjectures.
The metric on the four-dimensional vectormultiplet moduli space that arises from Type IIA/B strings on Calabi-Yau threefolds, can be readily calculated and the physics associated to the various limiting points is relatively well understood.
In this way the distance conjectures have been tested in the K\"ahler moduli space~\cite{Blumenhagen:2018nts,Corvilain:2018lgw,Joshi:2019nzi,Erkinger:2019umg} and complex structure moduli space~\cite{Bastian:2020egp}.
An intriguing connection has also been made between the distance of limiting points in the complex structure moduli space, the associated limiting Hodge structure and monodromy orbits of brane charges~\cite{Grimm:2018ohb,Grimm:2018cpv}.
While the conjectured identification between the brane orbits and the inifinite towers of states still faces some puzzles, the relation to mixed Hodge structures exemplifies the deep mathematical nature of the conjectures and might hint towards a general proof in this setup.
Using mirror symmetry these results have also been extended to the K\"ahler moduli space~\cite{Corvilain:2018lgw}.
The hypermultiplet moduli space has been less studied but results in the context of Type IIB compactifications can be found in~\cite{Marchesano:2019ifh,Baume:2019sry}.

The infinite tower of states at points of maximal unipotent  monodromy (MUM), that exists generically in Calabi-Yau 3-folds 
compactification at infinite distance in their complex moduli space~\footnote{Very rare one parameter models called
orphans do not exhibit a MUM point and are currently the only known exceptions~\cite{MR3951103}.}  ${\cal M}_{cs}$  has been explained as being mirror dual  to the BPS 
bound states of D2-D0 branes as counted by the topological string on the mirror by the unrefined BPS indices~\cite{Joshi:2019nzi} in the large volume limit.
The states that are counted by the standard Fourier expansion of the Jacobi forms $Z_\beta(\tau, \bm{z})$, the elliptic genera, or the NSI in the fibered  CY-3 folds 
are also such large volume states. While by the nature of the expansion the base curves  are 
always assumed to  have large  volume, the  information encoded the modular and the elliptic 
argument in $Z_\beta(\tau, \bm{z})$ is exact at least for each $\beta$.
This  allows  stringent and relatively easy checks of some of the weak gravity conjectures.
For example, as in 6d F-theory  the volume of the base sets the Planck scale  ${\rm vol} \sim m_{\text{Pl}}^4$ and the 
volume of a base curve $C_{\beta^2}$ supporting a 7-brane Yang-Mills theory sets the Yang-Mills coupling to
$ {\rm vol} (C^{\text{YM}}_{\beta})\sim g_{\text{el}}^{-2}\sim {\rm vol}$, one can attempt to challenge the weak gravity 
conjecture by holding  ${\rm vol} (B)$ finite, while sending 
${\rm vol} (C^{\text{YM}}_{\beta})\rightarrow \infty$. It was meticulously shown in~\cite{Lee:2018urn} that in this scenario $\beta^2=0$ and the $K3$ in the limit leads to a light Heterotic string spectrum with small string tension,  i.e. scenario $2)$.
Using the properties of lattice Jacobi forms and a careful analysis of the modified extremality bound for dilatonic Reissner-Nordstr\"om black holes, this led to a proof of the sLWGC for F-theory on elliptically fibered Calabi-Yau threefolds~\cite{Lee:2018urn,Lee:2018spm}.

The original sLWGC applies to the charges of particles under Abelian factors of the gauge group and thus in particular to non-Abelian gauge groups on the Coulomb branch.
However, a subsequent refinement~\cite{Heidenreich:2017sim}, which is again based on a proof in perturbative string theory, makes an even stronger claim for theories with non-Abelian gauge groups:\\\\
\noindent \textbf{The non-Abelian sLWGC} For any quantum gravity in $d\geq 5$ dimensions with zero cosmological constant and unbroken gauge group $G$, there is a finite-index Weyl-invariant sublattice $\Gamma$ of the weight lattice $L_{\text{w}}(G)$ such that for every dominant weight $\bm{\lambda}_{R} \in \Gamma$ there is a superextremal resonance transforming in the $G$ irreducible representation $R$ with highest weight $\bm{\lambda}_R$.\\\\
While for Abelian gauge groups this reduces to the sLWGC, the implications for non-Abelian theories are stronger and suggest a unified strong coupling scale for the gauge theory and gravity~\cite{Heidenreich:2017sim}.
In this paper we build upon the proof of the sLWGC by~\cite{Lee:2018urn,Lee:2018spm} and, deriving a novel relation among lattice theta functions and Weyl characters, extend it to show that F-theory on Calabi-Yau threefolds indeed satisfied the non-Abelian sLWGC.
This is done in Section~\ref{sec:nawg} and illustrated at the hand of multiple examples in Section~\ref{sec:examples}.
We discuss the validity for genus one fibrations in~\ref{sec:genusone}.

\vskip 3 mm
\noindent
{\bf Acknowledgement:}  It is a pleasure to  thank Min-xin Huang, Sheldon Katz and Rahul Pandharipande 
for discussions and collaborations on questions regarding the refinement of BPS indices for 
compact Calabi-Yau 3-folds.  We especially  thank Don Zagier for his insights and suggestions to manipulate theta functions and sums 
over dominant weights, Jie Gu for providing us with a Mathematica program constructing Weyl invariant Jacobi forms 
and Timo Weigand for explaining  aspects of the sublattice weak gravity conjecture in the context of the Heterotic/F-theory duality.  
We further like to thank Fabian Fischbach, Babak Haghighat,  Amir Kashani-Poor, Georg Oberdieck, Kaiwen Sun, Haowu Wang and Xin Wang for 
discussion on related subjects. The work of Thorsten Schimannek is supported by the Austrian Science Fund (FWF):P30904-N27. Cesar Fierro Cota would 
like to thank the financial support from the fellowship  ``Regierungsstipendiaten CONACYT-DAAD mit Mexiko'' under the grant 
number 2014 (50015952) and the Bonn-Cologne Graduate School of Physics and Astronomy for their generous support.

\section{BPS saturated amplitudes in Heterotic-Type IIA duality}
\label{sec:basics}
A key  quantity for our tests of the weak gravity conjectures and the starting point for our refinement  
is the BPS saturated one loop amplitude  that leads to the effective gravitional couplings
 \begin{equation}
        \label{eqn:gravcouplings}
                 I_g =\int_{M_4} \mathrm{d}^4x F_g(\bm{t},\bar{\bm{t}}) T_-^{2g-2} R_-^2\,, \quad g\in \mathbb{Z}_{\geq 0} \, 
        \end{equation}
in the 4d $\mathcal{N}=2$ supergravity action~\cite{Bershadsky:1993cx,Antoniadis:1993ze,Antoniadis:1995zn}.
Here $T_-$ is the anti-self-dual part  of the graviphoton field strength, $R_-$ is the anti-self-dual part of the 
Riemann curvature tensor, and $F_g(\bm{t},\bar{\bm{t}})$ is an effective gravitational coupling that  depends at generic  points in the moduli 
space\footnote{A mixing  of the complex structure (hyper multiplet) and K\"ahler structure (vector multiplet) 
moduli dependence is expected e.g. at singular points that allow transitions between Calabi-Yau spaces.}          
only on the vector multiplet scalars $\bm{t},\bar{\bm{t}}$.  
 
Conjectural Heterotic/Type IIA dual pairs in 4d with $\mathcal{N}=2$ supersymmetry 
involve the Heterotic string on $K3 \times T^2$ and the Type IIA string on a $K3$ fibered 
Calabi-Yau 3-fold $M$. The  $I_g$ can be calculated at least with partial moduli dependence   
on both sides and serve as an important check of the duality. In the perturbative regime of 
the Heterotic string $g_{\text{het}}\to 0$,  all 
$F_g$ can  be calculated by a BPS saturated one-loop amplitude, which depends in general on 
all vectors multiplet moduli except, the ones in supergravity  multiplets and the Heterotic dilaton 
\begin{equation}
S_{\text{het}} = \frac{4\pi}{g^2_{\text{het}}}+ i \theta\,.
\end{equation}
Typical moduli from Abelian Heterotic vector multiplets  are  the K\"ahler structure $T$ and the 
complex structure $U$ of the two torus $T^2$.  Depending on the gauge bundle 
configurations on the  Heterotic side, one can have up to 15 more perturbative Abelian vector 
multiplets $\bm{V} = (V_1,\ldots, V_r)$ from the unbroken gauge group (we consider mainly  the $E_8\times E_8$ version) 
$G \subset E_8 \times E_8$, where $r = \text{rk} (G)$, whose moduli  correspond to 
Wilson lines  along the cycles of $T^2$.
The Heterotic one loop computation involves an integration of the worldsheet complex structure over the fundamental domain, that can be solved 
by the unfolding trick~\cite{Dixon:1990pc}, which is more systematically implemented by the lattice reduction method 
of Borcherds~\cite{MR1625724,Marino:1998pg}. Examples of these calculations can be found in~\cite{Marino:1998pg,Klemm:2005pd,Weiss:2007tk,Enoki:2019deb}, 
as well as recent extensions to CHL-Heterotic orbifolds on $(K3\times T^2)/\mathbb{Z}_N$ \cite{Chattopadhyaya:2017zul,Banlaki:2019bxr,Chattopadhyaya:2020qqq}. In the 
perturbative Heterotic limit, the Borcherds lift calculation implies that $F_g(\bm{t},\bar{\bm{t}})$ are automorphic forms under the $T$-duality group $SO(2+r,2;\mathbb{Z})$. The latter group acts on the combined moduli space
\begin{equation}
\label{eqn:modspace}
\mathcal{M}_{T,U,\bm{V}}= \frac{SO(2,2+r)}{SO(2)\times SO(2+r)}\Big/SO(2+r,2;\mathbb{Z})\,,
\end{equation}
which is spanned by the moduli $(T,U,\bm{V})$ \cite{CARDOSO199468,de_Wit_1995}.

On the Type IIA side, the holomorphic limits of the $F_g(\bm{t}) = \lim_{\bar{\bm{t}} \to \infty} F_g(\bm{t},\bar{\bm{t}})$ 
are calculated by the  genus $g$ contribution of the topological string expansion. The latter localizes on holomorphic 
maps and  the $F_g(\bm{t})=\sum_{\kappa\in H_2(M,\mathbb{Z})} r_g^\beta e^{2 \pi i \bm{t}\cdot \beta}$   are generating 
functions of Gromov-Witten invariants $r_g^\kappa\in \mathbb{Q}$ for these holomorphic maps $\Phi: \Sigma_g \rightarrow  C_\beta \subset M$. 
Here $\Sigma_g$ denotes  a closed oriented  world-sheet of genus $g$ that is mapped by $\Phi$ into 
a holomorphic curve $C_\kappa$ in the homology class $\kappa\in H_2(M,\mathbb{Z})$. Note that  in order to calculate the $F_g$,
$M$ can be a generic Calabi-Yau threefold, not necessarily $K3$ fibered. We insist however that it has  a mirror 
manifold $W$, in our case given by the Batyrev(--Borisov) construction. This is to ensure that part of the 
actual calculation of the $F_g(\bm{t},\bar{\bm{t}} )$ can be done by mirror symmetry, i.e. using the period geometry 
and the holomorphic anomaly equations on $W$. In this  way we can get fairly easily the $F_0(\bm{t})$ and 
$F_1(\bm{t},\bar{\bm{t}} )$ contributions. The higher genus $F_g$ can be calculated using the holomorphic 
anomaly equations~\cite{Bershadsky:1993cx} and boundary conditions~\cite{Huang:2006hq} at the singular 
divisors in the moduli space.
However, automorphic symmetries that can the established for $Z=\exp(F(\lambda,\bm{t}))$ for Calabi-Yau manifolds 
with   genus one --- and  $K3$ fibration structures on the Type II side allow to use Jacobi-form ans\"atze for base degree 
expansion as in (\ref{eq:baseexpansion}) of the topological string partition function  $Z(\lambda,\bm{t})$, that together with $F_0,F_1$  
and vanishing conditions determines at least the first terms in this expansion exactly to all genus~\cite{Huang:2015ada}.

The Schwinger-loop calculation takes place in supergravity and is therefore universal on the Heterotic and on the Type II side. On the latter it  relates the topological string  
free energy $F(\lambda,\bm{t})$ to the BPS saturated amplitude~\cite{Gopakumar:1998ii,Gopakumar:1998jq} \footnote{We omitted in (\ref{eqn:multicovering}) the classical terms coming from $F_0(\bm{t})$ and $F_1(\bm{t})$. For an updated review on topological string theory and mirror symmetry, we refer the reader to the survey \cite{MR3967072}.} 
 \begin{equation}
      \label{eqn:multicovering}
               F(\lambda,\bm{t}) =\sum_{g=0}^\infty F_g (\bm{t}) \lambda^{2g-2}=\sum_{\kappa \in H_2(X,\mathbb{Z})} \sum_{g=0}^\infty \sum_{m=1}^\infty \frac{n^g_\kappa}{m}\left(2\sin\left(\frac{m\lambda}{2}\right)         \right)^{2g-2} \mathrm{e}^{2\pi i m  \bm{t} \cdot \kappa  }\,
      \end{equation}
 and allows therefore to extract the BPS indices $n^g_\kappa\in \mathbb{Z}$. Geometrically,  the latter
 have on the Type IIA side a definition in terms of the dimensions  of cohomology groups of the moduli space $\widehat {\cal M}_\kappa$ 
 of an M2 brane that wraps the curve $C_\kappa$. On this cohomology  $H^*_\kappa $ one has a $\mathrm{Spin}(4)\sim\mathrm{SU}(2)_L\times \mathrm{SU}(2)_R$ 
Lefschetz action.   It splits  the cohomology $H^*_\kappa$   into irreducible representations of this group $\mathrm{SU}(2)_L\times \mathrm{SU}(2)_R$,  
 denoted by their highest spin  $[j_L]$ and  $[j_R]$, as  $H^*_\kappa=\bigoplus_{j_L,j_R}N_\kappa^{j_L j_R} [j_L] [j_R]$. 
 The corresponding multiplicities  are denoted by  $N_\kappa^{j_L j_R}\in \mathbb{N}$. Note that the group $\mathrm{SU}(2)_L\times \mathrm{SU(2)}_R$  
 can be identified with the little  group of the 5d Lorentz representations of the effective theory of M-theory compactification on $X$ and 
 the  $N_\kappa^{j_L j_R}$ are multiplicities of 5d BPS states with indicated spin and a mass proportional to the charge  $[\kappa]$.
 In particular, these are the relevant states that run in the loop of the  amplitude that leads to  (\ref{eqn:gravcouplings}).
 Notice further that  the right spin representations of the states do not exhibit a spin dependent  coupling to $R_+$ and $F_+$  in this amplitude. 
 Therefore their contribution is obtained  by merely summing over their multiplicities with alternating sign due to the spin statistic factor, i.e. what 
 actually contributes is the trace        
 \begin{equation} 
{\rm Tr}_{[j_R]\subset H_\kappa} (-1)^{F_R} =  \sum_{j_R,j_L} (-1)^{2 j_R}(2 j_R+1) N_{\kappa}^{j_L,j_R}[ j_L]=\sum_{g=0}^\infty n^g_\kappa \left[L_g \right]\, . 
\label{eq:rightsum}
\end{equation}
A more detailed analysis reveals that the relevant $\mathrm{SU}(2)_L$ representation that contributes canonically 
is indeed $[L_g]=\left(2 [0_L]+\left[\frac{1}{2}_L\right]\right)^{\otimes g}$. This explains 
the way in which the BPS indices $n^g_\kappa$ contribute to (\ref{eqn:multicovering}) and (\ref{eqn:gravcouplings}). 

For important  physical considerations, as the counting of states that contribute to the microscopic entropy 
of ${\cal N}=2$ black  holes or the existence of superextremal states as demanded by  the weak 
gravity conjecture, the actual multiplicities $N_\kappa^{j_L j_R }\in \mathbb{N}$ of states in the charge 
sector defined by $\kappa$ is the canonical quantity. However  for generic compact Calabi-Yau 3 folds 
this quantity depends not only on the K\"ahler structure --- where it is defined for large radius --- but also on the complex structure as explained in~\cite{Huang:2020dbh}.
A lot of effort has been made to give a  general mathematical definition of  $N_\kappa^{j_L j_R }\in \mathbb{N}$, their piecewise stability 
as well as calculation techniques in certain mostly non-compact or $N=4$ supersymmetric 
circumstances~\cite{Huang:2020dbh}. Certainly  the knowledge of  $N_\kappa^{j_L j_R }\in \mathbb{N}$ degeneracies as 
contrasted with  their index $n_\kappa^g$ on compact Calabi-Yau manifolds is important for 
quantum gravity questions. In this  paper we follow a conjecture of~\cite{MR3524167} to 
extract the  $N_\kappa^{j_L j_R }\in \mathbb{N}$ for $K3$ fiberations with higher rank of the 
Picard group and for Type IIA theories, which are dual to $\mathbb{Z}_N$ CHL Heterotic strings, i.e.
with $N$-sections. This sheds light on the question, whether the cancellation is small, i.e. 
whether the  $n_\kappa^g$ are a good measure for the actual degeneracies and the effect
of the cancellation is sub leading for high $[\kappa]^2$.               

In the base expansion the volumes of base curves  ${\rm vol}(C_\beta)\sim \rm{ Im} (T_i)$  are assumed to be large, i.e. $Q_i=e^{2 \pi i T_i}$  is small. 
The large volume BPS states are therefore encoded in the Fourier expansion  of small $q^{2 \pi i \tau}$ and 
$\zeta_i=e^{2 \pi i z_i}$ of the Jacobi forms. I.e. an expansion where also the genus one fiber, its sections  
over the base curves and the curves whose volumes correspond to the exceptional curve classes that 
correspond to the  Cartan K\"ahler parameters of non-Abelian  gauge groups are large.  In this case 
$[\kappa]=[\beta,\phi]\in H_2(M,\mathbb{Z})$ denotes now the classes in $H_2(M,\mathbb{Z})$ distinguished  
w.r.t. to $[\beta]\in H_2(M,\mathbb{Z})$ while $[\phi]=[\tau, \zeta]$ consists  of the fiber class $[\tau]$, its sections 
and the  Cartan K\"ahler classes  $[\zeta]$ of $M$. For an  in-depth discussion of these curve classes 
and their dual divisor classes, see~\cite{Schimannek:2019ijf,Cota:2019cjx}.

A strong non-trivial check of Heterotic-Type II duality, with respective compactifications $K3\times T^2 \leftrightarrow M$, is comparing the Heterotic 1-loop amplitude with the topological free energy (\ref{eqn:multicovering}). This is done by choosing an appropriate basis  of K\"ahler moduli in $M$, such that $\bm{t} = (S_{\text{het}},T,U,\bm{V})$. In particular, we identify $S_{\text{het}} = \text{Vol}_{\mathbb{C}}\left( \mathbb{P}_{\mathrm{b}}^1\right)$. Thus, in the holomorphic limit, the assertion of Heterotic-Type II duality reduces  in the perturbative regime to
    \begin{equation}
     \label{eqn:HetTypII}
       \lim_{S_{\text{het}}\to \infty} F(\lambda,S_{\text{het}},T, U,\bm{V}) = F^{\text{1-loop}}_{\text{het}}(\lambda,T,U,\bm{V})\,.
    \end{equation}
The Heterotic one loop calculation from the worldsheet point has been first described in~\cite{Antoniadis:1995zn} and concretely 
evaluated for the STU-model~\cite{Kachru:1995wm} in~\cite{Marino:1998pg}.  The topological string zero sector for the
corresponding STU  Calabi-Yau 3-fold $M$, the standard elliptic fibration  over the Hirzebruch  surface $\mathbb{F}_2$ realised as resolved hypersurface 
of degree 24  in the weighted projective space $\mathbb{P}^4(1,1,2,8,12)$, has been solved  by mirror symmetry in~\cite{Hosono:1993qy}.

The Heterotic one-loop amplitude is calculated typically for an orbifold of an  $T^4\times T^2$  Heterotic worldsheet theory and  
splits, according to the number of insertions of the anti-self-dual graviphoton  $T_-^{2g}$ into  
(\ref{eqn:gravcouplings}), into integrals over  $SL(2,\mathbb{Z})$ invariants integrands over it's fundamental region 
${\cal F}$ ~\cite{Antoniadis:1995zn,Marino:1998pg,Klemm:2005pd}
\begin{equation} 
F_g=\int_{\cal F} d^2\tau \tau_2^{2g-3} \frac{{\hat {\cal P}}_g(\tau)}{Y^{g-1}} \sum_J  \overline \Theta_{\Gamma_J}^g(\tau) f_J(\tau) . 
\label{eq:Narain}
\end{equation}   
Here $J$ is the label of the orbifold sectors,  $\tau_2={\rm Im}(\tau)$, $Y$ is the space time modular real expression, which goes into the Heterotic K\"ahler potential 
as $K=-\log(Y)$,  ${\cal P}_g(\tau)$ are combinations of almost holomorphic Eisenstein series generated by  
\begin{equation}
\frac{e^{-\pi \lambda^2 \tau_2}}{\phi_{-2,1}(\tau,\lambda)} =\sum_{g=0}^\infty (2 \pi \lambda)^{2g-2} \hat {\cal P}_g(\tau),     
\end{equation}      
$f_J(\tau)$ is twisted  and twined oscillator sum depending on the orbifold sector. Here $\phi_{-2,1}(\tau,\lambda)$ is the weak Jacobi form defined in (\ref{eqn:weakJac}). This sector  dependence is also reflected in the twisted and twined  
Narain theta function $\Theta_{\Gamma_J}^g(\tau)$  defined for the invariant Narain Lattice $\Gamma_J$ of  signature $(b^+,b^-)$  with an insertion of the right 
momentum as 
\begin{equation} 
\label{eqn:SiegelNarain}
\Theta_{\Gamma_J}^g(\tau)=\sum_{p\in \Gamma_J} p_R^{2g-2} \exp\left(\pi i (p+\beta_J/2)_+^2+ \pi i \bar \tau (p+\beta_J)_-^2+\pi i(p+\beta_J/2,\alpha_J)\right)\ ,       
\end{equation}
where the $\pm$ indices denote the projection into an orthogonal decomposition of the $\Gamma_J$ into 
$P:\Gamma_J\otimes \mathbb{R}\simeq \mathbb{R}^{b_+}\perp \mathbb{R}^{b_-}$.   It is important to 
note  that integrand in (\ref{eq:Narain}) is also spacetime modular invariant. Combining this spacetime 
modular invariance  with the worldsheet modular invariance leads to the unfolding trick that results in 
a simpler integrand over a simpler domain and makes the integration 
possible~\cite{Harvey:1995fq,Marino:1998pg,MR1625724}.

It has been realised~\cite{Klemm:2004km}  that in the standard cases the all genus information on the right hand side of  (\ref{eqn:HetTypII})  
can be obtained from base degree zero genus zero sector of the topological string on $M$  encode  in $n^{(0)}_{\varphi,0}$, where $\varphi$ denotes all
classes in the $K3$ fiber,  combined with the product formulas for the Betti numbers of the Hilbert  scheme for symmetric product of the 
$K3$ fiber~\cite{Katz:1999xq,MR1838446} and information of the embedding of the second homology classes of the Calabi-Yau 3-form in the 
Picard lattice of the generic fiber. 

Moreover, the genus zero information has been noticed in~\cite{Klemm:2004km}  to be equivalent to the information  
in the loop amplitude that yields the threshold correction in the Heterotic string.   This amplitude has been calculated 
for the STU model in~\cite{Harvey:1995fq}. It is quite similar to the amplitude (\ref{eq:Narain}) and is integrated with the 
same methods discussed above. This yields  indeed the genus zero BPS numbers of~\cite{Hosono:1993qy}.  
The point of~\cite{Harvey:1995fq} was to relate this BPS saturated amplitude to an BPS index, which is in this case  
the so-called new supersymmetric  index of the Heterotic string~\cite{Cecotti_1992,Enoki:2019deb}
\begin{equation}
\mathcal{Z}(\tau,\bar{\tau}) = \frac{1}{\eta^2(\tau)} \text{Tr}_R (-1)^F F q^{L_0-\frac{c}{24}} \bar{q}^{\bar{L}_0 -\frac{\bar{c}}{24}}\ .
	\label{eqn:nssi1}
\end{equation}
Geometrically the genus zero base degree  zero information is equivalent to the Noether-Lefschetz generators of the $K3$ fibration~\cite{MR3114953}  
and this fact has been used to proof the SYZ conjecture~\cite{MR1398633}  in~\cite{MR2669707} as will explained in more detail 
in the next section.  

However there is a counter example to the claim that the genus zero base zero BPS invariants 
allow always to reconstruct the r.h.s. of~\ref{eqn:HetTypII}  namely  the Enriques Calabi-Yau 
manifold with $SU(2) \times \mathbb{Z}_2$ holonomy  which has  trivial genus zero instanton 
sector and so is the new supersymmetric index (\ref{eqn:nssi1})  due to the enhanced super symmetry in 
the genus zero amplitude. Nevertheless, the  Heterotic calculation has been performed 
in~\cite{Klemm:2005pd,Grimm:2007tm} for all ten fiber classes of the  Enriques Calabi Yau and an iteration  
for higher  base wrapping has been given. The results agree with the higher genus BPS numbers for 
a reduced Type II Calabi-Yau 3-fold compactification with two K\"ahler classes in the fiber.

Of course, the left hand side of (\ref{eqn:HetTypII}) as obtained by the all $n^{(g)}_{\varphi,0}$  is still a holomorphic function, while the  full one loop amplitude is real, 
but it has been shown in~\cite{Antoniadis:1993ze} and more completely in \cite{Grimm:2007tm} that the Heterotic amplitude fullfills the  holomorphic anomaly 
equations of~\cite{Bershadsky:1993cx}, so that the full information can be reconstructed on both sides of  (\ref{eqn:HetTypII}).

In the next section we argue that a simple calculation in mirror symmetry provides the necessary 
information to reconstruct the automorphic forms for $SO(2,2+r;\mathbb{Z})$ in (\ref{eqn:HetTypII}). There, the computations 
rely purely on the geometrical information provided by the $K3$ fibrations. In the section \ref{sec:refined} we follow up with
the refinement of the formulas  (\ref{eqn:gravcouplings}, \ref{eqn:multicovering}) to (\ref{eqn:refamps}, \ref{eqn:refmulticovering}),
and discuss the extension of the Heterotic Type IIA   duality at the basis of the refined invariants.  

\section{Higher Rank Noether-Lefschetz loci and Lattice Jacobi forms}
\label{sec:NLLJ}  

As we reviewed above, on the Heterotic side the BPS protected observables are encoded in the so-called new supersymmetric index which is a vector-valued modular form.
From the Type II perspective this can be identified with the generator of Noether-Lefschetz numbers that is associated to a lattice polarized $K3$ surfaces.
The lattice polarization is determined by the embedding of the vertical cohomology (see below) of the $K3$ fibered Calabi-Yau threefold into the cohomology of the $K3$ fiber.
The information contained in the Noether-Lefschetz numbers is equivalent to the knowledge of the Gopakumar-Vafa invariants of degree one  
with respect to the $\mathbb{P}^1$ base of the $K3$ fibration.

When the Calabi-Yau exhibits also an elliptic or genus one fibration structure, the Gopakumar-Vafa invariants are in turn encoded in lattice Jacobi forms.
More precisely, the topological string partition function $Z_{\text{top.}}$ admits an expansion
\begin{align}
\label{eq:baseexpansion}
	Z_{\text{top.}}(\lambda,\tau,\bm{z},\bm{t})=Z_0(\lambda,\tau,\bm{z})\cdot\left[1+\sum\limits_{\beta\in H_2(B)}Z_\beta(\tau,\lambda,\bm{z})\exp\left(2\pi i\beta_i t^i\right)\right]\,,
\end{align}
where $\lambda$ is the topological string coupling, $\tau$ is proportional to the complexified volume of the generic fiber, $\bm{z}$ are complexified volumes of components of reducible fibers that on the Heterotic side correspond to Wilson line parameters and $\bm{t}$
are the complexified volumes of curves in the base of the fibration.
The coefficients $Z_\beta(\tau,\lambda,\bm{z})$ are lattice Jacobi forms for a congruence subgroup of $\text{SL}(2,\mathbb{Z})$ and $\tau$ acts as the modular parameter while $\lambda$ and $\bm{z}$ are elliptic parameters.
At least for low base degrees they can be fixed with the knowledge of genus zero Gopakumar-Vafa invariants which can in turn be calculated using mirror symmetry.
This has been called the \textit{modular bootstrap}.

If the genus one fibration is compatible with the $K3$ fibration, the base of the genus one fibration has to be a Hirzebruch surface.
In case the fibration exhibits a section and is therefore elliptic, the elliptic genus of the Heterotic string corresponds to the base degree one topological string partition function 
$Z_{\beta}$, where $\beta=F$ is the fiber of the Hirzebruch base.
However, we will show in Section~\ref{sec:genusone} that if the $K3$ fibration only exhibits a compatible genus one fibration with $N$-sections, different twisted sectors that contribute to the elliptic genus are captured by the partition functions $Z_{n\cdot F},\,n=1,\dots,N$.

In this section we will review Noether-Lefschetz theory as well as the theory of lattice Jacobi forms and clarify the relation between the new supersymmetric index (Noether-Lefschetz generator), which is a vector-valued modular form, and the elliptic genus (that for elliptic fibration corresponds to the topological string partition function $Z_F$) which can be expressed in terms of lattice Jacobi forms.
Our discussion in this paper will not only accomodate multiple Wilson line/elliptic parameters but also incorporate the following two generalizations:
\begin{itemize}
\item We allow twists of the coroot lattice. In the language of~\cite{Bershadsky:1996nh} these signal non-perturbative gauge symmetry on the Heterotic side.
\item In Section~\ref{sec:genusone} we include Type II compactifications on genus one fibrations that are not elliptic. On the Heterotic side this corresponds to a compactification on $(K3\times T^2)/\mathbb{Z}_N$.
\end{itemize}
For the elliptic cases we demonstrate how the Noether-Lefschetz generator can easily be obtained using the modular bootstrap.
Note that a direct calculation of the new supersymmetric index using Heterotic techniques is not available when the gauge symmetry is not perturbative.
Moreover, Heterotic compactifications on $(K3\times T^2)/\mathbb{Z}_N$ have only been considered for $N\le 3$.

First we are going to review the Noether-Lefschetz theory associated to polarized $K3$ surfaces and $K3$ fibrations.
We will then use the theory of Jacobi forms and theta blocks to obtain the Noether-Lefschetz generators efficiently from the topological string partition functions.

\subsection{Noether-Lefschetz theory}
A $K3$ surface with generic complex structure has a Picard lattice
\begin{align}
\text{Pic}(S)= H^2(S,\mathbb{Z}) \cap H^{1,1}(S,\mathbb{C})\,,
\end{align}
of rank zero and therefore does not contain any holomorphic curves. We are mainly interested 
in enumerative problems associated to $K3$ manifolds and review the important ones. Since the Gromov-Witten invariants 
are not sensitive to complex structure deformations, one expects that the ordinary Gromov-Witten 
theory associated to $K3$ surfaces is trivial. This picture is confirmed  by standard application of mirror 
symmetry to $K3$ surfaces, which reveals that at the MUM points in the mirror dual family the double 
logarithmic solutions receive no instanton corrections. A beautiful BPS number count  has been 
associated to projective $K3$ surface  $S$ with an h-dimensional linear system of genus h curves that is 
generically degenerate to a rational curve with $h$  simple nodes~\cite{MR1398633}. The numbers of 
such degenerate rational curves $C$ with self intersection  $C^2=2h-2$ of formal genus $h$ is the $q^h$ 
coefficient of the generating functions  $q=e^{2 \pi i \tau}$ 
\begin{equation} 
\sum_{n\ge 0} \chi({\rm Hilb}^n(S)) q^n =\prod_{n=1}^\infty\frac{1}{(1-q^n)^{24}}=\frac{q}{\eta^{24}(q)} \ .        
\label{YZ} 
\end{equation}
In the five dimensional theory that is obtained by compactifying M-theory on $K3\times T^2$  the 
BPS states contributing to Gopakumar-Vafa invariants  or the BPS index vanish due a multiplicative factor 
$[L_R]=\left(2 [0_R]+\left[\frac{1}{2}_R\right]\right)$  that comes  form the $T^2$ factor and  vanishes in 
the trace (\ref{eq:rightsum}) over  the Lefschetz decompositions of the cohomology of the moduli 
space ${\widehat  {\cal M}_\kappa}$~\cite{Katz:1999xq}. Removing this zero and  using 
the Lefschetz decomposition of the cohomology of  ${\rm Hilb}^n(S)$  as well  as the sum over the 
right spins as in (\ref{eq:rightsum}) one can associate invariants for higher genus curves  
$n^g_{\varphi,0}$ to the fiber $S$~\cite{Katz:1999xq}.  This way one obtains  with\footnote{Here $2i\sin(ix) =-2 \sinh(x)= y^{-1}-y$ and $\chi_y(X)=\sum_{p,q} =y^p (-1)^q h^{pq}(X)$ is the Hirzebruch $y$ genus.}    $y=e^{i \lambda}=e^{ 2\pi i z}$ as an extension of (\ref{YZ}) the generating function  
\begin{equation} 
\sum_{n\ge 0} \chi_y({\rm Hilb}^n(S)) q^n = \prod_{n=1}^\infty \frac{1}{(1- y q^n)^2 (1-q^n)^{20}(1-y^{-1} q^n)^2}=\sum_{\kappa=0\atop g=0}^\infty n_{\varphi,0}^g \left(2 \sin\left(\frac{\lambda}{2}\right) \right)^{2g} q^\kappa \ ,
\label{KKV}
\end{equation} 
which can be suitably rewritten in terms of weak Jacobi forms (\ref{eqn:weakJac}) as in (\ref{eqn:KKV}).
This can be extended in two ways. First to include the wrappings of the  $T^2$. With the elliptic genus of $X$ 
\begin{equation} 
{\rm Ell}(X,\tau,z)={\rm Tr}_{H^*(X)}(-1)^F y^{F_L} q^H=\sum_{m\ge 0,l\in \mathbb{Z}} c_X(m,l) q^m y^l
\end{equation}
one can calculate ${\rm Ell}(S,\tau,z)=2 \phi_{0,1}(\tau,z)=2 y+20 +2/y+{\cal O}(q)$ and further with  the orbifold formula for the elliptic genus of the  
desingularised symmetric product~\cite{Dijkgraaf:1989hb}  denoting $p=e^{2 \pi i u}$ one gets 
\begin{equation} 
G(\tau,z,u)=\sum_{m=0}^\infty {\rm Ell}({\rm Sym}^m(S), \tau,z) p^m=\prod_{m>0,n\ge0,l\in \mathbb{Z}}\frac{1}{(1-p^nq^my^l)^{c_S(m n, l)}}
\end{equation}  
 one obtains the generating function  
\begin{equation} 
\sum_{\varphi,\beta=0\atop g=0}^\infty n_{\varphi,\beta}^g \left(2 \sin\left(\frac{\lambda}{2}\right) \right)^{2g-2} q^\varphi p^\beta 
=-\frac{ G(\tau,z,u)}{\phi_{-2,1}(\tau,z)} \ .
\label{KKV2}
\end{equation}  
for the reduced BPS indices on the $K3\times T^2$~\cite{Katz:1999xq}. 

A second extension of (\ref{KKV})  is the so called refinement which keeps now track  also  of the right spin in the 
$SU(2)_L\times  SU(2)_R$ Lefschetz decomposition of $H^*({\rm Hilb}^n(S))$. The effect is to replace 
(\ref{KKV}) or more conveniently  its modular version (\ref{eqn:KKV})  by  (\ref{eqn:RefKKV}). 

Lets consider a  $K3$ surfaces that appear as the  fiber in Calabi-Yau  3-folds $M$ as in $S \hookrightarrow M\rightarrow \mathbb{P}_{\mathrm{b}}^1$. 
Given such an embedding in $M$ the Noether-Lefschetz theory for a  $K3$ family  describes  a minimal rank 
Picard lattice in the generic member of the $K3$ fiber, which supports holomorphic  curves in the latter indepently of the complex 
structure  of $M$. Therefore  one gets nontrivial ordinary Gromov-Witten theory in the A-model localization to holomorphic maps. 
Moreover, by mirror symmetry the latter can be reconstructed from the period geometry of the  B-model on the mirror $W$  to $M$.  The 
theory hence contains a fixed Picard lattice and enumerative invariants  for curves $C_\varphi$ with $\varphi$ in this  Picard lattice 
that still depend only  the intersection $C_\varphi^2=\varphi\cdot \varphi=2g-2$ and whose higher genus GW theory is up to some
technical subtleties concerning the realizable  intersection $\varphi\cdot \varphi$ in the lattice  determined  by  (\ref{KKV}). Moreover the 
information encoded in the genus zero Gromov-Witten invariants, which is for hypersurfaces  or complete intersections  in toric varieties due to Batyrev's~\cite{1993alg.geom.10003B} 
or  Batyrev/Borisov's~\cite{Batyrev:1994pg} construction and the formalism for multi parameter models developed 
in~\cite{Hosono:1993qy,Hosono:1994ax} easily  available, determines the Noether-Lefschetz generators.   More precisely one 
need  only the numbers genus zero invariants $n_{\varphi,0}^0$ with zero degree w.r.t. $\mathbb{P}^1_{\mathrm{b}}$. This idea has been 
implicelly used in~\cite{Klemm:2004km}, in order to check  known Heterotic/Type II pairs of find new one loop results for 
candidates on the Type II side that  have no known  Heterotic dual in~\cite{Klemm:2004km}.  This leads to invariants that are 
physically captured by the new supersymmetric index in Heterotic strings on $K3\times T^2$, as well as by certain 
Gopakumar-Vafa invariants associated to the dual Type II strings on a $K3$ fibered Calabi-Yau threefold.
Generically as discussed in~\cite{Klemm:2004km}, the  $n_{\varphi,\beta}^0$ for  non-zero wrapping $\beta$ of the base  $\mathbb{P}^1_b$  
do depend on the fiber class $\varphi$ not only via $\varphi\cdot \varphi=2 h-2$ as in (\ref{KKV2}). Nevertheless some 
close  formulas that share light of the non-perturbative Heterotic string for non trivial base wrapping  can be obtained 
in special situations~\cite{Klemm:2004km} and in particular for the Enriques Calabi-Yau 3-fold~\cite{Klemm:2005pd}  
and with meromorphic Jacobi form Ansatz.

We now review the aspects of Noether-Lefschetz theory that will be necessary for our computations in the upcoming sections. 
In our presentation we closely follow the references~\cite{MR3114953,MR2669707,MR3508473}. 
For a very useful review on $K3$ compactifications from the physical point of view we refer the reader to~\cite{Aspinwall:1996mn}. 

\paragraph{$\Lambda$-polarized lattices} Let $S$ be a $K3$ surface. We say that a primitive class $L \in \text{Pic}(S)= H^2(S,\mathbb{Z}) \cap H^{1,1}(S,\mathbb{C})$ is a quasi-polarization if 
             \begin{equation}
              \label{eqn:quasipolarization}
                   \int_{S} L^2 > 0 \,, \qquad
                  \int_C L \geq 0 \text{ for all } C \in H_2(S,\mathbb{Z})\,.
             \end{equation}
Let $\Lambda$ be a rank $r$ lattice of signature $(1\,,r-1)$ with a torsion-free embedding of the following form
             \begin{equation}
                      \Lambda \hookrightarrow U \oplus U \oplus U \oplus E_8(-1) \oplus E_8(-1) \cong H^2(S,\mathbb{Z})\,.
             \end{equation}
Here $U$ is the hyperbolic lattice of rank $2$ and signature $(1\,,1)$, whereas $E_8(-1)$ is the $E_8$ lattice with intersection form defined by the negative Cartan matrix of the exceptional Lie group $E_8$.  
We say that $S$ is a $\Lambda$-polarized $K3$ surface if there is a torsion-free embedding 
             \begin{equation}
                      \jmath : \Lambda \hookrightarrow \text{Pic}(S)\,,
              \end{equation}
such that
        \begin{itemize}
                  \item The embeddings of  $\Lambda$ in  $U^3 \oplus E_8(-1)^3$ and $H^2(S,\mathbb{Z})$ are isomorphic via an isometry that restricts to the identity on $\Lambda$.
                  \item $\jmath(\Lambda)$ contains a quasi-polarization.
        \end{itemize}
The polarization lattice $\Lambda$ has an intersection form $I_\Lambda: \Lambda \times \Lambda \rightarrow \mathbb{Z}$.
Let $\left\{ v_1 , \ldots,v_r\right\}$ be an integral basis of $\Lambda$.
It is useful to consider $I_\Lambda$ as the matrix with entries $\left(I_\Lambda\right)^i\,_j = I_\Lambda( v_i \,, v_j)$.
In this fashion, the discriminant of $\Lambda$ reads from $\Delta(\Lambda) = \vert \text{det}(I_\Lambda)\vert$.

We will denote the moduli space of $\Lambda$-polarized $K3$ surfaces by $\mathcal{M}_\Lambda$, see~\cite{MR2669707} and~\cite{MR2306149} for a math and  
\cite{Aspinwall:1996mn} for a physics review.

\paragraph{Polarized $K3$ surfaces from $K3$ fibrations}
Our main interest will be $\Lambda$-polarized $K3$ surfaces that arise from $K3$ fibrations
\begin{equation}
\label{eqn:$K3$fib}
	\begin{tikzcd}
		S \arrow[r,hook]  &M  \arrow{d}{\pi}\\ 
		 &\mathbb{P}_\mathrm{b}^1
	\end{tikzcd}\,,
\end{equation}
equipped with holomorphic line bundles $\left(L_1\,, \ldots\,, L_r\right)$. The tuple $\left( M \,,L_1\,, \ldots\,, L_r\,, \pi \right)$ is a one-parameter family of $\Lambda$-polarized $K3$ surfaces if
        \begin{itemize}
                \item The fibers $\left(S_p,  L_{1,p} ,\ldots , L_{r,p} \right)$ of $\left(M, L_1, \ldots, L_r \right)$ at $p \in \mathbb{P}^1_{\mathrm{b}}$ define $\Lambda$-polarized $K3$ surfaces via the replacement $ v_i \mapsto L_{i,p}$ 
                for every $p\in \mathbb{P}_{\mathrm{b}}^1$.
	\item There is a quasi-polarization $\lambda^\pi_p=\sum_{i=1}^r \lambda_i^\pi L_{i,p} \in \Lambda$ that satisfies (\ref{eqn:quasipolarization}) with respect to the $K3$ fiber $S_p$ 
                                 for all $p\in \mathbb{P}_{\mathrm{b}}^1$.    
        \end{itemize}
The quasi-polarization vector $\lambda^\pi$ defines a notion of positivity. We say that a vector $( d_1, \ldots , d_r) \in \mathbb{Z}^r$ is positive if 
$\sum_{i=1}^r \lambda^\pi_i d_i > 0 \,.$

We are interested in the enumerative geometry of curves $\varphi \in H_2(M,\mathbb{Z})$ that live in the $K3$ fibers of $M$, i.e. that project to points in $\mathbb{P}_{\mathrm{b}}^1$.  These classes of curves are furnished by the $\pi$-vertical cohomology $H_2(M,\mathbb{Z})^\pi$ which is defined by the following short exact sequence
 \begin{equation}
\begin{tikzcd}
    0\arrow{r} & H_2(M,\mathbb{Z})^\pi \arrow{r} & H_2(M,\mathbb{Z})\arrow{r}{\pi_*} & H_2(\mathbb{P}^1_{\mathrm{b}},\mathbb{Z})\arrow{r} & 0 \,.
\end{tikzcd}
\end{equation}
For a set of divisors $(L_1, \ldots L_r) $ the degree of $\varphi \in H_2(M,\mathbb{Z})^\pi$ is obtained via the projection
\begin{equation}
\varphi \mapsto \left(\int_\varphi L_1, \dots , \int_\varphi L_r \right) = (d_1,\ldots, d_r)\,.
\end{equation}

\paragraph{Noether-Lefschetz divisors} Let $\mathbb{L}$  be a rank $r+1$ lattice with an even symmetric bilinear form $\langle \cdot \,, \cdot\rangle$ and a primitive embedding $\iota:\Lambda \hookrightarrow \mathbb{L}$.
Consequently, the lattice $\mathbb{L}$ has an additive struture of the following form
\begin{equation}
\mathbb{L} = \iota\left(\Lambda\right) \oplus \mathbb{Z} v\,, \qquad v \in \mathbb{L}\,. 
\end{equation} 
Consider the extended basis $\{v_1,\dots, v_r,v\}$, where $\Lambda = \text{Span}_{\mathbb{Z}} \{v_1,\dots, v_r\}$. The bilinear pairing $\langle \cdot\,, \cdot\rangle$ of $\mathbb{L}$  is encoded in the following matrix
\begin{equation}
\mathbb{L}_{h,d_1,\ldots,d_r}=\left(\begin{array}{rrrr}	\langle v_1\,, v_1 \rangle&\dots &\langle v_1\,, v_r\rangle & d_1  \\
	\phantom{aa}\vdots \quad \phantom{a} &\ddots &\phantom{aa}\vdots \quad  \phantom{a} & \vdots \phantom{.}\\
	       \langle v_r \,, v_1\rangle &\dots & \langle v_r\,,v_r\rangle &d_r\\
	       d_1\quad \phantom{.} & \dots & d_r\quad \phantom{.}& 2h-2
	\end{array}\right)\,.
	\label{eqn:Lmat}
\end{equation}
There are two invariants for the data $(\mathbb{L},\iota)$, namely 
\begin{itemize}
\item the discriminant $\Delta(h,d_1,\ldots, d_r) = (-1)^r \det (\mathbb{L}_{h,d_1,\ldots, d_r})$,
\item the coset, defined as the image of map $\delta_v$ represented by $v_i \mapsto d_i$,  is denoted $\delta (h,d_1,\ldots d_r) \in G_\Lambda/ \pm$,  
where $G_\Lambda = \Lambda^*/ \Lambda$  is the discriminant group, an abelian group of order $\Delta(\Lambda)$.
\end{itemize}

Generically  a Noether-Lefschetz divisor $\left[P_{\Delta,\delta}\right] \subset \mathcal{M}_\Lambda$ is the closure 
of the locus of $\Lambda$-polarized $K3$ surfaces  where $\left(\text{Pic}(S),\jmath\right)$ has rank $r+1$, discriminant $\Delta$, and coset $\delta$.
We defined {\sl the Noether-Lefschetz divisor} $[D_{h,(d_1,\ldots,d_r)}]$ as
             \begin{equation}
             \label{eqn:NLdiv}
                        \left[D_{h,(d_1,\ldots, d_r)} \right]= \sum_{\Delta, \delta} m\left(h,d_1,\ldots , d_r \mid \Delta, \delta\right) \cdot \left[ P_{\Delta,\delta}\right] \subset \mathcal{M}_\Lambda\,.
              \end{equation}
Here $m(h,d_1,\ldots d_r \mid \Delta, \delta)$ denotes the number of elements $\varphi \in \mathbb{L}$ of type $(\Delta,\delta)$ such that 
               \begin{equation}
                     \langle \varphi , \varphi \rangle = 2h-2 \,, \qquad \langle \varphi , v_i\rangle = d_i\,.
               \end{equation}
              
\paragraph{Noether-Lefschetz numbers} The Noether-Lefschetz number $NL^\pi_{h,(d_1,\ldots,d_r)}$ is defined as the classical intersection of the Noether-Lefschetz divisor (\ref{eqn:NLdiv})  with the image of $\mathbb{P}_{\mathrm{b}}^1$ in $\mathcal{M}_\Lambda$, i.e. 
                \begin{equation}
                    \label{eqn:NL}
                       NL_{h,(d_1,\ldots ,d_r)}^\pi = \int_{\mathbb{P}^1_{\mathrm{b}}} \imath^*_{\pi} \left[D_{h,(d_1,\ldots,d_r)}\right]\,,
                \end{equation}
where $\imath_\pi : \mathbb{P}_{\mathrm{b}}^1 \rightarrow \mathcal{M}_\Lambda$ is the morphism associated to a one-parameter family of $\Lambda$-polarized $K3$ surfaces $\left(M, L_1,\ldots,L_r,\pi\right)$.
                
The key fact of Noether-Lefschetz theory in this context observed is that the generator of Noether-Lefschetz numbers is a vector-valued modular 
form of weight $k=11 -\frac{r}{2}$ and type $\rho_{\Lambda}^*$ of the form~\cite{MR1682249},
               \begin{equation}
               \label{eqn:NLgen}
                       \Phi^\pi (q) = \sum_{\gamma \in G} \Phi_\gamma^\pi (q) \mathrm{e}_\gamma  \in \mathbb{C}\left[\left[q^{\frac{1}{2\Delta(\Lambda)}}\right]\right] \otimes \mathbb{C} \left[G_\Lambda\right]\,,
               \end{equation}
where $\{\mathrm{e}_\gamma\}$ is a formal basis in $\mathbb{C}[G_\Lambda]$.
The coefficients of $\Phi^\pi_\gamma(q)$ are determined by the Noether-Lefschetz numbers via
               \begin{equation}
                \text{Coeff}\left(\Phi_\gamma^\pi, q^{\Delta_{NL}}\right)   =   NL_{h,(d_1,\ldots ,d_r)}^\pi \,, \text{ where }\,  \Delta_{NL}=\frac{\Delta(h,d_1,\ldots,d_r)}{2\Delta(\Lambda)}\,.
                              \label{eqn:NLcoeff}
               \end{equation}
For a short review on vector-valued modular forms and the definition of the representation $\rho^*_\Lambda$ we refer the reader to Appendix~\ref{app:VVMF}.
\begin{table}[tbp]
\centering
\begin{tabular}{|c|cccccc|}
\hline
$r^g_h$&$h=0$& $$1$$ & $2$ & $3$ & $4$ & $5$ \\
\hline 
$g=0$ & $1$ & $24$ & $324$ & $3200$ & $25650$ & $176256 $\\
$1$ &  & $-2$ & $-54$ & $-800$ & $-8550$ & $-73440$ \\
$2$ &  & &  $3$ & $88$ & $1401$ & $15960$\\
$3$ &  & & & $-4$ & $-126$ & $-2136$\\
$4$ & & & & & $5$ & $168$\\
$5$ & & & & & & $-6$\\
\hline
\end{tabular}
\caption{Non-vanishing BPS invariants $r_h^g$ for $K3$ surfaces with $h \leq 5$. For a physical interpretation of these numbers, we refer the reader to \cite{Katz:1999xq}.}
\label{tab:KKV}
\end{table}

\paragraph{The GW-NL correspondence theorem}~\cite{MR3114953,MR3508473} \textit{For degrees $(d_1,\ldots, d_r)$ positive with respect to a quasi-polarization $\lambda^\pi$},
              \begin{equation}
                \label{eqn:GWNL}
              n^g_{(d_1,\ldots,d_r)} = \sum_{h=0}^\infty r^g_h \cdot NL_{h,(d_1,\ldots, d_r)}^\pi  \,.
              \end{equation}             
Here $n^g_{(d_1,\ldots,d_r)}$ is the Gopakumar-Vafa invariant associated to a curve class $\varphi \in H_2(M,\mathbb{Z})^\pi$ with positive degree $(d_1,\ldots, d_r)$. Moreover, $r_h^g$ are the coefficients of the \textit{KKV formula} expansion that reads \cite{Katz:1999xq}
             \begin{equation}
             \label{eqn:KKV}
                 \sum_{h= 0}^\infty \sum_{g = 0 }^\infty (-1)^g r^g_h \left(y^{\frac{1}{2}} - y^{-\frac{1}{2}}\right)^{2g-2}  q^{h-1} = \frac{1}{\eta^{24}(\tau) \phi_{-2,1}(\tau,\lambda)}\,,
             \end{equation}               
with $q = \exp\left(2\pi i \tau\right)$, and $y = \exp\left(2\pi i \lambda\right)$.
For concreteness we show some $r_h^g$ numbers in Table~\ref{tab:KKV}. 
  
Note  that the invariants that the invariants $n^g_\varphi$ depend only on the genus $g$ and the intersection number $\varphi^2$ 
as pointed out in~\cite{Klemm:2004km} and mathematically more rigorously in~\cite{MR3114953}\cite{MR2669707}. 
As a last remark, it suffices to fix the Noether-Lefschetz generator~\eqref{eqn:NLgen} in order to reproduce all Gopakumar-Vafa invariants for curve classes in $H_2(M,\mathbb{Z})^\pi$.
This is a consequence of Gromov-Witten/Noether-Lefschetz correspondence together with the symmetry of the quadratic form $\varphi^2$.

\subsection{Jacobi forms of lattice index and elliptic genera}
Our second objects of interest that, at least in the case of elliptic $K3$ fibrations, exhibit a more direct relation with the topological string partition function, are the elliptic genera of the six-dimensional Heterotic strings, that are dual to limits of F-theory on the $K3$ fibration.
The elliptic genera are lattice Jacobi forms and we will first introduce some of the associated theory that will also be crucial to make the connection with the Noether-Lefschetz generators and the new supersymmetric index.

Let $\underline{L}$ denote an integral lattice $L$ equipped with a symmetric non-degenerate bilinear form $\left( \cdot\,, \cdot \right) : L \times L \rightarrow \mathbb{Z}$. A Jacobi form of weight $k$ and index $\underline{L}$ is a holomorphic function $\phi(\tau,\bm{z})$ of variables $\tau \in \mathbb{H}$ and $\bm{z} \in L\otimes\mathbb{C}$ which satisfies the following properties:
\begin{itemize}
\item \textbf{Modular transformation:} For all $\gamma = {\small\left(\begin{array}{rr} a \,\, & \,\, b \\ c \,\,& \,\,d \end{array}\right)} \in \mathrm{SL}(2,\mathbb{Z}) $ it satisfies
          \begin{equation}
              \phi\left( \frac{a \tau + b}{ c\tau + d}\,, \frac{\bm{z}}{c\tau + d}\right) = (c\tau + d)^{k }
               \mathrm{e}^{\frac{2\pi ic \left(\bm{z},\bm{z}\right)}{c\tau +d}} 
               \phi(\tau,\bm{z})\,.
          \end{equation}
\item \textbf{Elliptic transformation:} For all $\bm{\lambda}, \bm{\mu} \in L $ it satisfies
           \begin{equation}
                \phi(\tau, \bm{z} + \bm{\lambda}\tau + \bm{\mu}) = 
                  \mathrm{e}^{-2\pi i \left( \frac{1}{2}(\bm{\lambda},\bm{\lambda}) +  (\bm{\lambda},\bm{z})\right)  } \phi(\tau,\bm{z})\,.
           \end{equation}
\item \textbf{Fourier expansion:} The Fourier expansion of $\phi$ is of the form
           \begin{equation}
             \label{eqn:Fourier1}
                \phi(\tau,\bm{z}) = \sum_{n \geq n_0} \sum_{\bm{\lambda} \in L^* } c(n,\bm{\lambda}) q^n \mathrm{e}^{2\pi i \left(\bm{\lambda},\bm{z}\right)} \,,
           \end{equation}
\end{itemize}
where $q=\exp(2\pi i\tau)$, $n_0 \in \mathbb{Z}$ and $L^*$ is the dual lattice of $L$.

If the Fourier coefficients $c(n,\bm{\lambda})$ of $\phi$ vanish for $n<0$, we say that $\phi$ is a weak Jacobi form. 
If on the other hand $c(n,\bm{\lambda})$ vanishes unless $2n - \left(\bm{\lambda}\,,\bm{\lambda} \right)\geq0$ ($2n - \left(\bm{\lambda}\,,\bm{\lambda} \right) > 0$), we say that $\phi$ is a holomorphic Jacobi form (cusp Jacobi form).
Otherwise, we say that $\phi$ is a weakly holomorphic Jacobi form.  
We respectively denote by
      \begin{equation}
         J^!_{k,\underline{L}} \supseteq   J^w_{k,\underline{L}} \supseteq   J_{k,\underline{L}} \supseteq   J^{\text{cusp}}_{k,\underline{L}} \,.
               \end{equation}
the vector spaces of weakly holomorphic, weak, holomorphic and cusp Jacobi forms of weight $k$ and index $\underline{L}$.
It is possible to extend this definition of Jacobi forms  by including characters or by considering odd lattices.
We refer the reader to the reference \cite{gritsenko2019theta} for a broader view of the theory of lattice index Jacobi forms. 
\\\\
\noindent\textbf{The theta expansion of lattice Jacobi forms} The Jacobi theta functions associated to an integral lattice $\underline{L}$ are defined as
      \begin{equation}
         \label{eqn:JacTheta}
               \vartheta_{\underline{L},\bm{\mu}} (\tau, \bm{z}) = \sum_{\substack{\bm{\lambda} \in L^* \\ \bm{\lambda} \equiv \bm{\mu} \text{ mod } L}} 
                   q^{\frac{1}{2}(\bm{\lambda},\bm{\lambda})} \exp\left( 2\pi i (\bm{\lambda},\bm{z})\right)\,,
        \end{equation}
and they span a $\mathbb{C}$-vector space
\begin{equation}
 \label{eqn:spaceTheta}
     \Theta(\underline{L}) = \text{Span}_{\mathbb{C}} \left\{ \vartheta_{\underline{L}, \bm{\mu}} \, \mid \,  \bm{\mu} \in L^* / L \right\}\,,
\end{equation}
which has dimension $\vert L^*/L\vert$.
Every lattice index Jacobi form $\phi \in J_{k,\underline{L}}$ has a theta expansion of the following form \cite{MR3309829,gritsenko2019theta}
      \begin{equation}
         \label{eqn:ThetaExp}
              \phi(\tau,\bm{z}) = \sum_{\bm{\mu} \in L^*/L} h_{\bm{\mu}}(\tau) \vartheta_{\underline{L},\bm{\mu}}(\tau,\bm{z})\,.
      \end{equation}
Moreover, the space of Jacobi theta functions $\Theta(\underline{L})$ follows invariance under the metaplectic group $\text{Mp}(2,\mathbb{Z})$  \cite{MR3309829}. This enables to establish an isomorphism via the theta expansions of Jacobi forms (\ref{eqn:ThetaExp}), which reads \cite{MR2512363,gritsenko2019theta}
    \begin{equation}
       \label{eqn:JacVVMF}
        J_{k,\underline{L}} \cong \left( M_{k-\frac{r}{2}} \otimes \Theta(\underline{L})\right)\,.
    \end{equation}
Here $M_{k-\frac{r}{2}}$ denotes the $\text{Mp}(2,\mathbb{Z})$-module generated by all spaces $M_{k-\frac{r}{2}}(4N)$, where the latter denotes the space of holomorphic functions $f : \mathbb{H}\rightarrow \mathbb{C}$ that transform as $f(\gamma\tau) = w_\gamma(\tau)^{2k - r} f(\tau)$ under $\gamma\in \Gamma(4N)$. See appendix \ref{app:VVMF}. This means, there exists an $N \in \mathbb{N}$ such that $h_{\bm{\lambda}} \in M_{k-\frac{r}{2}}(4N)$ for each $\bm{\lambda} \in L^*/L$. In fact, the projection 
\begin{equation}
\label{eqn:hmap}
h : J_{k,\underline{L}} \rightarrow \text{Mod}\left( \text{Mp}(2,\mathbb{Z}),k-\frac{r}{2}, \rho^*_L\right)\,,\quad \phi(\tau,\bm{z})   \mapsto h(\phi)\equiv\sum_{\bm{\mu}\in L^*/L} h_{\bm{\mu}}(\tau) \mathrm{e}_{\bm{\mu}}\,.
\end{equation}
yields an isomorphism with between Jacobi forms and vector-valued modular forms~\cite{MR2512363,MR3309829,Oberdieck:2017pqm}.

\subsection{Elliptic genera and Noether-Lefschetz theory}
\label{sec:EGandNL}

We discussed in the previous section that there is a correspondence between the Noether-Lefschetz numbers and the Gromov-Witten invariants in $K3$ fibrations.   
 In a natural manner, there is also a correspondence between Noether-Lefschetz theory and lattice Jacobi forms defined by the $\Lambda$-polarized lattice of $K3$ fibrations. 
To make this concrete, we will consider $K3$ fibrations with polarization lattice $\Lambda$ of the form
         \begin{equation}
              \label{eqn:LatticeDec}
                 \Lambda = U \oplus L_1(-m_1) \oplus \cdots \oplus L_a(-m_a)\,,
           \end{equation}
where each lattice $L_I(-m_I)$ is given by the coroot lattice of a simple Lie algebra $\mathfrak{g}_I$ with a twist determined by a number $m_I \in \mathbb{N}$.

To obtain the $\Lambda$ lattice (\ref{eqn:LatticeDec}) we first regard Calabi-Yau threefolds $M$ that  
 admit an elliptic fibration 
 \begin{equation}
\label{eqn:$K3$fib}
	\begin{tikzcd}
		\mathcal{E} \arrow[r,hook]  &M  \arrow{d}{\pi_{\mathrm{e}}}\\ 
		 & \mathbb{F}_n
	\end{tikzcd}\,,
\end{equation}
which develops Kodaira singularities  over curves in the base given by the Hirzebruch surface $B = \mathbb{F}_n$.
The latter is defined as a $\mathbb{P}^1$-fibration $\mathrm{p}: \mathbb{F}_n \rightarrow \mathbb{P}_{\mathrm{b}}^1$ and the fiber $F \cong \mathbb{P}^1$ has vanishing self-intersection. 
It is this last property that enables us to find an elliptic $K3$ surface $S$ with fibration structure
\begin{equation}
\label{eqn:$K3$fiber}
   	\begin{tikzcd}
		\mathcal{E} \arrow[r,hook]  &S  \arrow{d}{\varpi}\\ 
		 & F
	\end{tikzcd}\,.
\end{equation}
  In this way, the Calabi-Yau threefold $M$ has also a $K3$ fibration $\pi : M \rightarrow \mathbb{P}_{\mathrm{b}}^1$ with fibers given by $S$ in (\ref{eqn:$K3$fiber}). With this construction in mind, let us comment now on the appearance of the $\Lambda$-polarized lattice (\ref{eqn:LatticeDec}). 

On the one hand, for an elliptic $K3$ surface $S$ with section there exists always an embedding $U \hookrightarrow \text{Pic}(S,\mathbb{Z})$ \cite{MR3586372}. On the other hand, the sublattices $L_I(-m_I)$ are spanned by fibral divisors obtained by resolving singularities 
over components of the discriminant locus $\mathcal{S}_{\mathfrak{g}_I}^{\mathrm{b}} \subset B$.  Each  irreducible curve $\mathcal{S}_{\mathfrak{g}_I}^{\mathrm{b}}$ corresponds to a singularity of an ADE simple Lie algebra $\widetilde{\mathfrak{g}}_I$, as the fiber splits into $\text{rk}(\widetilde{\mathfrak{g}}_I)+1$ curves that form the topology of the affine Dynkin diagram associated to $\widetilde{\mathfrak{g}}_I$. If the monodromies along $\mathcal{S}_{\mathfrak{g}_I}^{\mathrm{b}}$ act on these fibral curves, their invariant orbits result in a folded affine Dynkin diagram associated to a non-simply laced Lie algebra $\mathfrak{g}_I$. Otherwise $\mathfrak{g}_I = \widetilde{\mathfrak{g}}_I$. Fibering the monodromy invariant orbits of fibral curves over $\mathcal{S}_{\mathfrak{g}_I}^{\mathrm{b}}$, where $I =1, \dots, a$ yields a set of fibral divisors $\{E_{i_I}\}_{i_I=1,\ldots, \text{rk}(\mathfrak{g}_I),I=1,\ldots,a}$ that follow the intersection rules~\cite{Weigand:2018rez} 
\begin{equation}
E_{i_I} \cdot E_{j_J} \cdot \pi_{\mathrm{e}}^* D^{\mathrm{b}} = -m_I\left( \bm{\alpha}_{i_I}^\vee, \bm{\alpha}_{j_J}^\vee\right) \delta_{IJ}\,, \quad m_I = \mathcal{S}_{\mathfrak{g}_I}^{\mathrm{b}}\cdot D^{\mathrm{b}} \in \mathbb{N}\,,
\end{equation}
where $D^{\mathrm{b}}$ is the divisor associated to the volume
of the base $\mathbb{P}_{\mathrm{b}}^1$, $\{\bm{\alpha}_{i_I}^\vee\}_{{i_I}=1,\ldots, \text{rk}(\mathfrak{g}_I),I=1,\ldots, a}$ are the coroots of $\mathfrak{g}_I$, and $( \cdot\,, \cdot)$ is the Killing form of the Lie algebra $\mathfrak{g} =\bigoplus_{I=1}^a \mathfrak{g}_I$. Hence, we associate the fibral divisors $\{E_{i_I}\} \subset H^{1,1}(M,\mathbb{Z})$ with the generators of the coroot lattice $L^\vee(\mathfrak{g}_I)$ up to a twist factor.  In other words, $L_I(-m_I) \equiv \text{Span}_{\mathbb{Z}}\{E_{i_I}\}$.
 
 Let us discuss now the Noether-Lefschetz theory for the lattice configuration (\ref{eqn:LatticeDec}). It is convenient to choose the basis of curves for $H_2(M,\mathbb{Z})^\pi$, such that it projects the degree of a curve $\beta\in H_2(M,\mathbb{Z})^\pi$ as follows
 \begin{equation}
\beta =  \ell C_{U}+n C_{T}+ \sum_{I=1}^a\sum_{i_I=1}^{\text{rk}(\mathfrak{g}_I)}\lambda_I^{i_I}C_{i_I} \mapsto \left(\ell,n,\bm{\lambda}_1,\ldots,\bm{\lambda}_a\right) \,,
\end{equation}
where we introduced the curves  $C_U \equiv F + \mathcal{E}$, $C_T\equiv \mathcal{E}$, and $\{C_{i_I}\}$ is the set of curves dual to the fibral divisors $\{E_{i_I}\}$. In this way, this basis of curves follows the intersection relations
 \begin{equation}
 \label{eqn:curvesdeg}
C_U \cdot C_T = 1 \,, \quad C_{i_I} \cdot C_{i_J} = -m_I^{-1} \left(\bm{\omega}_{i_I} , \bm{\omega}_{j_J} \right) \delta_{IJ}\,,
\end{equation}
whereas the rest of the intersections among curves vanish. Note that the vectors $\{\bm{\omega}_{i_I}\}$ denote the fundamental weights, which are dual to the simple coroots $\{\bm{\alpha}^\vee_{i_I}\}$. Hence, we identify $\bm{\lambda}_I$ with an element in the weight lattice $L_{\text{w}}(\mathfrak{g}_I)$ with twist $m_I^{-1}$. 
 With this information, we are able to compute from (\ref{eqn:NLcoeff}) that
\begin{equation}
\label{eqn:NLdisc}
\Delta_{NL} = n \ell -\sum_{I=1}^a \frac{1}{2m_I}\left( \bm{\lambda}_I,\bm{\lambda}_I\right) +1 -h\,,
\end{equation}
which encodes the Noether-Lefschetz discriminant. The other Noether-Lefschetz invariants are the cosets determined by the discriminant group of the lattice $\oplus_I^a L_I(-m_I)$.
 Thus, we observe the {1:1} correspondence
\begin{equation}
\text{NL generator: } \Phi^\pi(\tau) \text{ }\longleftrightarrow \text{ } \text{Elliptic genus: } Z_{F}(\tau,\bm{z},\lambda), 
\end{equation}
where the elliptic genus is that of a string that results from a  D3 brane wrapping the base curve $\beta = F\in H_2(B)$ in  the F-theory picture. More precisely, this is given by the respective coefficient in the expansion (\ref{eq:baseexpansion}) that follows a modular Ansatz
\begin{equation}
\label{eqn:EG}
Z_F(\tau,\bm{z},\lambda) = \frac{\Phi(\tau,\bm{z})}{\eta^{24}(\tau) \phi_{-2,1}(\tau,\lambda)}\,, \quad \Phi(\tau,\bm{z}) \in J_{10,\underline{L}}\,,
\end{equation}
 where the $\underline{L}$ is the lattice given by $L = \bigoplus L^\vee(\mathfrak{g}_I)(m_I)$ equipped with the Killing form in $\mathfrak{g} = \bigoplus_{I}^a \mathfrak{g}_I$.  
  This equivalence is obtained via  the isomorphism (\ref{eqn:hmap}) $ h : \Phi(\tau,\bm{z}) \mapsto \Phi^\pi(\tau)$. In the language of Jacobi forms, the Noether-Lefschetz discriminant form (\ref{eqn:NLdisc}) is translated into the Jacobi form discriminant 
  \begin{equation}
  \Delta_J = 2n -\sum_{I=1}^a  \frac{1}{m_I}(\bm{\lambda}_I,\bm{\lambda}_I)\,.
  \end{equation}
Moreover, the basis of theta functions (\ref{eqn:spaceTheta}) takes the role of the discriminant group $G_\Lambda$. We elaborate further on this last piece in the upcoming paragraph.

For simplicity of the discussion, we restrict first to the case that $\mathfrak{g}$ is a simple Lie algebra. Then the lattice Jacobi forms of interest are those associated to the coroot lattice $L = L^\vee (\mathfrak{g})(m)$ with $m$-twist, i.e. the lattice is equipped with the $m$-scaled Killing form $m\left( \cdot \,, \cdot\right)$ of $\mathfrak{g}$.
The discriminant group of $L^\vee(\mathfrak{g})(m)$ can be defined by the following quotients~\cite{MR3859399}
\begin{equation}
G_L =  (L^\vee(\mathfrak{g})(m))^*/L^\vee(\mathfrak{g})(m) \cong ( L_{\text{w}}(\mathfrak{g})/ \sqrt{m}) /(\sqrt{m}L^\vee(\mathfrak{g})) \cong L_{\text{w}}(\mathfrak{g}) / mL^\vee(\mathfrak{g})\,. 
\end{equation} 
It is convenient to use the last quotient form to express the theta functions  (\ref{eqn:JacTheta}) for $L^\vee(\mathfrak{g})(m)$ in terms of the weight lattice $L_{\text{w}}(\mathfrak{g})$ and  the $m$-scaled coroot lattice $mL^\vee (\mathfrak{g})$. In this way, we arrive to the following expression for each $\bm{\mu} \in G_{L}$  theta function of $L^\vee(\mathfrak{g})(m)$~\cite{MR750341,MR1672053}
\begin{equation}
\label{eqn:LieTheta}
\vartheta_{m,\bm{\mu}}^{\mathfrak{g}}(\tau,\bm{z}) = \sum_{ \substack{{\bm{\omega}} \in  L_{\text{w}}(\mathfrak{g}) \\ {\bm{\omega}} \equiv \bm{\mu} \text{  mod } m L^\vee(\mathfrak{g})}} q^{\frac{1}{2m}({\bm{\omega}},{\bm{\omega}})} \mathrm{e}^{2\pi i  ({\bm{\omega}},\bm{z})}\,.
\end{equation}
The extension for a semisimple Lie algebra $\mathfrak{g} = \bigoplus_{I=1}^a \mathfrak{g}_I$ generalizes as follows
\begin{equation}
\label{eqn:semisimpleProduct}
\vartheta_{\bm{m},\bm{\mu}}^{\mathfrak{g}}(\tau,\bm{z}) = \prod_{I =1}^a \vartheta_{m_I,\bm{\mu}_I}^{\mathfrak{g}_I}(\tau,\bm{z}_I)\,,
\end{equation}
where $\bm{m}=(m_1,\ldots, m_a)$, $\bm{z}_I \in L^\vee(\mathfrak{g}_I) \otimes \mathbb{C}$, and 
\begin{align}
   \bm{\mu}=(\bm{\mu}_1, \ldots, \bm{\mu}_a) \in L_{\text{w} }(\mathfrak{g}_1) /m_1 L^\vee(\mathfrak{g}_1) \oplus \cdots \oplus L_{\text{w} }(\mathfrak{g}_a) /m_a L^\vee(\mathfrak{g}_a)  \,.
\end{align}

Let us finally  relate the Noether-Lefshetz generators to the new supersymmetric index (\ref{eqn:nssi1}) introduced in 
Section  \ref{sec:basics}, which corresponds to the 4d $\mathcal{N}=2$ dual Heterotic theory on $K3 \times T^2$. In~\cite{Enoki:2019deb} it was argued that
\begin{equation}
\mathcal{Z}(\tau,\bar{\tau}) = \sum_{\gamma \in G_\Lambda} \theta_{\Lambda'(-1) +\gamma}(\tau,\bar{\tau}) \frac{\Phi^\pi_\gamma(\tau)}{\eta^{24}(\tau)}\,,
\end{equation}
where $\theta_{\Lambda'(-1) +\gamma}$ are the vector components of the Siegel theta function (\ref{eqn:SiegelTheta}) for the extended polarization lattice $\Lambda' = U(-1) \oplus \Lambda$, where $U(-1) \cong H^0(S,\mathbb{Z}) \oplus H^4(S,\mathbb{Z})$.
In our configuration (\ref{eqn:LatticeDec}) the Siegel theta function follows the form
\begin{align}
	\theta_{\Lambda'(-1)+\bm{\mu}}(\tau,\bar{\tau})=\sum\limits_{\bm{\lambda}_1\in L_1(m_1)+\bm{\mu}_1} \cdots \sum\limits_{\bm{\lambda}_a\in L_a(m_a)+\bm{\mu}_a} \sum\limits_{(k_0,w_0)\in U}\sum\limits_{(k,w)\in U(-1)}q^{\frac{p_L^2(\bm{v})}{2}}\bar{q}^{\frac{p_R^2(\bm{v})}{2}}\,,
	\label{eqn:siegeltheta1}
\end{align}
where $\bm{v} = (\bm{\lambda}_1 , \ldots \bm{\lambda}_a; k_0,w_0; k, w)$. Replacing $\mathrm{e}_{\bm{\mu}} \mapsto\theta_{\Lambda'(-1)+\bm{\mu}}(\tau,\bar{\tau})$~\eqref{eqn:NLgen} maps the Noether-Lefschetz generator into the 
new supersymmetric index of the Heterotic string  (\ref{eqn:nssi1}). Note that in both~\eqref{eqn:siegeltheta1} and~\eqref{eqn:nssi1} we have 
suppressed the dependence on the four-dimensional vector moduli $T,U,\bm{V}$, which are encoded in the left-right moving momenta $p_{L/R}$.
We will discuss the generalization of this map for $K3$ fibrations that are not elliptic but only exhibit a genus one fibration in Section~\ref{sec:nssiandtwistedelliipticgenus}.

\section{Refined fiber invariants for $K3$ fibered CY 3-folds }
\label{sec:refined}

In this section we shortly review the proposal for defining refined amplitudes for the heterotic string on $K3\times T^2$ 
following~\cite{Nakayama:2011be,Antoniadis_2013}, as well as the proposal of \cite{MR3524167} to refine the
counting function  for BPS states in M-theory compactifications  on K3 fibered Calabi-Yau 3-folds by using 
Noether-Lefschetz theory in the K3 fiber. We will apply this formalism to K3 fibers with higher Picard rank 
at the end of Sections \ref{subsec:A1}, \ref{subsec:A2} and \ref{subsec:G2},  as well as for K3 fibers with genus 
one fibrations at the end of Section   \ref{sec:4sec}. Furthermore, we specialize the Heterotic amplitude 
to an $SU(2)$ enhancement point of the STU model to make a check on the physical consistency of the proposal of ~\cite{MR3524167}.   

\subsection{Refined Heterotic string theory on $K3\times T^2$}
Attempts for defining a worldsheet description for the refined topological string theory have been addresed in the works \cite{Antoniadis_2010,Nakayama:2011be,Antoniadis_2013,Antoniadis_2014,Antoniadis_2015}.
There, the authors realized the $\Omega$-background in Heterotic string theory on $K3\times T^2$. In this configuration, the previous authors consider one-loop BPS amplitudes of the form
\begin{equation}
\label{eqn:refamps}
 I_{g,n} = \int_{M_4} \mathrm{d}^4 x F_{g,n}(\bm{t},\bar{\bm{t}}) R_-^2 T_-^{2g-2} T_+^{2n} + \cdots\,,
\end{equation}
which is the refinemenet of (\ref{eqn:gravcouplings}). In the same fashion, $R_-$ and $T_-$ are defined as in the unrefined case, wheras $T_+$ is a self-dual insertion that differs in the proposals \cite{Nakayama:2011be,Antoniadis_2013}. See both references for details and comparison. 

As a reminder, we review briefly the $\Omega$-background next. We can realize this background from a 6d $N = (1,0)$ theory with topology $\mathbb{R}^4 \times T^2$ and metric \cite{MR2181816}
\begin{equation}
\label{eqn:OmBack}
\mathrm{d}s^2 = \left(\mathrm{d}x^\mu + \Omega^\mu \mathrm{d}z + \bar{\Omega}^\mu \mathrm{d}\bar{z}\right)^2 + \mathrm{d}z \mathrm{d} \bar{z}\,,
\end{equation} 
where $(z,\bar{z})$ are coordinates on $T^2$ and the $\Omega^\mu$ entries in (\ref{eqn:OmBack}) fulfil the equation
\begin{equation}
T= \mathrm{d}\Omega = \epsilon_1 \mathrm{d}x^1 \wedge \mathrm{d}x^2 + \epsilon_2 \mathrm{d}x^3\wedge \mathrm{d}x^4\,.
\end{equation}
Here the linear combination $\epsilon_\pm = \frac{1}{2} (\epsilon_1  \pm \epsilon_2)$ define equivariant rotation parameters of $\mathbb{C}^2 \cong \mathbb{R}^4$, which can be expressed in the spinor notation as
\begin{equation}
\epsilon_-^2 = - \text{det} \Big(T_{\alpha \beta}\Big) \,, \quad \epsilon_+^2 = - \det\Big( T_{\dot{\alpha}\dot{\beta}}\Big) \,,
\label{eqn:epmSpinor}
\end{equation}
where $\alpha, \beta, \dot{\alpha}, \dot{\beta} = 1,2$ are spinor indices for $SU(2)_L\times SU(2)_R$. In addition to the non-trivial metric (\ref{eqn:OmBack}), another $SU(2)_I$ $R$-symmetry is necessary to preserve the amount of supersymmetry. The effect of the latter symmetry is to compensate metric induced holonomies on the spinors (\ref{eqn:epmSpinor}), which must be covariantly constant. 
If gravity can be decoupled, a  further $SU(2)_I$ $R$-symmetry emerges. In this case the degeneracies of BPS states 
  are protected. Consequently, it makes sense to introduce the BPS supertrace \cite{Nekrasov:2003af}
\begin{equation}
\mathcal{Z}_{\text{Nek}}(\epsilon_-,\epsilon_+,\bm{t}) = \text{Tr}_{\text{BPS}}\left(-1\right)^{2(J_L+J_R)} \mathrm{e}^{-2 \epsilon_- J_L} \mathrm{e}^{-2 \epsilon_+\left(J_R + J_I\right)} \mathrm{e}^{-\beta H}\,,
\end{equation}
where $J_*$ is the Cartan generators of each $SU(2)_*$ symmetry group. Moreover, the combination $J_R'\equiv J_R +J_I$ enables a twist of $SU(2)_R$ such that the BPS degeneracies $N_{j_L,j_R'}^\beta$ are invariant~\cite{Choi:2012jz}. From now on, we will denote by $j_R$ the spin of the twisted $J_R'$ generator. Furthermore, the refined Schwinger loop calculation yields the multicovering formula for the refined topological free energy $\mathcal{F}(\epsilon_-,\epsilon_+,\bm{t}) = \log \mathcal{Z}_{\text{Nek}}(\epsilon_-,\epsilon_+,\bm{t})$~\cite{Iqbal:2007ii,Choi:2012jz}:

\makeatletter
    \def\tagform@#1{\maketag@@@{\normalsize(#1)\@@italiccorr}}
\makeatother
{\footnotesize
\begin{align}
\begin{split}
\mathcal{F} ( \epsilon_1,\epsilon_2, \bm{t}) =
\sum_{\beta \in H_2(M,\mathbb{Z})}  \sum_{j_L,j_R \in \frac{1}{2} \mathbb{Z}_{\geq0}}  \sum_{ m =1 }^\infty  (-1)^{2j_L + 2j_R}  \frac{N_{j_L\,,j_R}^\beta}{m} \frac{\left[j_L\right]_{x^m} \left[ j_R\right]_{y^m}}{(X^{\frac{m}{2}} -X^{-\frac{m}{2}})(Y^{\frac{m}{2}} -Y^{-\frac{m}{2}})} \mathrm{e}^{2\pi i m\bm{t} \cdot \beta}\,. 
\end{split}
\label{eqn:refmulticovering}
  \end{align}
 }
 
\noindent{where} 
\begin{equation}
 X = \exp\left( 2\pi i \epsilon_1\right) = x y  \,, \quad Y = \exp\left( 2\pi i \epsilon_2\right) = \frac{y}{x}\,,
\end{equation}
and for $j \in \frac{1}{2}\mathbb{Z}_{\geq0}$
\begin{equation}
\left[ j\right]_u \equiv u^{-2j} + u^{-2j +2}+ \cdots +u^{2j-2} + u^{2j}\,.
\end{equation}

Coming back to the refined worldsheet proposals~\cite{Nakayama:2011be,Antoniadis_2013}, we highlight two remarkable properties of such theories. In the first place, the latter authors are able to recover the unrefined worldsheet theory by switching off $\epsilon_+ = 0$. Secondly, at a point of $SU(2)$ enhanced gauge symmetry  they reproduce the perturbative Nekrasov partition function below \cite{MR2181816,Nakayama:2011be}
\begin{equation}
\mathcal{Z}_{\text{Nek}} (\epsilon_1,\epsilon_2) \Big\vert_{\text{pert.}} \sim \int \frac{ds}{s} \frac{- 2 \cos(2 \epsilon_+)}{ \sin\left(\frac{s \epsilon_1}{2}\right)\sin\left(\frac{s \epsilon_2}{2}\right)}\mathrm{e}^{-m^2 s}\,. 
\label{eqn:Nek}
\end{equation}

In the next section, we will consider the Type IIA dual side on a $K3$ fibered CY threefold $M$, where we argue the decoupling of gravity $m_{\text{el}} \to \infty$ by taking the limit $\text{Vol}_{\mathbb{C}}(\mathbb{P}_{\mathrm{b}}^1)\to \infty$~\cite{Katz:1996fh,Klemm:1997gg}. With this in mind, we will examine the refinement proposal of \cite{MR3524167}.  We will argue that this proposal satisfies the requirement (\ref{eqn:Nek}). In addition to that, it provides a solution to the refined amplitudes (\ref{eqn:refamps}) in the Heterotic weak coupling limit, which gives back the unrefined theory when $\epsilon_+=0$. 

\subsection{The KKP conjecture}
\label{subsec:kkp}
Following the formulation of \cite{MR3524167}, we introduce refinements for the Noether-Lefschetz numbers (\ref{eqn:NL}), which are representations of $SU(2)_L\times SU(2)_R$ lying in the space
\begin{equation}
\mathbb{Z}_{\geq0} \left[0\,, 0\right] \oplus \mathbb{Z}_{\geq0} \left[0\,, \frac{1}{2}\right]\,.
\end{equation}
There are two kind of refined Noether-Lefschetz numbers to be considered: $\mathsf{RNL}_{h,(d_1,\ldots, d_n)}^{\pi,\circ} $ and $\mathsf{RNL}_{h,(d_1,\ldots, d_n)}^{\pi,\diamond} $. The former ones are defined as follows
\begin{equation}
\mathsf{RNL}_{h,(d_1,\ldots, d_n)}^{\pi,\circ} = NL_{h,(d_1,\ldots,d_n)}^\pi \cdot \left[0 \,, 0\right]\,.
\end{equation}
The other Noether-Lefschetz refinement requires more explanation. Let an effective divisor $\left[D_{h,(d_1,\ldots, d_n)}\right]$ be divided in two components
\begin{equation}
\label{eqn:NLdivdiv}
D_{h,(d_1,\ldots, d_n)}  = S_\imath + T_\imath\,,
\end{equation}
where $S_\imath$ is the sum of divisors not containing $\imath_\pi( \mathbb{P}_{\mathrm{b}}^1)$ and $T_\imath$ is the sum of divisors containing $\imath_\pi( \mathbb{P}_{\mathrm{b}}^1)$. Having said this, we define the refinement $\mathsf{RNL}_{h,(d_1,\ldots d_n)}^{\pi,\diamond}$ in the following way
\begin{itemize}
\item If $\Delta(h,d_1,\ldots, d_n) < 0$, then $\mathsf{RNL}_{h,(d_1,\ldots d_n)}^{\pi,\diamond}=0$.
\item If $\Delta(h,d_1,\ldots, d_n) = 0 $, then $\mathsf{RNL}_{h,(d_1,\ldots d_n)}^{\pi,\diamond} = \left[0\,,\frac{1}{2}\right]$.
\item If $\Delta(h,d_1,\ldots, d_n) > 0$, then
\begin{equation}
\mathsf{RNL}_{h,(d_1,\ldots d_n)}^{\pi,\diamond} = \int_{\mathbb{P}_{\mathrm{b}}^1} \imath_\pi^* S_\imath \cdot \left[0\,,0\right] - \frac{1}{2} \int_{\mathbb{P}_{\mathrm{b}}^1} \imath_\pi^* T_\imath \cdot \left[ 0\,, \frac{1}{2}\right]\,.
\label{eqn:RNLpos}
\end{equation}
\end{itemize}

As we will see, the KKP conjecture is a refined version of the GW-NL correspondence theorem (\ref{eqn:GWNL}). Therefore, an important ingredient will be the refinement of the KKV formula (\ref{eqn:KKV}), which we introduce next.

	\begin{table}[h!]
	\centering
	\centering
	{\small
	\begin{tabular}{|r|c|}\hline
			$\mathsf{R}_{j_L,j_R}^{0}$ &$2j_+=$0\\ \hline
			$2j_-=$0& $1$\\ 
			\hline \end{tabular}
		\begin{tabular}{|r|cc|}\hline
			$\mathsf{R}_{j_L,j_R}^{1}$ &$2j_+=$0 & 1\\ \hline
			$2j_-=$0& $20$ & \\ 
			1 & & $1$\\
			\hline \end{tabular}
		\begin{tabular}{|r|ccc|}\hline
			$\mathsf{R}_{j_L,j_R}^{2}$ &$2j_+=$0&1&2\\ \hline
			$2j_-=0$& $231 $ & & \\
			1 &  &$21 $&  \\
			2 & &  &$1$\\
			\hline \end{tabular}
                \begin{tabular}{|r|cccc|}\hline
			$\mathsf{R}_{j_L,j_R}^{3}$ &$2j_+=$0&1&2& 3 \\ \hline
			$2j_-=0$ & $1981 $ &  &$1$ &  \\ 
			1 &  &$252 $&   & \\ 
			2 & $1$ &   &$ 21$ &    \\  
			3&  &  &  &  $1$   \\
			\hline \end{tabular}
                \begin{tabular}{|r|ccccc|}\hline
			$\mathsf{R}_{j_L,j_R}^{4}$ &$2j_+=$0&1&2& 3&4 \\ \hline
			$2j_-=0$ & $13938 $ &  &$21$ & &  \\ 
			1 &  &$2233 $&   & $1$& \\ 
			2 & $21$ &   &$ 253$ &  &  \\  
			3&  & $1$  &  & $21$  &   \\
			4& & & & & $1$\\
			\hline \end{tabular}
						}
			\caption{Non-vanishing refined BPS invariants $\mathsf{R}_{j_L,j_R}^{h}$ for $K3$ surfaces with $h \leq 4$. }
	\label{tab:KKVref}
\end{table}

The refined BPS invariants $\mathsf{R}_{j_L,j_R}^{h}$ for $K3$ surfaces, where $j_L,j_R \in \frac{1}{2} \mathbb{Z}_{\geq0}$, were defined in \cite{MR3524167} via the refined KKV formula:
{\small
\begin{equation}
\sum_{h=0}^\infty \sum_{j_L,j_R \in \frac{1}{2}\mathbb{Z}_{\geq 0}}  \mathsf{R}^h_{j_L\,,j_R} \frac{\left[ j_L\right]_y}{ X^{\frac{1}{2}} -X^{-\frac{1}{2}}} \frac{\left[ j_R \right]_x}{Y^{\frac{1}{2}}-Y^{-\frac{1}{2}}} q^{h-1} = \frac{1}{\eta^{24}(\tau)}\frac{1}{\phi_{-1,\frac{1}{2}}(\tau,\epsilon_1 )\phi_{-1,\frac{1}{2}}(\tau,\epsilon_2 )} \,.
\label{eqn:RefKKV}
\end{equation}}

\noindent{For} illustration, we show a few BPS invariants $\mathsf{R}_{j_L,j_R}^{h}$ in table \ref{tab:KKVref}. Let the $K3$ invariants $\mathsf{R}^h_{j_L,j_R}$ be divided into two parts
\begin{equation}\label{eqn:Rdiv}
\mathsf{R}^h_{j_L,j_R} = \mathsf{R}^{h,\circ}_{j_L,j_R} +\mathsf{R}^{h,\diamond}_{j_L,j_R} \,,
\end{equation}
where the $\mathsf{R}^{h,\diamond}_{j_L,j_R}$ contributions read from the formula
{\small
\begin{equation}
\sum_{h=0}^\infty \sum_{j_L,j_R \in \frac{1}{2}\mathbb{Z}_{\geq 0}}  \mathsf{R}^{h\,,\diamond}_{j_L\,,j_R} \frac{\left[ j_L\right]_y}{ X^{\frac{1}{2}} -X^{-\frac{1}{2}}} \frac{\left[ j_R \right]_x}{Y^{\frac{1}{2}}-Y^{-\frac{1}{2}}} q^{h-1} = \frac{q^{-\frac{5}{6}}}{\eta^{4}(\tau)}\frac{1}{\phi_{-1,\frac{1}{2}}(\tau,\epsilon_1 )\phi_{-1,\frac{1}{2}}(\tau,\epsilon_2 )} \,.
\label{eqn:RefKKVDiamond}
\end{equation}}With this information at hand, we can compute refined BPS numbers along $K3$ fibers via the KKP conjecture, which we state now. \\\\
\textbf{Refined P-NL correspondence conjecture:} \cite{MR3524167}  \textit{A 1-parameter family of $\Lambda$-polzarized $K3$ surfaces}
\begin{equation}
\pi: M \rightarrow \mathbb{P}_{\mathrm{b}}^1\,,
\end{equation} 
\textit{with Calabi-Yau total space determines a division (\ref{eqn:Rdiv}) satisfying the following property. For degrees $(d_1,\cdots, d_r)$ positive with respect to the quasi-polarization $\lambda^\pi$}, 
\begin{align}\label{eqn:KKP}
   \begin{split}
              \sum_{j_L,j_R} \mathsf{N}_{j_L,j_R}^{(d_1,\cdots, d_n)}[j_L,j_R] = & \sum_{j_L,j_R} \sum_{h = 0}^\infty \mathsf{R}^{h,\circ}_{j_L,j_R} \otimes \mathsf{RNL}^{\pi,\circ}_{h,(d_1,\cdots,d_n)}\\
                                                                                                          &+  \sum_{j_L,j_R} \sum_{h = 0}^\infty \mathsf{R}^{h,\diamond}_{j_L,j_R} \otimes \mathsf{RNL}^{\pi,\diamond}_{h,(d_1,\cdots,d_n)}\,.
    \end{split}
\end{align}

We give a few remarks about the numbers  $\mathsf{N}_{j_L,j_R}^{(d_1,\cdots, d_n)}$ that we obtain from (\ref{eqn:KKP}).
 First, recall the 5d Hilbert space of BPS states due to M2-branes wrapping a curve $\kappa \in H_2(M,\mathbb{Z})$, whose structure reads
\begin{equation}
 \sum_{j_L,j_R \in \frac{1}{2}\mathbb{Z}_{\geq0}} N_{j_L,j_R}^{\beta} \left(\left[\frac{1}{2} \,, 0   \right] \oplus 2 \left[0\,, 0\right]\right) \otimes \left[ j_L \,, j_R\right]\,.
\end{equation}
Here $N_{j_L,j_R}^\kappa$ count the multiplicities of the $SU(2)_L\times SU(2)_R$ representations $\left[j_L,j_R\right]$. This physical constraint implies that $N_{j_L,j_R}^\kappa \in \mathbb{Z}_{\geq0}$. For a positive degree $(d_1,\ldots, d_n) \in H_2(M,\mathbb{Z})^\pi$, the refined BPS number $N_{j_L,j_R}^{(d_1,\ldots d_r)}$ can be calculated via the stable pairs of moduli spaces of $M$~\cite{Choi:2012jz,MR3524167}. As argued from previous section, we expect these BPS counts to be invariant under complex structure deformations of $M$, as well as $\mathsf{N}_{j_L,j_R}^{(d_1,\ldots, d_n)} =N_{j_L,j_R}^{(d_1,\ldots, d_n)}$.  In this work, we give plenty of evidence for $\mathsf{N}_{j_L,j_R}^{(d_1,\ldots, d_n)} \in \mathbb{Z}_{\geq0}$. Furthermore, we observe that our computations always fulfil the additional  constraint
\begin{equation}
\label{eqn:weightedTr}
\sum_{j_L,j_R \in \frac{1}{2}\mathbb{Z}_{\geq0}} (-1)^{2j_R}(2j_R+1) \mathsf{N}_{j_L,j_R}^{(d_1,\ldots, d_n)} \left[j_L\right] = \sum_{g \in \mathbb{Z}_{\geq0}} n_{(d_1,\ldots,d_n)}^g I^g_L\,,
\end{equation}
where $n_{(d_1,\ldots,d_n)}^g$ are the unrefined Gopakumar-Vafa invariants in (\ref{eqn:GWNL}) and $I_L^g$ denotes the $SU(2)_L$ representation that reads
\begin{equation}
I_L^g = \left( 2\left[0\right] + \left[ \frac{1}{2}\right] \right)^{\otimes g}\,.
\end{equation}

Now, we argue that the KKP conjecture leads to a consistent refinement for the type II theory side, whose dual is the Heterotic on $K3\times T^2$ with $\Omega$-background.  In the reference \cite{Antoniadis_2013}, it was shown that an $SU(2)$ enhancement point leads to a symmetrized perturbative Nekrasov partition function. 
 The latter authors found out that in the 5d limit
\begin{align}
\mathcal{F}(\epsilon_1,\epsilon_2, \bm{t})\Big\vert_{SU(2)}  \sim\left( \gamma_{\epsilon_1,\epsilon_2}(x \vert \beta) + \gamma_{-\epsilon_1,-\epsilon_2}(x \vert \beta) \right)\,,
\label{eqn:pertNek}
\end{align}
where~\cite{MR2181816}\footnote{As mentioned in \cite{Antoniadis_2013}, we only regard the cut-off independent contribution.}
\begin{equation}
\label{eqn:pertNek2}
\gamma_{\epsilon_1,\epsilon_2}(x\vert\beta) = \sum_{m=1}^\infty \frac{1}{m} \frac{\mathrm{e}^{- m \cdot \beta x} }{\left( \mathrm{e}^{\beta m \epsilon_1}-1\right)\left( \mathrm{e}^{\beta m \epsilon_2}-1\right)}\,.
 \end{equation}
  
 To compare this result with the KKP conjecture,  we take the STU model as an example. The latter contains an $SU(2)$ enhancement at the conifold locus $T-U=0$. This has been identified with the lattice points $(k,w) \in U$ that have an intersection $ k w = -1$, i.e. $(k, w) = (1,-1)$ or $(k,w) = (-1,1)$~\cite{Marino:1998pg}. As a first step, we calculate the topological free energy contributions associated to these lattice points. For this, we introduce the notation $x \equiv \pm (1,-1) \cdot(T,U)= \pm (T-U)$, $\beta \equiv 2\pi i$  and recall that the only non-trivial GV invariant for these lattice points is $n_{\pm(1,-1)}^0 =-2$. Thus, the corresponding contributions read
 
 {\small
 \begin{align}
 \begin{split}
 F(\lambda,T,U) \Big\vert_{SU(2)}  & =  -2 \sum_{g=0}^\infty \sum_{m =1}^\infty \frac{ \mathrm{e}^{- \beta x }}{m}\left(2 \sin \left( \frac{m \lambda}{2} \right) \right)^{2g-2}  \\
  & = -2  \lambda^{-2} \text{Li}_3\left( \mathrm{e}^{-\beta x} \right) -2 \sum_{g=1}^\infty \lambda^{2g-2} (-1)^{g+1} \frac{B_{2g}}{2g(2g-2)!} \text{Li}_{3-2g}\left( \mathrm{e}^{-\beta x}  \right)\,,
  \end{split}
 \end{align}}
 
\noindent{where} $B_{2g}$ are Bernoulli numbers. Notice that this result coincides with (\ref{eqn:pertNek}) and (\ref{eqn:pertNek2}) in the case $\epsilon_2 = -\epsilon_1$, as described in Appendix A.0.3 in~\cite{MR2181816}.
The refinement of this computation amounts to replacing $n_{\pm(1,-1)}^0 =-2$ with $\mathsf{N}_{0,\frac{1}{2}}^{\pm(1,-1)}= 1$ as the only non-vanishing contribution. 
With this in mind, we calculate the refined multicovering formula (\ref{eqn:refmulticovering}) at the $SU(2)$ enhancement point
\begin{align}
\begin{split}
\mathcal{F}(\epsilon_1,\epsilon_2,T,U) \Big\vert_{SU(2)} & = \sum_{m = 1 }^\infty \frac{1}{m} \frac{y^m + y^{-m}}{\left(X^{\frac{m}{2}} - X^{-\frac{m}{2}}\right)\left(Y^{\frac{m}{2}} - Y^{-\frac{m}{2}}\right)} \mathrm{e}^{-m \cdot \beta x}\\
& =  -\left( \gamma_{\epsilon_1,\epsilon_2}(x \vert \beta) + \gamma_{-\epsilon_1,-\epsilon_2}(x \vert \beta)  \right)\,.
\end{split}
\end{align}
In other words, with only  Noether-Lefschetz information and the KKP conjecture we obtain the expression (\ref{eqn:pertNek}), which was shown in~\cite{Antoniadis_2013} to be equivalent to (\ref{eqn:Nek}).

\section{The non-Abelian sublattice weak gravity conjecture}
\label{sec:nawg}
 
Perhaps the oldest and most widely accepted swampland conjecture states that a complete theory of 
quantum gravity can not exhibit any global symmetries~\cite{Banks:2010zn}.
There are several arguments supporting this claim, see~\cite{Palti:2019pca} for a review of 
this and also the other swampland conjectures that will be discussed below, but the most intuitive reason is that
black holes can swallow objects that carry arbitrary amounts of global charge and subsequently evaporate.
Quantum gravity effects will only become relevant at the very end of this process to potentially alter the classical ``no-hair'' behaviour.
However, at that point the black hole does not have enough mass to radiate particles that carry a sufficiently large amount of global charges.
The only alternative is that the black hole does not evaporate completely but turns into a remnant.
For this to be the possible fate for all states that can form such black holes the number of distinct remnant species  
in the theory would need to be infinite. While there are convincing arguments that this leads to a thermodynamic 
catastrophe~\cite{Susskind:1995da} it is difficult to turn them into a general proof~\footnote{For a derivation in the context of AdS/CFT duality see~\cite{Harlow:2018tng}.}.
Nevertheless, many refinements of the conjecture have been proposed and they are generally 
motivated from properties of string compactifications, that seem to be universal in that no counterexamples are known.

One of these refinements is the weak gravity conjecture~\cite{ArkaniHamed:2006dz}.
It states that for a theory of gravity that is coupled to a $U(1)$ gauge symmetry there should always
exist a stable superextremal particle, i.e. one of mass $m$ and electric charge $q$ such that
\begin{align}
g_{\text{el}}\, q\ge\frac{m}{\sqrt{2}\, m_{\text{Pl}}}\,,
\label{eqn:wgc}
\end{align}
where $g_{\text{el}}$ is the electric gauge coupling constant and $m_{\text{Pl}}$ is the Planck mass.
The analogous constraint on magnetic charges implies that there exists a cutoff scale $\Lambda\lesssim g_{\text{el}}\, m_{\text{Pl}}$ 
above which the effective field theory description breaks down and additional degrees of freedom
must be taken into account.
Together this ensures that all electrically or magnetically charged black holes in the theory can decay.
The weak gravity conjecture implies the absence of global symmetries in the sense that a global symmetry can be interpreted as a gauge symmetry with vanishing gauge coupling.
Plenty of arguments have been brought forward to support this conjecture and we again refer to~\cite{Palti:2019pca} for a review of the literature.

The generalization of the weak gravity conjecture to theories with multiple $U(1)$ factors leads to a so-called convex hull condition~\cite{Cheung:2014vva} which ensures that superextremal states exist in every rational direction of the charge lattice.
However, it was subsequently observed that in all known constructions an even stronger statement holds true~\cite{Heidenreich:2015nta,Heidenreich:2016aqi}.
The latter authors used Kaluza-Klein reduction of Einstein-Maxwell theory on tori and toroidal orbifolds as well as modular invariance in string theory to show that, at least in those examples,
not only one superextremal particle exists but there is a sublattice $\Gamma_{\text{ext}}$ of the lattice $\Gamma$ of charges that are compatible with the Dirac quantization condition such that at every
site of the sublattice there exists a (possibly unstable) superextremal state.
Furthermore, it was suggested that the mass scale associated with this sublattice gives rise to the cutoff scale $\Lambda$ and in the limit of small gauge coupling the tower of particles in the corresponding direction of the charge lattice becomes massless
and leads to a breakdown of the effective theory.

A striking check of this ``sublattice weak gravity conjecture'' (sLWGC) has been performed in F-theory on elliptically fibered Calabi-Yau threefolds~\cite{Lee:2018urn,Lee:2018spm}.
First the authors have shown that in the moduli space of the corresponding six-dimensional supergravity a limit in which a gauge symmetry becomes global while keeping the Planck mass finite is necessarily at infinite distance.
Moreover, in order for such a limit to exist in the first place, the base of the fibration has to be itself fibered over a $\mathbb{P}^1$, and therefore be a Hirzebruch surface, or it is a blow-up thereof.
The volume of the fiber of this surface is then inversely proportional to the (squared) gauge coupling constant and a tensionless Heterotic string arises in the decoupling limit.
At least for theories with purely Abelian gauge group, the particle-like excitations of this string were then found to contain the sublattice of superextremal states that is required by the sLWGC.
To verify superextremality, the condition~\eqref{eqn:wgc} had to be modified to take into account the Yukawa forces that are present in the dilatonic Einstein-Maxwell theories that arise from F-theory~\cite{Lee:2018spm}.
The presence of the sublattice was then shown to be a consequence of the properties of the Jacobi and lattice Jacobi forms that appear in the elliptic genus of the tensionless string.

An extension of the sLWGC for non-Abelian gauge groups has been formulated in~\cite{Heidenreich:2017sim}.
Non-Abelian gauge theories are already constrained by the sLWGC applied to the Cartan torus of the gauge group.
However, it was observed in~\cite{Heidenreich:2017sim} that a stronger statement, namely that superextremal particles should exist in every representation of the gauge group, is supported by perturbative string theory and stable under compactification.

In this section we will show that the non-Abelian sublattice weak gravity conjecture~(nAsLWGC) can, for F-theory on genus one fibered Calabi-Yau threefolds, again be derived as a consequence of the properties of
lattice Jacobi forms.

\subsection{The non-Abelian sLWGC and lattice Jacobi forms}
\label{subsec:najac} 
Let us first repeat the precise proposal for a nAsLWGC that has been made in~\cite{Heidenreich:2017sim}:\\\\
\noindent \textbf{The non-abelian sLWGC} For any quantum gravity in $d\geq 5$ dimensions with zero cosmological constant and unbroken gauge group $G$, there is a finite-index Weyl-invariant sublattice $\Gamma$ of the weight lattice $L_{\text{w}}(G)$ such that for every dominant weight $\bm{\lambda}_{R} \in \Gamma$ there is a superextremal resonance transforming in the $G$ irreducible representation $R$ with highest weight $\bm{\lambda}_R$.\\\\
Note that for purely Abelian gauge groups this reduces to the ordinary sLWGC.
In particular, the condition that a resonance is superextremal means that its charge to mass ratio is greater than that of a large, extremal Reissner-Nordstr\"om black hole
with parallel charge vector.

Let us now consider this proposal in the context of F-theory on a Calabi-Yau threefold such that the effective theory admits a limit in which the gauge coupling would go to zero.
It follows from the general results in~\cite{Lee:2018urn} that the base of the elliptic fibration contains a curve of self-intersection zero and geometrically the limit amounts to this curve shrinking to zero volume.
The string that arises from a D3-brane that wraps this curve becomes tensionless and is dual to a, not necessarily perturbative, critical Heterotic string.
More precisely, the string is weakly coupled if the base of the fibration is a Hirzebruch surface.
In general additional tensor multiplets signal the presence of NS5-branes and therefore of non-perturbative effects.
Nevertheless, in each case we can calculate the elliptic genus and this will be a modular or, in the non-perturbative case, a quasi-modular Jacobi form.
Let us stress that the non-perturbative effects due to the presence of NS5-branes are different from those that lead to a non-perturbative gauge group.
In those cases where the gauge group is not perturbative but the base of the Calabi-Yau threefold on the F-theory side is a Hirzebruch surface, the elliptic genus
of the dual Heterotic string will still be a modular Jacobi form.

The elliptic genus encodes a subset of the particle-like string excitations via a trace over the Ramond-Ramond sector along a torus
\begin{align}
	Z(\tau,\bm{z})=\text{Tr}_{RR}(-1)^FF^2q^{H_L}\bar{q}^{H_R}\prod^{n_V}_{a=1}\left(\zeta^a\right)^{J_a}\,,
	\label{eqn:ellipticgenus1}
\end{align}
where $F$ is the Fermion number, $H_{L/R}$ are the left- and right-moving Hamiltonians and $q=\exp(2\pi i \tau)$ is the modular parameter of the torus.
Moreover, the rank of the gauge group is $n_V$ and it acts as a global symmetry with generators $J_a$ on the worldsheet theory.
The corresponding fugacities are denoted by $\zeta^a=\exp(2\pi i z^a)$.
In terms of modular objects the elliptic genus takes the general form
\begin{align}
	Z(\tau,\bm{z})=\frac{\Phi_{10,\bm{m}}(\tau,\bm{z})}{\eta^{24}(\tau)}\,,
	\label{eqn:hetegenus}
\end{align}
where $\Phi_{10,\bm{m}}(\tau,\bm{z})$ is a holomorphic, Weyl invariant lattice Jacobi form of weight $10$ with index matrix $\bm{m}$.
It therefore admits an expansions~\eqref{eqn:ThetaExp}
\begin{align}
	\Phi_{10,\bm{m}}(\tau,\bm{z})=\sum\limits_{\bm{\mu}\in L^*/L}h_{\bm{\mu}}(\tau)\vartheta_{\underline{L},\bm{\mu}}(\tau,\bm{z})\,,
	\label{eqn:elgenthetaexp}
\end{align}
with $\vartheta_{\underline{L},\bm{\lambda}}(\tau,\bm{z})$ being Jacobi theta functions associated to the, in general twisted, coroot lattice $\underline{L}$ of the gauge group.

The elliptic genus essentially encodes only the left-moving excitations and the states have to be paired with right moving excitations, taking into account the level matching condition, to lead to actual states of the theory.
In particular, the numerator $\Phi_{10,\bm{m}}$ in~\eqref{eqn:hetegenus} always contains a non-zero constant term that arises from the left-moving tachyon.
This implies that there is always a contribution~\footnote{We thank Timo Weigand for explaining this point to us.}
\begin{align}
	\Phi_{10,\bm{m}}=h_{\bm{0}}(\tau)\vartheta_{\underline{L},\bm{0}}(\tau,\bm{z})+\dots\,,
\end{align}
with $h_0(\tau)=-2+\mathcal{O}(q)$~\cite{Lee:2019tst}.
It was shown in~\cite{Lee:2018spm} that the states from this sector, with the corresponding Jacobi theta function given by
\begin{align}
       \vartheta_{\underline{L},\bm{0}} (\tau, \bm{z}) = \sum_{\substack{\bm{\lambda} \in L^* \\ \bm{\lambda} \equiv \bm{0} \text{ mod } L}} 
	   q^{\frac{1}{2}(\bm{\lambda},\bm{\lambda})} \exp\left( 2\pi i (\bm{\lambda},\bm{z})\right)\,,
   \label{eqn:theta0}
\end{align}
are superextremal with respect to a dilatonic Reissner-Nordstr\"om of the corresponding Cartan $U(1)$ charges.
Moreover, they form a sublattice of the $U(1)$ charge lattice and therefore are sufficient to satisfy the original sublattice weak gravity conjecture.

We will now argue that, at least for simple gauge groups, this sector in fact satisfies the stronger claim of the non-Abelian sLWGC.
To this end we need to see how the states encoded in~\eqref{eqn:theta0} arrange into representations of the gauge group $G$ with Lie algebra $\mathfrak{g}$.
More precisely, we will rewrite the right-hand side in terms of Weyl characters.
Let us first recall some definitions\footnote{For further guidance on Lie algebras and representation theory, we refer the reader to Appendix \ref{app:LART}.}.
Let $(R,V_{\bm{\lambda}})$ be an irreducible, finite-dimensional representation of a complex semisimple Lie algebra $\mathfrak{g}$ with Cartan subalgebra $\mathfrak{h}$.
Here $V_{\bm{\lambda}}$ is a highest weight module with highest weight $\bm{\lambda}$, and $R$ is the $\mathfrak{g}$-representation associated to $V_{\bm{\lambda}}$.
The Weyl character of $(R,V_{\bm{\lambda}})$ is the function $\chi_{\bm{\lambda}}: \mathfrak{h} \rightarrow \mathbb{C}$ with
\begin{equation}
\label{eqn:WeylCh1}
	\chi_{\bm{\lambda}}(\bm{z}) = \text{tr}_{V_{\bm{\lambda}}} \left(\exp\left(2\pi i R(\bm{z})\right)\right)\,.
\end{equation} 

From now on we assume that the gauge group is a simple Lie group with algebra $\mathfrak{g}$.
However, we remark that the upcoming derivations can be generalized for semisimple Lie algebras via the product formula (\ref{eqn:semisimpleProduct}).
Thus, for a simple Lie algebra the index $\bm{m}$ of the elliptic genus as a lattice Jacobi form will then be $m$ times the negative of the coroot lattice intersection form, where $m$ is some positive integer.
The numerator of the elliptic genus admits an expansion in terms of Weyl invariant lattice theta functions
\begin{align}
	\Phi_{10,m}(\tau,\bm{z}) = \sum_{\bm{\lambda} \in L_{\text{w}}(\mathfrak{g})/ m L^\vee(\mathfrak{g})} h_{\bm{\lambda}} (\tau) \vartheta^{\,\mathfrak{g}}_{m,\bm{\lambda}}(\tau,\bm{z})\,,
\end{align}
with the states encoded in $-2\cdot\vartheta^{\,\mathfrak{g}}_{m,\bm{0}}(\tau,\bm{z})$ again forming the superextremal sublattice.
The expansion
\begin{align}
	\vartheta_{m,\bm{0}}^{\,\mathfrak{g}}(\tau,\bm{z})=\sum_{ \substack{w \in W(\mathfrak{g})}}\sum_{ \substack{w\cdot \bm{\lambda} \in m L^\vee(\mathfrak{g}) \\ \bm{\lambda} \in P_+(\mathfrak{g})}} \text{sign}(w) q^{\frac{1}{2m}(w\cdot \bm{\lambda},w\cdot \bm{\lambda})} \chi_{\bm{\lambda}}(\bm{z})\,,
\end{align}
follows from a more general relation that we are going to prove in the next section.
Here $W(\mathfrak{g})$ is the Weyl group and the shifted Weyl reflection
\begin{align}
	\cdot: W(\mathfrak{g}) \times L_{\text{w}}(\mathfrak{g}) \rightarrow L_{\text{w}}(\mathfrak{g})\,,
	\label{eqn:shiftedreflection}
\end{align}
acts as $w\cdot \bm{\lambda} = w(\bm{\lambda} + \bm{\rho}) - \bm{\rho}$.
Moreover, $P_+(\mathfrak{g}) = L_{\text{w}}(\mathfrak{g}) \cap \mathcal{W}(\mathfrak{g})$ is the set of dominant weights, i.e. those in the fundamental Weyl chamber $\mathcal{W}(\mathfrak{g})$. 

We can further decompose the expansion into
\begin{align}
  \label{eqn:thetaChambersDec}
	\begin{split}
		\vartheta_{m,\bm{0}}^{\,\mathfrak{g}}(\tau,\bm{z})=&\sum_{\bm{\lambda} \in m L^\vee(\mathfrak{g})\cap P_+(\mathfrak{g})}  q^{\frac{1}{2m}(\bm{\lambda}, \bm{\lambda})} \chi_{\bm{\lambda}}(\bm{z})\\
		&+\sum_{ \substack{w \in W(\mathfrak{g})\\ w\ne\text{id}}}\sum_{ \substack{w\cdot \bm{\lambda} \in m L^\vee(\mathfrak{g}) \\ \bm{\lambda} \in P_+(\mathfrak{g})}} \text{sign}(w) q^{\frac{1}{2m}(w\cdot \bm{\lambda},w\cdot \bm{\lambda})} \chi_{\bm{\lambda}}(\bm{z})\,,
	\end{split}
\end{align}
It is easy to show that $( w\cdot \bm{\lambda} , w \cdot \bm{\lambda}) > ( \bm{\lambda} , \bm{\lambda}) $ for $\bm{\lambda} \in P_+(\mathfrak{g})$ if $w \neq \text{id}$. This follows from the inequality
\begin{align}
\begin{split}
	(w\cdot \bm{\lambda} , w\cdot \bm{\lambda} )  & = \left( w(\bm{\lambda}+\bm{\rho}) - \bm{\rho}, w(\bm{\lambda} + \bm{\rho}) -\bm{\rho}\right) \\
	& = (\bm{\lambda}, \bm{\lambda}) + \underbrace{\left( \bm{\lambda} + \bm{\rho} , \bm{\rho} - w^{-1}(\bm{\rho}) \right)}_{\geq0} + \underbrace{\left( \bm{\rho}, \bm{\lambda} + \bm{\rho} - w^{-1}(\bm{\lambda} + \bm{\rho}) \right)}_{\geq0} \geq (\bm{\lambda}, \bm{\lambda}) \,,
\end{split}
	\label{eqn:thetacharexpansion}
\end{align}
where $\bm{\rho}$ is the Weyl vector
\begin{align}
	\bm{\rho}=\frac12\sum\limits_{\bm{\alpha}\in\Phi^+(\mathfrak{g})}\bm{\alpha}\,,
\end{align}
with $\Phi^+(\mathfrak{g})$ being the set of positive roots.
The inequality is in turn a consequence of proposition 2.4 and note 4.14 of~\cite{MR781344}: \textit{If $\bm{\gamma} \in \mathcal{W}(\mathfrak{g})$ and $w\in W(\mathfrak{g})$, then $\left( \bm{\gamma} - w(\bm{\gamma}) , \tau \right) \geq0 $ for every} $\tau \in \text{Int}\left( \mathcal{W}(\mathfrak{g})\right)$; $\bm{\gamma} + \bm{\rho} \in \text{Int}\left( \mathcal{W}(\mathfrak{g})\right)$ \textit{iff $\bm{\gamma} \in \mathcal{W}(\mathfrak{g})$}. Here $\text{Int}\left(\mathcal{W}(\mathfrak{g})\right)$ denotes the interior of $\mathcal{W}(\mathfrak{g})$.

The second line of~\eqref{eqn:thetacharexpansion} ensures no cancelations for the first line in (\ref{eqn:thetaChambersDec}). This implies there is a superextremal resonance transforming in the irreducible representation associated to every dominant weight $\bm{\lambda}$ in the sublattice $mL^\vee(\mathfrak{g})\subset L_{\text{w}}(\mathfrak{g})$.

\subsection{Weyl invariant character sums over dominant weights}
\label{subsec:sumsdominant} 

We now introduce a generalized version of the identity~\eqref{eqn:theta0} and prove it:\\\\
\noindent \textbf{Claim I:} Let $\mathfrak{g}$ be a complex simple Lie algebra.
For any positive integer $m\in\mathbb{N}$ and Weyl invariant subset $\{\bm{\mu}_{i,\,i=1,\dots,k}\} \subseteq L_{\text{w}}(\mathfrak{g})/mL^\vee(\mathfrak{g})$ we can define the lattice subset
\begin{equation}
\label{eqn:Ksubset}
K = \left(\bm{\mu}_1 + mL^\vee(\mathfrak{g})\right) \oplus \cdots \oplus \left(\bm{\mu}_k + mL^\vee(\mathfrak{g}) \right) \,,
\end{equation}
and the corresponding sum of theta functions satisfies the relation
\begin{equation}
\label{eqn:Claim}
\vartheta_{m,\bm{\mu}_1}^{\mathfrak{g}}(\tau,\bm{z})+ \cdots +\vartheta_{m,\bm{\mu}_k}^{\mathfrak{g}}(\tau,\bm{z}) = \sum_{w \in W(\mathfrak{g})}  \sum_{\substack{w\cdot \bm{\lambda} \in K \\ 
\bm{\lambda} \in P_+(\mathfrak{g})}}\text{sign}(w) q^{\frac{1}{2m}\left(w\cdot \bm{\lambda},w\cdot\bm{ \lambda}\right)} \chi_{\bm{\lambda}}(\bm{z})\,.
\end{equation}

In order to prove the \textbf{Claim I}, we need to discuss the \textit{Weyl character formula}.
The trace (\ref{eqn:WeylCh1}) that defines the Weyl character $\chi_{\bm{\lambda}}(\bm{z})$ results into a weighted sum over the highest weight module $V_{\bm{\lambda}}$.
We recall that the latter decomposes as a direct sum of weight spaces, i.e. $V_{\bm{\lambda}} = \oplus_{\bm{\omega}} V_{\bm{\omega}}$.
This means that the Weyl character $\chi_{\bm{\lambda}}(\bm{z})$ admits an expansion of the form 
\begin{equation}
\label{eqn:multiplicities}
\chi_{\bm{\lambda}}(\bm{z}) = \sum_{\bm{\omega}\in V_{\bm{\lambda}}} m_{\bm{\omega}} \mathrm{e}^{2\pi i ( {\bm{\omega}}, \bm{z})}\,, 
\end{equation} 
where $m_{\bm{\omega}}\in\mathbb{N}$ is the multiplicity of each weight space $V_{\bm{\omega}} \subset V_{\bm{\lambda}}$ associated to a weight ${\bm{\omega}}$. 
Alternatively,  we can expand the characters using the famous Weyl character formula~\cite{MR1153249}
\begin{equation}
\label{eqn:WeylFormula}
\chi_{\bm{\lambda}}(\bm{z}) = \frac{1}{
\Delta_W(\bm{z})}\sum_{w \in W(\mathfrak{g})} \text{sign}(w) \exp\left[2\pi i \left(w({\bm{\lambda}} + {\bm{\rho}}) , \bm{z}\right) \right]\,,
\end{equation}
where $\bm{\rho}$ is again the Weyl vector and $\Delta_{W}(\bm{z})$ is defined by
\begin{equation}
\Delta_W(\bm{z}) \equiv  \prod_{\bm{\alpha} \in \Phi^+(\mathfrak{g})}\left( \mathrm{e}^{\pi i (\bm{\alpha}, \bm{z}) }-  \mathrm{e}^{-\pi i (\bm{\alpha}, \bm{z}) } \right)= \sum_{w \in W(\mathfrak{g})} \text{sign}(w) \exp\left[2\pi i \left(w( {\bm{\rho}}) , \bm{z}\right)\right]\,.
\end{equation}
Here $\Phi^+(\mathfrak{g})$ denotes the set of positive roots in $\mathfrak{g}$.
Note that $w(\bm{\lambda})$ is the Weyl reflection $w$ applied to the weight $\bm{\lambda}$ and should not be confused with the shifted Weyl reflection $w\cdot\bm{\lambda}$ defined in~\eqref{eqn:shiftedreflection}.
With this information at hand, we proceed to prove \textbf{Claim I}.

\paragraph{Proof of Claim I}  
First, we prove the relation
 \begin{equation}
 \sum_{{\bm{\omega}} \in W_{\bm{\lambda}}} \mathrm{e}^{2\pi i({\bm{\omega}},\bm{z})} = \sum_{{\bm{\omega}} \in W_{\bm{\lambda}}} \chi_{\bm{\omega}}(\bm{z})\,,
 \label{eqn:Id1}
 \end{equation}
 where we introduced $W_{\bm{\lambda}} \equiv \{ w({\bm{\lambda}})\}_{w\in W(\mathfrak{g})}$ for a given ${\bm{\lambda}} \in L_{\text{w}}(\mathfrak{g})$.
 This is equivalent to showing that
  \begin{equation}
 \sum_{{\bm{\omega}} \in W_{\bm{\lambda}}} \Delta_W(\bm{z}) \cdot \mathrm{e}^{2\pi i({\bm{\omega}},\bm{z})}  = \sum_{{\bm{\omega}} \in W_{\bm{\lambda}}} A_{{\bm{\omega}} + {\bm{\rho}}} (\bm{z})\,,
 \label{eqn:Id2}
 \end{equation}
 where $A_{{\bm{\omega}} +{\bm{\rho}}}$ is the numerator of the Weyl character formula (\ref{eqn:WeylFormula}), i.e. $\chi_{\bm{\omega}} (\bm{z}) = \Delta_W^{-1} A_{{\bm{\omega}}+ {\bm{\rho}}}(\bm{z})$.
It directly follows from expanding the left-hand side of the equation~\eqref{eqn:Id2}
 \begin{align}
 \begin{split}
  \sum_{{\bm{\omega}} \in W_{\bm{\lambda}}} \Delta_W(\bm{z}) \cdot  \mathrm{e}^{2\pi i({\bm{\omega}},\bm{z})}  & =  \sum_{w'\in W(\mathfrak{g})}  \sum_{w\in W(\mathfrak{g})} \text{sign}(w') \mathrm{e}^{2\pi i \left(w'({\bm{\rho}}) + w({\bm{\lambda}})) ,\bm{z}\right) }\\
  & = \sum_{w\in W(\mathfrak{g})}  A_{ w({\bm{\lambda}}) + {\bm{\rho}}}(\bm{z})\\ 
    & = \sum_{{\bm{\omega}} \in W_{\bm{\lambda}}}  A_{{\bm{\omega}} + {\bm{\rho}}}(\bm{z})\,,
  \end{split}
  \label{eqn:Id2Exp}
 \end{align}
where we have used the fact
 \begin{equation}
 \sum_{w\in W(\mathfrak{g})} \mathrm{e}^{2\pi i \left(w({\bm{\lambda}}),\bm{z}\right)}  =  \sum_{w\in W(\mathfrak{g})} \mathrm{e}^{2\pi i \left(w'w({\bm{\lambda}}),\bm{z}\right) }\,,  \quad w'\in W(\mathfrak{g})\,.
 \end{equation}

Now we proceed to prove the formula~\eqref{eqn:Claim}.
Recall that the left-hand side reads
\begin{equation}
\vartheta_{K}(\tau ,\bm{z}) \equiv \vartheta^{\mathfrak{g}}_{m,\bm{\mu}_1}(\tau,\bm{z})+ \cdots +\vartheta^{\mathfrak{g}}_{m,\bm{\mu}_k}(\tau,\bm{z})\,.
\end{equation}
Using the identity~\eqref{eqn:Id1} this can be rewritten as
\begin{align}
\label{eqn:calc1}
\begin{split}
	\vartheta_{K} (\tau,\bm{z} ) =& \sum_{{\bm{\omega}} \in K} q^{\frac{1}{2m} ({\bm{\omega}},{\bm{\omega}})} \mathrm{e}^{2\pi i ({\bm{\omega}},\bm{z})} \\
	=& \sum_{{\bm{\lambda}} \in K\cap\mathcal{W}(\mathfrak{g})}\sum\limits_{\bm{\omega}\in W_{\bm{\lambda}}} q^{\frac{1}{2m} ({\bm{\omega}},{\bm{\omega}})} \mathrm{e}^{2\pi i ({\bm{\omega}},\bm{z})} \\
	=& \sum_{{\bm{\omega}} \in K} q^{\frac{1}{2m} ({\bm{\omega}},{\bm{\omega}})}  \chi_{\bm{\omega}} (\bm{z})\\
	=& \sum_{w \in W(\mathfrak{g})}  \sum_{\substack{w\cdot {\bm{\lambda}} \in K \\ {\bm{\lambda}} \in P_+(\mathfrak{g})}} q^{\frac{1}{2m} (w\cdot {\bm{\lambda}},w \cdot {\bm{\lambda}} )}  \chi_{w\cdot {\bm{\lambda}}} (\bm{z})\,. 
 \end{split}
\end{align}
where we used the fact the $K\subset L_{\text{w}}(\mathfrak{g})$ is Weyl invairant.

Moreover, a simple calculation reveals the following identity
\begin{align}
\label{eqn:id2}
\begin{split}
\chi_{w\cdot {\bm{\lambda}}}(\bm{z})  &= \frac{1}{\Delta_W (\bm{z})} \sum_{w'\in W(\mathfrak{g})} \text{sign}(w') \mathrm{e}^{2\pi i \left(w'(w \cdot {\bm{\lambda}} + {\bm{\rho}}),\bm{z}\right)}\\
& = \frac{\text{sign}\left(w^{-1}\right)}{\Delta_W(\bm{z})} \sum_{w'\in W(\mathfrak{g})} \text{sign}(w'w) \mathrm{e}^{2\pi i \left(w'w( {\bm{\lambda}} + {\bm{\rho}}), \bm{z}\right)}\\
& = \text{sign}(w) \chi_{\bm{\lambda}} (\bm{z})\,.
\end{split}
\end{align}
Inserting the identity~\eqref{eqn:id2} into the last expression in ~\eqref{eqn:calc1} gives the conjectured expression~\eqref{eqn:Claim}.

To prove the non-Abelian sublattice weak gravity conjecture, we were particularly interested in the Weyl invariant theta function $\vartheta_{m,\bm{0}}(\tau,z) = \vartheta_{m L^\vee(\mathfrak{g})}(\tau,\bm{z})$ which corresponds to the sublattice $m L^\vee(\mathfrak{g}) \subset L_{\text{w}}(\mathfrak{g})$~\cite{Heidenreich:2017sim}.
 However, our \textbf{Claim I} is more general.
 In the following we provide two examples to illustrate this. 
\begin{figure}[h!]
	\centering
	\begin{minipage}[b]{.35\textwidth}
	\includegraphics[width=\linewidth]{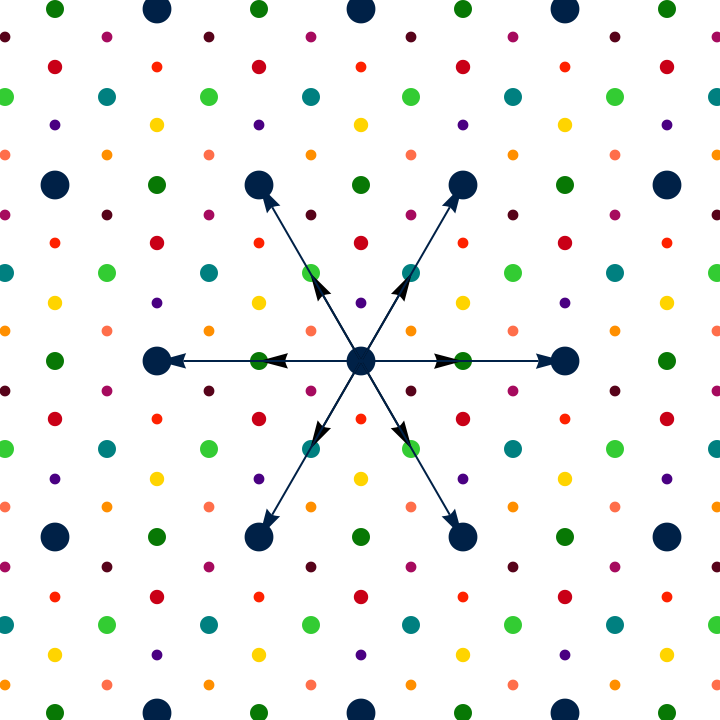}
	\end{minipage}
	\hspace{1cm}
	\caption{Schematic representation of the lattice $L^*(2)$ in terms of the weight lattice $L_{\text{w}}(A_2)$. Each coset in the discriminat group $G$ is represented by a color in (\ref{eqn:color}), e.g. the dark blue dots $\textcolor{oxfordblue}{\bullet}$ represent the sublattice $2L^\vee(A_2)$. The long arrows indicate the scaled root system $2\Phi(A_2)$ contained in $2L^\vee(A_2)$, whereas the short arrows indicate the root $\Phi(A_2)$ system contained in $L^\vee(A_2)$.} 
	\label{fig1}
\end{figure}

\noindent\textbf{Example 1:} Let us consider the gauge algebra $\mathfrak{g}=A_2$ and lattice $L(m)$ with twist given by the index $m=2$.
The discriminant group $G_L = L_{\text{w}}(\mathfrak{g})/mL^\vee(\mathfrak{g})$ has dimension $\text{dim}(G_L) = m^{\text{rk}(L^\vee(\mathfrak{g}))} \text{det}(L^\vee(\mathfrak{g}))$.
In this case we therefore find $\text{dim}(G_L) =12$.
With this in mind we take the following weights as representatives for the cosets $\{\bm{\mu}_i\}$ of $G_L$
\begin{align}
\label{eqn:color}
\begin{split}
&\textcolor{oxfordblue}{\bullet}\text{ }(0,0) \,,\quad \textcolor{indigo(web)}{\bullet}\text{ }(0,1) \,, \quad \textcolor{harvardcrimson}{\bullet}\text{ }(0,2) \,,\quad \textcolor{lasallegreen}{\bullet} \text{ } (2,-1)\,,  \\
&\textcolor{yellow(ncs)}{\bullet}\text{ }(2,0) \,,\quad  \textcolor{scarlet}{\bullet}\text{ }(2,1) \,, \quad
\textcolor{princetonorange}{\bullet} \text{ } (1,0)\,,\quad \textcolor{teal}{\bullet} \text{ } (1,1)\,, \\
&\textcolor{jazzberryjam}{\bullet} \text{ } (1,2)\,, \quad \textcolor{outrageousorange}{\bullet} \text{ }(3,-1)\,,\text{ }\textcolor{limegreen}{\bullet} \text{ } (-1,2)\,, \text{ } \textcolor{darkscarlet}{\bullet} \text{ }  (-1,3)\,,
\end{split}
\end{align}
where $ {\bm{\omega}}_1 = (1,0)$ and ${\bm{\omega}}_2 = (0,1)$ are the fundamental weights basis vectors in $L_{\text{w}}(A_2)$.
Each representative in~\eqref{eqn:color} is labelled with a color $\bullet$ that indicates the class of points associated to each coset $ [\bullet] \equiv \bm{\mu}_i$ in $G_L$.
The corresponding cosets are illustrated in Figure~\ref{fig1}. Let us take the Weyl invariant sublattice $[\textcolor{oxfordblue}{\bullet}] = 2L^\vee(A_2)$ to exemplify the formula (\ref{eqn:Claim}). For this we introduce the short-hand notation
\begin{equation}
 \quad \zeta_{\pm}^{( a,b )}\equiv \zeta_1^a \zeta_2^b + \zeta_1^{-a}\zeta_2^{-b}\,, 
\end{equation}
where $\zeta_i \equiv \exp(2\pi i (\bm{\omega}_i,\bm{z}))$. Having said this, we provide a few $q$-expansions for $\vartheta_{[\textcolor{oxfordblue}{\bullet}]}(\tau,\bm{z})$. First by uisng formula (\ref{eqn:LieTheta}), and then by using formula (\ref{eqn:Claim}). 
This reads

{\small
\begin{align}
\begin{split}
 \vartheta_{[\textcolor{oxfordblue}{\bullet}]}(\tau,\bm{z}) &= 1 + \left(\zeta_\pm^{(2,4)} + \zeta_\pm^{(2,2)} + \zeta_\pm^{(4,2)}\right)q^2 + \left( \zeta_\pm^{(6,0)} + \zeta_\pm^{(6,6)} + \zeta_\pm^{(0,6)} \right)q^6 +\mathcal{O}(q^8) \\
 & = \chi_{(0,0)} +\left(\chi_{(2, 2)}(\bm{z}) - \chi_{(0, 3)}(\bm{z}) - \chi_{(0, 0)} - \chi_{(3, 0)}(\bm{z})\right)q^2\\
 & \qquad+ \left(\chi_{(0, 3)}(\bm{z}) + \chi_{(0, 6)} (\bm{z})- \chi_{(1, 4)}(\bm{z}) + \chi_{(3, 
   0)}(\bm{z}) - \chi_{(4, 1)}(\bm{z}) + \chi_{(6, 0)}(\bm{z}) \right)q^6\\
   & \qquad \qquad +\mathcal{O}(q^8)\,,\\
   \end{split}
\end{align}}

\noindent{Comparison} of both expansions can be verified by using the Weyl character formula (\ref{eqn:WeylFormula}).
Moreover, we provide more Weyl invariant subsets of the type (\ref{eqn:Ksubset})
\begin{align}
\begin{split}
\label{eqn:KsWeyl}
&K_0 =[\textcolor{oxfordblue}{\bullet}]\,, \quad  K_1 = [\textcolor{harvardcrimson}{\bullet}]\,, \quad K_2 = [\textcolor{yellow(ncs)}{\bullet}]\,,\\
K_3 = [\textcolor{lasallegreen}{\bullet}] \oplus[\textcolor{teal}{\bullet}]\oplus&[\textcolor{limegreen}{\bullet}] \,, \quad K_4 =[\textcolor{indigo(web)}{\bullet}] \oplus[\textcolor{jazzberryjam}{\bullet}]\oplus[\textcolor{darkscarlet}{\bullet}]\,, \quad K_5 = [\textcolor{scarlet}{\bullet}] \oplus[\textcolor{outrageousorange}{\bullet}]\oplus[\textcolor{princetonorange}{\bullet}]\,.
\end{split}
\end{align}
We provide the Weyl character expansions for the rest of the theta functions $\vartheta_{K_i}(\tau,\bm{z})$ in (\ref{eqn:KsThetas}).  Essentially,  combinations of the Weyl invariant subsets in (\ref{eqn:KsWeyl})  fulfill \textbf{Claim I}.

\begin{figure}[h!]
	\centering
	\begin{minipage}[b]{.35\textwidth}
	\includegraphics[width=\linewidth]{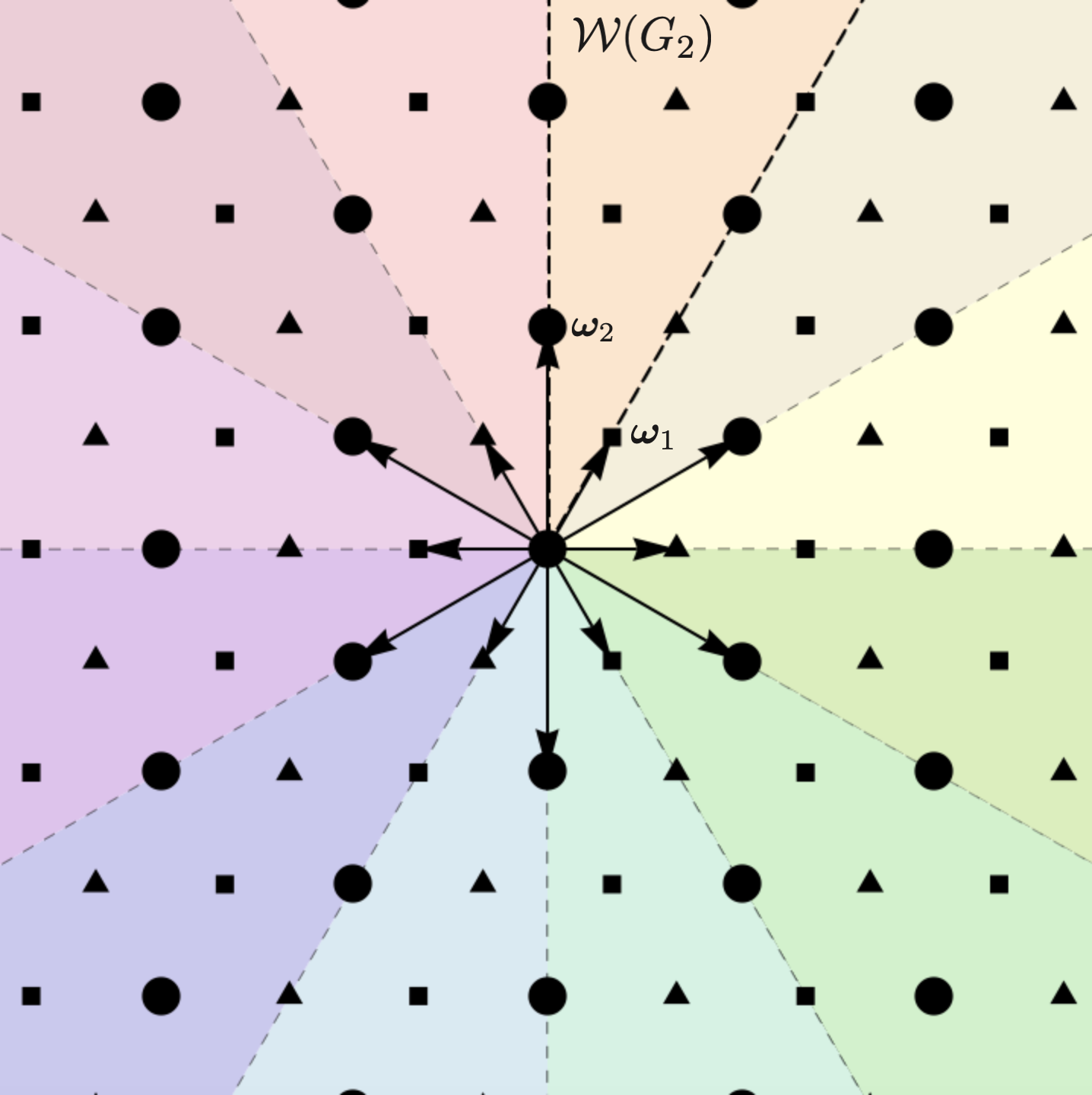}
	\end{minipage}
	\hspace{1cm}
	\caption{Weight lattice $L_{\text{w}}(G_2)$. We indicate each Weyl chamber by a colored shaded region, e.g. the orange region corresponds to the fundamental Weyl Chamber $\mathcal{W}(G_2)$. The fundamental weights are indicated by $\bm{\omega}_1$ and $\bm{\omega}_2$. The arrows indicate the root system $\Phi(G_2)$. We denote by {\Large$\bullet$}, {\tiny$\blacksquare$}, and $\blacktriangle$ those points associated to $L^\vee(G_2)$, $\bm{\mu}_1 +L^\vee(G_2)$, and $\bm{\mu}_2 + L^\vee(G_2)$ respectively.} 
	\label{fig2}
\end{figure}

\noindent\textbf{Example 2:} Let us now consider the gauge algebra $\mathfrak{g}=G_2$ with no twist.
In this case we have that $\text{dim}(G_L) =3$. We choose the following weights as representatives for the cosets $\{\bm{\mu}_i\}$ of $G_L$
\begin{align}
\bm{\mu}_{0} = (0,0) \,, \quad \bm{\mu}_1 = (4,-2),\quad \bm{\mu}_2 = (2,-1)\,.
\end{align}
When taking the combinations of theta functions
\begin{align}
L^\vee(G_2) &: \vartheta_{1,\bm{\mu}_0}(\tau,\bm{z})\,,\\
\left(\bm{\mu}_1 + L^\vee(G_2)\right)\oplus \left(\bm{\mu}_2 +L^\vee(G_2)\right)&: \vartheta_{1,\bm{\mu}_1}(\tau,\bm{z})+ \vartheta_{1,\bm{\mu}_2}(\tau,\bm{z})\,,
\end{align}
 the relation~\eqref{eqn:Claim} holds.
Note that the lattice subsets $L^\vee(G_2)$ and $\bm{\mu}_1 + L^\vee(G_2) \oplus \bm{\mu}_2 + L^\vee(G_2)$ are invariant under Weyl reflections in $W(G_2)$. See figure \ref{fig2}.

\section{Examples}
\label{sec:examples}
In this section we are illustrating the non-Abelian sublattice weak gravity conjecture, as well as the refinement of the Nother-Lefschetz invariants, at the hand of several examples.
First we are studying a family of elliptic fibrations with $I_2$ singular fibers that via F-theory lead to an $SU(2)$ gauge symmetry with matter in the fundamental representation.
Next we move on to analogous families with $I_3$ singular fibers and correspondingly an $SU(3)$ gauge symmetry in F-theory.
In both cases there will be one geometry where the gauge symmetry in the Heterotic string, that is dual to the corresponding Type IIA compactification, is realized perturbatively.
Most of the time, the gauge symmetry is non-perturbative and therefore the usual perturbative Heterotic techniques can not be applied.
Finally we consider one elliptic fibration where a family of  singular fibers is folded by monodromy
around points in the base such that the resulting gauge symmetry in F-theory is $G_2$.
Note that genus one fibrations without sections will be discussed separately in Section~\ref{sec:genusone}.

\subsection{$G=SU(2)$ geometries}
\label{subsec:A1}  

We study elliptic fibrations $M^{SU(2)}_{a,b}$ over the Hirzebruch surface $\mathbb{F}_1$ that exhibit one independent section $E_0$ and one fibral divisor $D_f$.
The fibrations exhibit families of $I_2$ singular fibers over a curve $\mathcal{S}_{A_1}^{\mathrm{b}}$ in the base that are resolved by $D_f$,
The points of the corresponding polytopes as well as the generic parts of the linear relations among them are listed in Table~\ref{tab:su2points}
\begin{table}[h!]
\begin{align*}
\begin{blockarray}{rrrrrrrrl}
	&&&&C_1&C_2&B&F\\
\begin{block}{|rrrr|rrrr|l}
	 1& 0& 0& 0& 1& 0& *& *&\\
	 0& 1& 0& 0& 0& 1& *& *&\\
	-3&-2& 0& 0& 1&-1& *& *&\leftarrow\text{holomorphic section }E_0\\
	-1&-1& 0& 0&-2& 3& *& *&\leftarrow\text{fibral divisor }D_f\\
	 0& 0& 1& 0& 0& 0& 0& 1&\\
	a&2+a&-1& 0& 0& 0&-1& 1&\leftarrow\text{vertical divisor }\pi^{-1}(B)\\
	 0& 0& 0& 1& 0& 0& 1& 0&\\
	b&3+b&-1&-1& 0& 0& 1& 0&\leftarrow\text{vertical divisor }\pi^{-1}(F)\\
	 0& 0& 0& 0& 0&-3& *& *&\\
\end{block}
\end{blockarray}
\end{align*}
	\caption{The points of a set of reflexive polytopes that lead to elliptically fibered Calabi-Yau manifolds with families of $I_2$ fibers.
	Some parts of the linear relations among the points are given in the last four columns.}
\label{tab:su2points}
\end{table}
and we consider the $20$ geometries for values $a=-5,\dots,-1$ and $b=a-3,\dots,a$.
Another three geometries correspond to $a=0$ and $b=-3,\dots,-1$. The case $(a,b)=(0,0)$ leads to a geometry with $h^{1,1}=5$ and will not be considered.

In every case there is a unique triangulation that is compatible with the projection onto the last two coordinates.
Indeed, the corresponding fibration of the associated toric variety induces the elliptic fibration of the Calabi-Yau hypersurface.
The intersection of the fibral divisor with the vertical divisors $\pi^{-1}(B)$ and $\pi^{-1}(F)$, that respectively stem from the base and the fiber of the Hirzebruch surface, are determined by
\begin{align}
	D_f^2\cdot \pi^{-1}(B)=2(a-b)-6\,,\quad D_f^2\cdot \pi^{-1}(F)=-2a-12\,.
\end{align}
Using the general results from~\cite{Klevers:2014bqa} we find that the Euler characteristic $\chi$ of $M^{SU(2)}_{a,b}$ is
\begin{align}
	\chi=-6(32+a-a^2+2b+2ab)\,,
\end{align}
while the genus $g$ of $\mathcal{S}_{A_1}^{\mathrm{b}}$ and the number of isolated $I_2$ fibers $n_{\bm{2}}$ are given by
\begin{align}
	g=\frac12(5+a)(10-a+2b)\,,\quad n_{\bm{2}}=2(24-2a+a^2-4b-2ab)\,.
\end{align}

The polarization lattice~\eqref{eqn:LatticeDec} of the $K3$ fibration has rank $\text{rk}(\Lambda)=3$.
Moreover, the theory of lattice index Jacobi forms reduces to that of ordinary Jacobi forms that has been introduced in~\cite{MR781735}.
The expansion of the topological string partition function~\eqref{eq:baseexpansion} encodes the elliptic genera of two types of strings.
Six-dimensional E-strings arise in F-theory from D3-branes that wrap the base of the Hirzebruch surface and the elliptic genus can be identified with $Z_{B}(\tau,\lambda,z)$.
In particular, the index of $Z_{B}(\tau,\lambda,z)$ as a Jacobi form with respect to the geometric elliptic parameter $z$ is
\begin{align}
	m_B=-\frac12D_f^2\cdot \pi^{-1}(B)=3+b-a\,.
\end{align}
It actually turns out that $Z_B(\tau,\lambda,z)$ depends only on $m_B$ and we find
\begin{align}
 \label{eqn:ZBEstring}
	Z_B^{(m_B)}(\tau,\lambda,z)=\frac{1}{\eta(\tau)^{12}}\frac{E_{4,m_B}(\tau,z)}{\phi_{-2,1}(\tau,\lambda)}\,, \quad m_B \in \{0,1,2,3\}\,.
\end{align}
Here $E_{k,m}$ are the Jacobi-Eisenstein series of weight $k$ and index $m$ introduced in appendix~\ref{app:ModOp}.
The expression (\ref{eqn:ZBEstring}) is a generalization of the elliptic genus of an E-string as first observed in \cite{Klemm_1997}. Here the $E_{4,m}$ can be identified with the Jacobi theta function of the $E_8$ root lattice as follows~\cite{MR4105717}
\begin{align}
E_{4,m}(\tau,z) = \vartheta_{\underline{E_8}}(\tau, z \bm{\lambda}_m)\,, \text{  for a  }  \bm{\lambda}_m \in E_8\,,   \text{  such that  }\left(\bm{\lambda}_m , \bm{\lambda}_m\right)_{E_8} = 2m\,,
\end{align}
where 
\begin{align}
\vartheta_{\underline{E_8}}(\tau,\bm{z}) =\sum_{\bm{\lambda}\in E_8} q^{ \frac{1}{2} \left(\bm{\lambda} , \bm{\lambda} \right)_{E_8}} \exp\left( 2\pi i  \left(\bm{\lambda} , \bm{z} \right)_{E_8}\right) \,,
\end{align}
and $(\cdot\,, \cdot)_{E_8}$ is the Killing form associated to the $E_8$ Lie algebra.

On the other hand, the six-dimensional strings that arise from D3-branes wrapping the fiber of the Hirzebruch surface can be identified with dual Heterotic strings.
The elliptic genus is given by $Z_{F}(\tau,\lambda,z)$ in the expansion~\eqref{eq:baseexpansion} of the topological string partition function and the index
\begin{align}
	m_F=-\frac12 D_f^2\cdot \pi^{-1}(F)=6+a\,,
\end{align}
of $Z_{F}(\tau,\lambda,z)$ only depends on $a$, but $Z_{F}(\tau,z,\lambda)$ itself also depends on the value of $b$.
The contribution of massless states to the Heterotic elliptic genus is given by the chiral index of the dual F-theory spectrum~\cite{Lee:2018urn} and this fixes
\begin{align}
	Z_{F}(\tau,\lambda,z) \Big\vert_{\lambda^{-2}}=-2q^{-1}+(2g-2)\left(\chi_{\lambda_{\bm{3}}}-1\right)+n_{\bm{2}}\left(\chi_{\lambda_{\bm{2}}}+\chi_{\lambda_{\bar{\bm{2}}}}\right)-\chi+\mathcal{O}(q)\,,
	\label{eqn:SU2Weyl}
\end{align}
where $\chi_{\lambda_R}$ is the Weyl character associated to the representation $R$ and the label $\lambda^{-2}$ indicates restriction to genus zero contributions.
Note that we substract the uncharged states from the adjoint representation since they are already contained in the contribution from the Euler characteristic.
This fixes the numerators of the elliptic genera
\begin{align}
 \label{eqn:ZF}
	Z_{F}^{(m_B,m_F)}(\tau,z,\lambda)=\frac{1}{\eta(\tau)^{24}}\frac{\Phi^{(m_B,m_F)}(\tau,z)}{\phi_{-2,1}(\tau,\lambda)}\,,
\end{align}
to take the form
\begin{align}
	\begin{split}
	\Phi^{(m_B,1)}(\tau,z)=&-\left(\frac{13}{12}+\frac{m_B}{6}\right)E_4E_{6,1} - \left(\frac{11}{12}-\frac{m_B}{6}\right)E_{4,1}E_6\,,\\
	\Phi^{(m_B,2)}(\tau,z)=&-\left(\frac{13}{12}+\frac{m_B}{12}\right)E_4E_{6,2} - \left(\frac{11}{12}-\frac{m_B}{12}\right)E_{4,2}E_6\,,\\
	\Phi^{(m_B,3)}(\tau,z)=&-\left(\frac{15}{12}+\frac{m_B}{6}\right)E_{4,1}E_{6,2} - \left(\frac{9}{12}-\frac{m_B}{6}\right)E_{4,2}E_{6,1}\,,\\
	\Phi^{(m_B,4)}(\tau,z)=&-\left(\frac{1}{12}+\frac{m_B}{24}\right)E_4\tilde{E}_{6,4} - 2E_{4,2}E_{6,2} + \left(\frac{1}{12}+\frac{m_B}{24}\right)\tilde{E}_{4,4}E_6\,.
	\end{split}
	\label{eqn:NumSU2ab}
\end{align}

Now, we proceed to apply the technology exposed in sections \ref{sec:refined} and  \ref{sec:nawg}  to address the refinement of 5d BPS invariants and the non-Abelian sLWGC for 6d (1,0) supergravity theories with $SU(2)$ gauge symmetry. The essential point of these computations is to regard the theta expansion of Jacobi forms
\begin{equation}
\phi(\tau,z) =\sum_{\mu \in \mathbb{Z}/2m\mathbb{Z}} h_\mu(\phi) \vartheta^{A_1}_{m,\mu}(\tau,z)\,, \quad \vartheta^{A_1}_{m,\mu} (\tau,z) = \sum_{\substack{r \in \mathbb{Z}\\ r \equiv \mu \text{ mod } 2m }} q^{\frac{r^2}{4m}} \zeta^r\,,  \quad \zeta \equiv \mathrm{e}^{2\pi i z}\,,
\label{eqn:ThetaExp1var}
\end{equation}
where $\phi \in J_{k,m}(A_1)$ is a generic Jacobi form of weight $k$ and index $m$. Here we introduced the notation $h_\mu(\phi)$ to denote the $h_\mu$ theta-coefficient of $\phi$. Having this expansion in mind, we look into the theta expansions for the holomorphic Jacobi forms (\ref{eqn:NumSU2ab}) that determine the elliptic genera (\ref{eqn:ZF}). Thus, we summarize our main applications in the following way:
\begin{itemize}
\item The $h$-map projection (\ref{eqn:hmap}) on $\Phi^{(m_B,m_F)}$ fixes the Noether-Lefschetz generator, i.e.
\begin{equation}
\Phi^\pi = \sum_{\mu \in \mathbb{Z}/2m_F\mathbb{Z}} h_\mu\left( \Phi^{(m_B,m_F)}\right) \mathrm{e}_{\mu}\,. 
\label{eqn:NLonevar}
\end{equation}
This object determines all Noether-Lefschetz numbers via the GW-NL theorem (\ref{eqn:NLcoeff}). 
Furthermore, we obtain refined BPS numbers for all curve classes in the $K3$ fiber  of $M_{a,b}^{SU(2)}$ by using the KKP conjecture formula (\ref{eqn:KKP}).
\item We consider the term associated to the Weyl invariant sublattice $2m_F \mathbb{Z}$
\begin{equation}
\Phi^{(m_B,m_F)}(\tau,z)\Big\vert_{2m_F \mathbb{Z}}  \equiv h_{0}\left(\Phi^{(m_B,m_F)}\right) \vartheta_{m_F,0}^{A_1}(\tau,z)\,,
\end{equation}
which is given by genus zero contributions of the elliptic genera (\ref{eqn:ZF}). As proved in section (\ref{subsec:najac}), we find that the non-trivial series contribution
\begin{equation}
\label{eqn:subser}
-2 \sum_{r \in 2m_F \mathbb{Z}_{\geq0} } q^{\frac{r^2}{4m_F}}\chi_{\lambda_{\bm{r+1}}}(z) \subset \Phi^{(m_B,m_F)}(\tau,z)\Big\vert_{2m_F \mathbb{Z}} \,, 
\end{equation}
as well as Weyl-invariance of $\vartheta_{m_F,0}^{A_1}(\tau,z)$, suffice to manifest the non-Abelian sLWGC. We provide a notebook~\cite{DataLink} in which we exemplify explicitly the appearance of the contribution (\ref{eqn:subser}) in the Weyl characters expansions (\ref{eqn:SU2Weyl}) for all the $SU(2)$ models associated with the Jacobi forms (\ref{eqn:NumSU2ab}).
\end{itemize}

To elaborate on explicit computations for the theta expansions, from now on, we keep the discussion to the elliptic genera $Z_{F}(\tau,\lambda,z)$ corresponding to the cases $(m_B,2)$. We remind the reader that this case corresponds to one with non-perturbative gauge symmetry. 
As a starting point, we reckon the theta expansions (\ref{eqn:ThetaExp1var}) of the Jacobi-Eisenstein series $E_{4,2}$ and $E_{6,4}$.  
The Taylor expansions thereof read

\makeatletter
    \def\tagform@#1{\maketag@@@{\normalsize(#1)\@@italiccorr}}
\makeatother
{\scriptsize
\begin{align}
\label{eqn:m2ser}
\begin{split}
E_{4,2}(\tau,z)& = 1+q \left(14 \zeta^{\pm2}+64\zeta^\pm+84\right)  +q^2
   \left(\zeta^{\pm4}+64 \zeta^{\pm3}+280 \zeta^{\pm2}+448
   \zeta^\pm +574\right)+\mathcal{O}\left(q^3\right)\,,\\
E_{6,2}(\tau,z)& = 1+q \left(-10 \zeta^{\pm2}-128 \zeta^{\pm}-228\right)+q^2
   \left(\zeta^{\pm4}-128 \zeta^{\pm3}-1496 \zeta^{\pm2}-3968
   \zeta^{\pm}-5450\right)+\mathcal{O}\left(q^3\right) \,.
   \end{split}
\end{align}
}
Next, we introduce the following basis 
\begin{align}
\label{eqn:m2basis}
\alpha(\tau) \equiv \vartheta_{2,0}(\tau,0)\,, \quad  \beta(\tau) \equiv \vartheta_{2,1}(\tau,0)\,, \quad \gamma(\tau) \equiv \vartheta_{2,2}(\tau,0)\,,
\end{align}
where $\alpha, \beta, \gamma \in M_{1/2}(8)$. See appendix \ref{app:VVMF}. An Ansatz for the Jacobi-Eisenstein series (\ref{eqn:m2ser}) in terms of the basis (\ref{eqn:m2basis}) leads to the following expresions
{\scriptsize
\begin{align}
\begin{split}
h_0(E_{4,2} ) & = \alpha ^7+21 \alpha ^5 \gamma ^2+35 \alpha ^3 \gamma ^4+7 \alpha  \gamma ^6 \,, \\
h_1(E_{4,2}) &  = 32 \alpha ^3 \beta ^3 \gamma +32 \alpha  \beta ^3 \gamma ^3\,,\\
h_2 (E_{4,2}) & = 7 \alpha ^6 \gamma +35 \alpha ^4 \gamma ^3+21 \alpha ^2 \gamma ^5+\gamma ^7\,, \\ 
h_3 (E_{4,2}) & = 32 \alpha ^3 \beta ^3 \gamma +32 \alpha  \beta ^3 \gamma ^3\ \,,\\
 h_0(E_{6,2}) & = \alpha ^{11}-57 \alpha ^9 \gamma ^2-342 \alpha ^7 \gamma ^4-434 \alpha ^5 \gamma ^6-187 \alpha ^3 \gamma ^8-5 \alpha  \gamma ^{10}\,,\\
 h_1(E_{6,2}) & = -64 \alpha ^7 \beta ^3 \gamma -448 \alpha ^5 \beta ^3 \gamma ^3-448 \alpha ^3 \beta ^3\gamma ^5-64 \alpha  \beta ^3 \gamma ^7\,,\\
 h_2 (E_{6,2}) &= -5 \alpha ^{10} \gamma -187 \alpha ^8 \gamma ^3-434 \alpha ^6 \gamma ^5-342 \alpha ^4  \gamma ^7-57 \alpha ^2 \gamma ^9+\gamma ^{11}\,,\\
 h_3(E_{6,2}) & = -64 \alpha ^7 \beta ^3 \gamma -448 \alpha ^5 \beta ^3 \gamma ^3-448 \alpha ^3 \beta ^3  \gamma ^5-64 \alpha  \beta ^3 \gamma ^7 \,.
 \end{split}
 \label{eqn:hEm2}
\end{align}
}In this way, we obtain the following vector-valued modular forms
{\footnotesize
\begin{align}
\label{eqn:vmfm2ser}
\begin{split}
h(E_{4,2}) =& \sum_{\mu \in \{0,1,2,3\} } h_\mu(E_{4,2}) \mathrm{e}_\mu \\
 =&  \left(1+84 q+574 q^2\right)
\mathrm{e}_0+\left(64 q^{\frac{7}{8}}+448 q^{\frac{15}{8}}+1344 q^{\frac{23}{8}}%
\right)\mathrm{e_1} \\
&+ \left( 14 q^{\frac{1}{2}}+280 q^{\frac{3}{2}}+840 q^{\frac{5}{2}}\right) \mathrm{e}_2 +\left(64 q^{\frac{7}{8}}+448 q^{\frac{15}{8}}+1344 q^{\frac{23}{8}}
\right)\mathrm{e_3} + \mathcal{O}(q^3)\,,\\
h(E_{6,2}) = & \sum_{\mu \in \{0,1,2,3\} } h_\mu(E_{6,2}) \mathrm{e}_\mu \\
                 =   & \left(1-228 q-5450 q^2\right)\mathrm{e}_0 + \left(-128 q^{\frac{7}{8}}-3968 q^{\frac{15}{8}}-27264 q^{\frac{23}{8}}\right)\mathrm{e}_1\\
                   &+ \left(-10 q^{\frac{1}{2}}-1496 q^{3/2}-14088 q^{5/2}\right)\mathrm{e}_2 + \left(-128 q^{\frac{7}{8}}-3968 q^{\frac{15}{8}}-27264 q^{\frac{23}{8}}\right)\mathrm{e}_3 + \mathcal{O}(q^3)\,.
\end{split}
\end{align}}Note that the series expansions of (\ref{eqn:m2ser}) and (\ref{eqn:vmfm2ser}) encode the same information in their Fourier coefficients. With these pieces of information, the calculation of the $h$-map projection (\ref{eqn:hmap}) for $\Phi^{(m_B,2)}(\tau,z)$ reduces to
\begin{equation}
\label{eqn:NLm2}
\Phi^\pi= -\left(\frac{13}{12} +\frac{g}{12}\right) E_4 \cdot h(E_{6,2}) - \left( \frac{11}{12} -\frac{g}{12}\right)h(E_{4,2} )\cdot E_6\,,
\end{equation}
which is a vector-valued modular from of weight $19/2$ that transforms under the Weil representation $\rho^*_{L^\vee(A_1)}$. Here we expressed (\ref{eqn:NLm2}) in terms of the genus $g$ of the curve $\mathcal{S}^{\mathrm{b}}_{A_1}$, as $m_B = g$ for the $(m_B,2)$ cases. 

To proceed with the computation of BPS invariants, we introduce the Jacobi form discriminant $\Delta_J$, equivalently the Noether-Lefschetz discriminant $\Delta_{NL}$, given by
\begin{equation}
\Delta_J = 2n\ell  - \frac{r^2 }{2m} \,, \quad \Delta_{NL} = \frac{\Delta_J}{2} + 1 - h \,, 
\end{equation}
where $(n,\ell,r)$ are the degrees of the curves (\ref{eqn:curvesdeg}) correponding to the $TUV$ parameters in the Heterotic language (\ref{eqn:HetTypII}), and $m=2$ for the current cases of discussion.
Making use of the GW-NL correspondence theorem (\ref{eqn:GWNL}), we obtain the GV invariants shown in table \ref{tab:$K3$gv2}.
\begin{table}[h!]
\centering\scalebox{.65}{
  \begin{tabular}{c|ccccccccc}
  $g\backslash \frac{\Delta_J}{2}$&$-1$& $-\frac{1}{2}$ & $-\frac{1}{8}$&$0$&$ \frac{1}{2}$ &$\frac{7}{8}$&$1$ & $ \frac{3}{2}$ & $\frac{15}{8}$   \\\hline
 $0$ & $-2 $ & $-2+2g$& $80+16 g$ & $324-36 g $ & $10384-192 g$ & $68656+624 g$ & $124652-896 g$& $1019376-3312 g$ & $4169808+8208 g$ \\
$1$ & $0$ & $0$ & $ 0$ & $4 $ & $4-4 g$ & $-160-32 g$ & $-636+72 g$ &$-20756+372 g$ &  $-137792-1344 g$\\$2$ & $ 0$ & $0$ & $ 0$ &$ 0$ &$ 0$  & $0$ & $-6$ &$-6+6 g$ & $240+48 g$ \\
 $3$ & $0$ & $0$ & $ 0$ & $0 $& $0 $& $ 0$& $0 $ &$0$ &$0$ \\
$4$ &  $0$ & $0$ & $0$ & $0$ & $0$ & $0$ & $0 $ & $0$ & $ 0$\\
       \end{tabular}
        \caption{$n^g_{[\Delta_J/2]}$ GV invariants of the $K3$ fiber with $\Lambda = U \oplus L^\vee(A_1)\left(-2\right)$. }
	\label{tab:$K3$gv2}. }
\end{table}
We note that all GV invariants $n_\varphi^g$ with the same curve class intersection form $\varphi^2 = \Delta_J$ are equivalent\footnote{Generally speaking, if the respective degrees of $\varphi, \varphi'\in H_2(M,\mathbb{Z})^\pi$ belong to different cosets in $G_\Lambda$, the numbers $n_\varphi^g$ and $n_{\varphi'}^g$ can differ in the case that $\varphi^2  = \varphi'^2$. However, in the examples we show here it occurs that  $n_\varphi^g = n_{\varphi'}^g$  when $\varphi^2  = \varphi'^2$. }, where $\varphi \in H_2(M,\mathbb{Z})^\pi$. Hence we denote by $n_{[\Delta_J/2]}^g$ that class of GV invariants.

The refined version of the latter computation is the KKP conjecture (\ref{eqn:KKP}). For the cases $g \in \{1,2,3\}$, we obtain the BPS numbers that appear in the table \ref{tab:Xref6}.
For the case $g = 0$, we note that the values $\frac{\Delta_J}{2} =-\frac{1}{2}$ and $h=0$ yields the Noether-Lefschetz numbers $NL_{h,(n,\ell,r)} =-2$. This indicates a Noether-Lefschetz divisor of the form $\left[ D_{h,(n,\ell,r)} \right] = T_\imath$. See (\ref{eqn:NL}). Taking into account this observation in (\ref{eqn:RNLpos}), we obtain the refined BPS number shown in the table \ref{tab:Xref}.

\begin{table}[h!]
	\centering
	{\scriptsize
	
	\begin{tabular}{|r|cc|}\hline
			$\mathsf{N}_{j_-,j_+}^{[-1]}$ &$2j_+=$0&1\\ \hline
			$2j_-=$0&& $1$\\ 
			\hline \end{tabular}
		\begin{tabular}{|r|c|}\hline
			$\mathsf{N}_{j_-,j_+}^{[-\frac{1}{2}]}$ &$2j_+=$0\\ \hline
			$2j_-=$0& $2g-2$\\ 
			\hline \end{tabular}
		\begin{tabular}{|r|c|}\hline
			$\mathsf{N}_{j_-,j_+}^{[-\frac{1}{8}]}$ &$2j_+=$0\\ \hline
			$2j_-=$0& $80+16 g$\\ 
			\hline \end{tabular}
			\\
                        \begin{tabular}{|r|ccc|}\hline
			$\mathsf{N}_{j_-,j_+}^{[0]}$ &$2j_+=$0&1&2\\ \hline
			$2j_-=$0&$332-36g  $&&\\ 
			1 &1&&1\\ 
			\hline \end{tabular}
			\\
	 \begin{tabular}{|r|cc|}\hline
			$\mathsf{N}_{j_-,j_+}^{[\frac{1}{2}]}$ &$2j_+=$0&1\\ \hline
			$2j_-=$0&$10392-200 g$&\\ 
			1 &&$2g-2$\\ 
			\hline \end{tabular}
	 \begin{tabular}{|r|cc|}\hline
			$\mathsf{N}_{j_-,j_+}^{[\frac{7}{8}]}$ &$2j_+=$0&1\\ \hline
			$2j_-=$0&$68336+560 g$&\\ 
			1 &&$80+16g$\\ 
			\hline \end{tabular}
         \\
		\begin{tabular}{|r|cccc|}\hline
			$\mathsf{N}_{j_-,j_+}^{[1]}$ &$2j_+=$0&1&2&3\\ \hline
			$2j_-=$0&$123352-752 g$&1&&\\ 
			1 &1&$332-36 g$&1&\\ 
			2 &&1&&1\\  
			\hline \end{tabular}
	 \begin{tabular}{|r|ccc|}\hline
			$\mathsf{N}_{j_-,j_+}^{[\frac{3}{2}]}$ &$2j_+=$0&1 & 2\\ \hline
			$2j_-=$0&$977834-2538 g$& & \\ 
			1 &&$10390-198 g$& \\ 
			2 & & & $2g-2$\\
			\hline \end{tabular}
						}
			\caption{Refined BPS invariants $\mathsf{N}_{j_-,j_+}^{[\Delta_J/2]}$ of the $K3$ fiber with $\Lambda_S = U \oplus L^\vee(A_1)\left(-2\right)$ and $g= 1,2,3$. }
	\label{tab:Xref6}
\end{table}

\begin{table}[h!]
	\centering
	\centering
	{\scriptsize
	
	\begin{tabular}{|r|cc|}\hline
			$\mathsf{N}_{j_-,j_+}^{[-1]}$ &$2j_+=$0&1\\ \hline
			$2j_-=$0&& $1$\\ 
			\hline \end{tabular}
        \begin{tabular}{|r|cc|}\hline
			$\mathsf{N}_{j_-,j_+}^{[-\frac{1}{2}]}$ &$2j_+=$0&1\\ \hline
			$2j_-=$0&& $1$\\ 
			\hline \end{tabular}
		\begin{tabular}{|r|c|}\hline
			$\mathsf{N}_{j_-,j_+}^{[-\frac{1}{8}]}$ &$2j_+=$0\\ \hline
			$2j_-=$0& $80$\\ 
			\hline \end{tabular}
			\\
                        \begin{tabular}{|r|ccc|}\hline
			$\mathsf{N}_{j_-,j_+}^{[0]}$ &$2j_+=$0&1&2\\ \hline
			$2j_-=$0&$332  $&&\\ 
			1 &1&&1\\ 
			\hline \end{tabular}
                        \begin{tabular}{|r|ccc|}\hline
			$\mathsf{N}_{j_-,j_+}^{[\frac{1}{2}]}$ &$2j_+=$0&1&2\\ \hline
			$2j_-=$0&$10392  $&&\\ 
			1 &1&&1\\ 
			\hline \end{tabular}
			\\
	 \begin{tabular}{|r|cc|}\hline
			$\mathsf{N}_{j_-,j_+}^{[\frac{7}{8}]}$ &$2j_+=$0&1\\ \hline
			$2j_-=$0&$68336$&\\ 
			1 &&$80$\\ 
			\hline \end{tabular}
		\begin{tabular}{|r|cccc|}\hline
			$\mathsf{N}_{j_-,j_+}^{[1]}$ &$2j_+=$0&1&2&3\\ \hline
			$2j_-=$0&$123352$&1&&\\ 
			1 &1&$332$&1&\\ 
			2 &&1&&1\\  
			\hline \end{tabular}
				\begin{tabular}{|r|cccc|}\hline
			$\mathsf{N}_{j_-,j_+}^{[\frac{3}{2}]}$ &$2j_+=$0&1&2&3\\ \hline
			$2j_-=$0&$977836$&1&&\\ 
			1 &1&$10392$&1&\\ 
			2 &&1&&1\\  
			\hline \end{tabular}	         
						}
			\caption{Refined BPS invariants $\mathsf{N}_{j_-,j_+}^{[\Delta_J/2]}$ of the $K3$ fiber with $\Lambda_S = U \oplus L^\vee(A_1)\left(-2\right)$ and $g=0$. }
	\label{tab:Xref}
\end{table}

As a complement, we provide further BPS numbers  $\mathsf{N}_{j_-,j_+}^{\varphi}$ in a notebook~\cite{DataLink} in which we regard the KKP conjecture for the rest of the $SU(2)$ models associated with the Jacobi forms (\ref{eqn:NumSU2ab}). 

\subsection{$G=SU(3)$ geometries}
\label{subsec:A2}  

We now want to study geometries $M^{SU(3)}_{a,b}$ where the corresponding gauge group via F-theory is $SU(3)$.
As base we choose again $B=\mathbb{F}_1$ and the points of the corresponding polytopes as well as the generic parts of the linear relations among them are listed in Table~\ref{tab:su3points}.
\begin{table}[h!]
\begin{align*}
\begin{blockarray}{rrrrrrrrrl}
	&&&&C_1&C_2&C_3&B&F\\
\begin{block}{|rrrr|rrrrr|l}
	 1& 0& 0& 0& 0& 0& 1& *& *&\\
	 0& 1& 0& 0& 0& 1& 0& *& *&\\
	-1& 0& 0& 0& 1&-2& 0& *& *&\leftarrow\text{fibral divisor }D_f^1\\
	-2&-1& 0& 0&-2& 1& 2& *& *&\leftarrow\text{fibral divisor }D_f^2\\
	-3&-2& 0& 0& 1& 0&-1& *& *&\leftarrow\text{holomorphic section }E_0\\
	 0& 0& 1& 0& 0& 0& 0& 0& 1&\\
	2a-2&a-2&-1& 0& 0& 0& 0&-1& 1&\leftarrow\text{vertical divisor }\pi^{-1}(B)\\
	 0& 0& 0& 1& 0& 0& 0& 1& 0&\\
	2b-3&b-3&-1&-1& 0& 0& 0& 1& 0&\leftarrow\text{vertical divisor }\pi^{-1}(F)\\
	 0& 0& 0& 0& 0& 0&-2& *& *&\\
\end{block}
\end{blockarray}
\end{align*}
	\caption{The points of a set of reflexive polytopes that lead to elliptically fibered Calabi-Yau manifolds with families of $I_3$ fibers.
	Some parts of the linear relations among the points are given in the last five columns.}
\label{tab:su3points}
\end{table}
We consider $(a,b)$ for $a\in\{-1,0,1\}$ and $b\in\{a-1,a,a+1\}$ as well as $(-2,-2)$, $(2,1)$, $(2,2)$ and $(4,6)$.
Using the general results from~\cite{Klevers:2014bqa} we find that the geometry indeed contains a curve of $I_3$ singularities as well as isolated reducible fibers that via F-theory
lead to matter in the fundamental representation of the associated $SU(3)$ gauge symmetry.
The genus $g$ of the curve $\mathcal{S}_{A_2}^{\mathrm{b}}$ and the multiplicity $n_{\mathbf{3}}$ of the matter loci are given by
\begin{align}
	g=1+\frac12(a - a^2) +  b +  a b\,,\quad n_{\mathbf{3}}=3 (16 + a + a^2 + 2 b - 2 a b)\,,
\end{align}
while the Euler characteristic of the Calabi-Yau is
\begin{align}
\chi=-12 (16 - 2 a - a^2 - 4 b + 2 a b)\,.
\end{align}

We denote the fibral divisor that corresponds to the point $(-1,0,0,0)$ by $D_f^1$ and find the intersections
\begin{align}
	\begin{split}
		\left(D_f^1\right)^2\cdot \pi^{-1}(B)=&2(a-b)-2\,,\\
		\left(D_f^1\right)^2\cdot \pi^{-1}(F)=&-2a-4\,.\\
	\end{split}
\end{align}
The corresponding indices with respect to the geometric elliptic parameter are
\begin{align}
	m_B=1+b-a\,,\quad m_F=a+2\,.
\end{align}
In order to perform the modular bootstrap we need rank two Jacobi forms that are invariant under the Weyl group of $A_2$~\cite{DelZotto:2017mee}.
A basis of the corresponding ring is given by the three Jacobi forms~\cite{MR1775220}
\begin{align}
	\begin{split}
	\label{eqn:Bertola}
		\phi_3=&-\frac{i}{\eta(\tau)^9}\prod\limits_{i=1}^3\theta_1(\tau,u_i)\bigg|_{u_*\rightarrow z_*}\,,\\
		\phi_2=&\left(\sum\limits_{i=1}^3\frac{\mathfrak{d}_{u_i}\theta_1(\tau,u_i)}{\theta_1(\tau,u_i)}\right)\cdot \phi_3\bigg|_{u_*\rightarrow z_*}\,,\\
		\phi_0=&\left(-2\mathfrak{d}_\tau - \frac{E_2}{2}+\frac23\left(\mathfrak{d}^2_{z_1}+\mathfrak{d}^2_{z_2}+\mathfrak{d}_{z_1}\mathfrak{d}_{z_2}\right)\right)\circ\phi_2\,,
	\end{split}
\end{align}
which are of index one and the weight of $\phi_{-k}$ is $k$.
In the definitions we introduced the derivative $\mathfrak{d}_x=(2\pi i)^{-1}\partial_x$, while $u_*\rightarrow z_*$ is the change of parametrization
\begin{align}
	u_1=z_1\,,\quad u_2=z_2-z_1\,,\quad u_3=-z_2\,.
\end{align}

We can again consider the elliptic genera of Heterotic strings that in F-theory arise from D3-branes that wrap the fiber of the Hirzebruch surface.
Using the generic form
\begin{align}
	\begin{split}
	\label{eqn:ZFA2}
		Z_F(\tau,\lambda,z_1,z_2) \Big\vert_{\lambda^{-2}}=&
		   \frac{\Phi^{(m_B,m_F)}(\tau,z_1,z_2)}{\eta^{24}(\tau)\phi_{-2,1}(\tau,\lambda)}\\
		=&-2q^{-1}+(2g-2)\left(\chi_{\bm{\lambda}_{\bm{8}}}-2\right)+n_{\bm{3}}\left(\chi_{\bm{\lambda}_{\bm{3}}}+\chi_{\bm{\lambda}_{\bar{\bm{3}}}}\right)-\chi+\mathcal{O}(q)\,,
	\end{split}
\end{align}
we can fix the numerators
\begin{align}
  \label{eqn:JacNums}
	\begin{split}
		\Phi^{(m_B,1)}=&E_4E_6\phi_0-12^{-2}\left[(13+2m_B)E_4^3+(11-2m_B)E_6^2\right]\phi_{2}\,,\\
		\Phi^{(m_B,2)}=&-2^{-1}E_4E_6\phi_0^2 +12^{-2}\left[(13+m_B)E_4^3+(11-m_B)E_6^2\right]\phi_0\phi_{2}\\
		&-72^{-1}E_4^2E_6\phi_{2}^2+2^{-1}\cdot12^{-3} (1+m_B) \left(E_6^2-E_4^3\right)E_4\phi_3^2\,, \\
		\Phi^{(m_B,3)}=&4^{-1}E_4E_6\phi_0^3-576^{-1}\left[(3\cdot13+2m_B)E_4^3+(3\cdot11-2m_B)E_6^2\right]\phi_0^2\phi_2\\
		&+48^{-1}E_4^2E_6\phi_0\phi_2^2-12^{-4}\left[(9-2m_B)E_4^3+(15+2m_B)E_6^2\right]E_4\phi_2^3\\
		&+2^{-2}\cdot 12^{-4}(3+2m_B)\left(E_4^3-E_6^2\right) \left(12E_4\phi_0
		-2 E_6\phi_2\right)\phi_3^2\,.
	\end{split}
\end{align}
Note that we are not aware of a higher rank generalization of Eisenstein Jacobi series and therefore used the generic generators for the ring of Weyl invariant $A_2$ lattice Jacobi forms. We remind the reader that each Jacobi form $\Phi^{(m_B,m_F)}$ in (\ref{eqn:JacNums}) corresponds to a $K3$ fibration with polarization lattice
\begin{equation}
\Lambda = U \oplus L^\vee(A_2)(-m_F)\,.
\end{equation} 
To make contact with the applications of sections \ref{sec:refined} and \ref{sec:nawg}, in the upcoming, we exemplify the theta expansion calculations for the elliptic genera (\ref{eqn:ZFA2}), or equivalent the numerators (\ref{eqn:JacNums}).
  For simplicity of the exposition we restrict to the cases when $m_F =1$. However,  similar computations are possible for non-perturbative gauge cases, in which $m_F >1$. 
  
    \begin{figure}[h!]
	\centering
	\begin{minipage}[b]{.4\textwidth}
	\includegraphics[width=\linewidth]{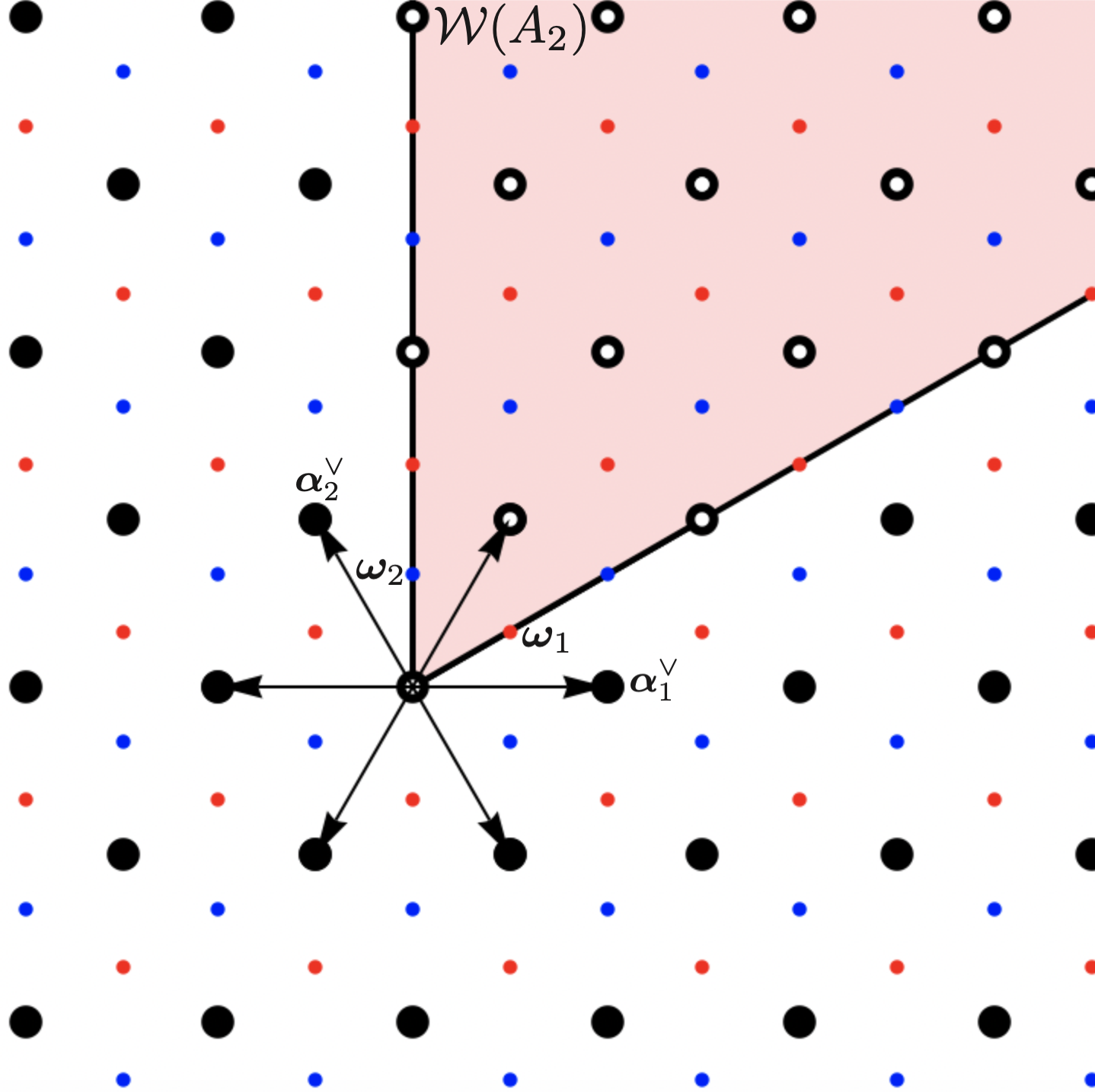}
	\end{minipage}
	\hspace{1cm}
	\caption{Weight lattice $L_{\text{w}}(A_2)$.  The black dots represent the coroot lattice $L^\vee (A_2)$. The red and blue dots represent the cosets associated to $\bm{\omega}_1$ and $\bm{\omega}_2$ respectively. The red shaded region represents the fundamental Weyl chamber $\mathcal{W}(A_2)$. The arrows point out the roots in $\Phi(A_2) \subset L^\vee(A_2)$. Moreover, we label the set of dominant weights $\bm{\lambda}_R$ in $L^\vee(A_2)$ by circles \Circpipe.  To each point \text{ }\Circpipe \text{ } there is an associated superextremal resonance transforming in the irreducible representation $R$.}
	\label{fig:A2lat}
\end{figure}

  We know that the discriminant group is of the form $G_{L^\vee(A_n)} \cong \mathbb{Z}/(n+1)\mathbb{Z}$ \cite{MR3586372}. Consequently, we can decompose the weight lattice $L_{\text{w}}(A_2)$ in terms of cosets. Without loss of generality,  we choose the fundamental weights as representatives of such cosets, i.e.
     \begin{equation}
     L_{\text{w}}\left(A_2\right) \cong L^\vee(A_2) \oplus \left(\bm{\omega}_1  + L^\vee(A_2) \right) \oplus \left(\bm{\omega}_2  + L^\vee(A_2) \right) \,.
     \end{equation}
     See figure \ref{fig:A2lat}. Given a $L^\vee(A_2)$-lattice Jacobi form $\phi$, we want to perform its theta expansion
     \begin{align}
     \phi(\tau,z_1,z_2) = h_{\bm{0}}(\phi) \vartheta_{1,\bm{0}}^{A_2}(\tau,z_1,z_2) + h_{\bm{\omega}_1}(\phi) \vartheta_{1,\bm{\omega}_1}^{A_2}(\tau,z_1,z_2)+ h_{\bm{\omega}_2}(\phi) \vartheta_{1,\bm{\omega}_2}^{A_2}(\tau,z_1,z_2)\,.
     \end{align}
   As an example, we regard the ring generators (\ref{eqn:Bertola}) of $W(A_2)$ invariant Jacobi forms. To calculate the theta coefficients $h_{\bm{\mu}}(\phi)$, we use the basis of vector-valued modular forms $M_{1/2}(24)$ in (\ref{eqn:VVMFbasis}). 
    For this, we introduce the notation
    \begin{equation}
    \theta_{N,\ell}(\tau) \equiv \vartheta^{A_1}_{N,\ell}(\tau,0)\,.
    \end{equation}
    Then, the theta coefficients for the $J_{*,*}(A_2)$ generators (\ref{eqn:Bertola}) read
    {\small
    \begin{align}
      \begin{split}
       h_{\bm{0}}(\phi_{3}) &= 0\,, \\
       h_{\bm{\omega}_1} (\phi_3) &=   \eta^{-9}(\theta_{6,1} - \theta_{6,5})\,,\\ 
        h_{\bm{\omega}_2} (\phi_3) &=  - h_{\bm{\omega}_1}(\phi_3)\,,\\ 
        h_{\bm{0}}( \phi_2) &=  3 \eta^{-9} \left(\theta_{6,0}^2\theta_{6,3}-\theta_{6,3} \theta_{6,6}^2\right) \,,\\
        h_{\bm{\omega_1}}(\phi_2)  &=  \frac{1}{2}\eta^{-9}\left(4\theta_{6,2}\theta_{6,3}\theta_{6,6}+\theta_{6,1} \theta_{6,6}^2 +\theta_{6,5} \theta_{6,6}^2 -4\theta_{6,0}\theta_{6,3}\theta_{6,4}-       
                                                              \theta_{6,0}^2\theta_{6,1}- \theta_{6,0}^2\theta_{6,5}\right)\,, \\
        h_{\bm{\omega_2}}(\phi_2)  &=  h_{\bm{\omega}_1}(\phi_2)\,,\\
        h_{\bm{0}}(\phi_0) &= -2\partial^{\text{RS}} h_{\bm{0}}(\phi_2)\,,\\
        h_{\bm{\omega_1}}(\phi_0) &= -2\partial^{\text{RS}} h_{\bm{\omega}_1}(\phi_2)\,,\\
         h_{\bm{\omega}_2}(\phi_0) &=-2 \partial^{\text{RS}} h_{\bm{\omega}_2}(\phi_2)\,.        
         \end{split}
        \label{eqn:ThetaDecBer}
    \end{align}
    }Here $\partial^{\text{RS}}: M_k(\Gamma) \rightarrow M_{k+2}(\Gamma)$ is the \textit{Ramanujan-Serre derivative}, which acts on a holomorphic modular form $f \in M_k(\Gamma)$ with $\Gamma \subseteq \mathrm{SL}(2,\mathbb{Z})$ as follows \cite{dabholkar2012quantum} 
\begin{equation}
\partial^{\text{RS}} f= \mathfrak{d}_\tau f - \frac{k}{12} E_2 f\,.
\end{equation}
To explain the Ramanujan-Serre derivative in (\ref{eqn:ThetaDecBer}), we notice that the heat differential operator $\mathcal{L} = -2 \mathfrak{d}_\tau + \Delta_{A_2}$ acts on the theta functions as $\mathcal{L} \vartheta_{1,\bm{\mu}}^{A_2} = 0$, where $\Delta_{A_2}$ is the Laplacian operator associated with the Killing form in $A_2$. Then, by the definition (\ref{eqn:Bertola}), we have that
\begin{equation}
     \phi_0 = \left(\mathcal{L} -\frac{1}{2}E_2\right)\circ \phi_2 = 
      -2 \sum_{\bm{\mu}}\partial^{\text{RS}}h_{\bm{\mu}}(\phi_2) \cdot \vartheta_{1,\bm{\mu}}^{A_2}\,,
\end{equation}
where we identified that each $h_{\bm{\mu}}(\phi_2)$ has weight $k=-3$. Notice that the theta coefficients (\ref{eqn:ThetaDecBer}) are all the information we need to obtain the theta expansions of $\Phi^{(m_B,1)}$. 

\begin{table}[h!]
{\scriptsize
\centering
\begin{tabular}{|c |c |} 
 \hline
$n$ & $p_{n-1}(\bm{z})$ \\ [0.5ex]  \hline
  $0$& $\textcolor{red}{-2\chi_{(0,0)}}$\\ 
  $1$ &  $(352 - 72 m_{B})\chi_{(0,0)} \textcolor{red}{-2 \chi_{(1,1)}(\bm{z})}$  \\
    $2$ &  $(120480-504m_B)\chi_{(0,0)} + (348-72 m_B) \chi_{(1,1)}(\bm{z})$ \\
      $3$ &  $(5444788 - 2592 m_B) \chi_{(0, 0)} +(121180 -648m_B)\chi_{(1,1)}(\bm{z}) \textcolor{red}{-2\chi_{(3,0)}(\bm{z}) -2\chi_{(0,3)}(\bm{z}) }$    \\
   $4$  & $123698662-10512 m_B + (5686448-3744 m_B)\chi_{(1,1)}(\bm{z}) +(350-72m_B)\left(\chi_{(3,0)}(\bm{z}) + \chi_{(0,3)}(\bm{z}) \right) \textcolor{red}{-2\chi_{(2,2)}(\bm{z})} $\\
 \hline
\end{tabular}
\caption{Weyl character expansions of $Z(\tau,\bm{z}) \vert_{L^\vee(A_2)}$. We indicate by red color those contributions associated with the superextremal resonances. Note that $m_B \in \{0,1,2\}$.}
\label{tab:WeylSU3}}
 \end{table}
 
Now, we consider the non-Abelian sLWGC. The contributions of the Weyl invariant sublattice $L^\vee(A_2)$ to the elliptic genus arise from the term
\begin{equation}
\label{eqn:EGsublat}
Z(\tau,z_1,z_2)\Big\vert_{L^\vee(A_2)} \equiv  \frac{1}{\eta^{24}(\tau)}h_{\bm{0}}\left(\Phi^{(m_B,1)}\right)\vartheta_{1,\bm{0}}^{A_2}(\tau,z_1,z_2)\,,
\end{equation}
Let us regard the Fourier expansions of (\ref{eqn:EGsublat}) as follows
\begin{align}
Z(\tau,z_1,z_2)\Big\vert_{L^\vee(A_2)}  = \sum_{n= 0}^\infty p_{n-1}(z_1,z_2) q^{n-1}\,,
\end{align}
where $p_n(z_1,z_2)$ is a Poincar\'e polynomial in $\mathbb{Z}[\zeta_1^\pm,\zeta_2^\pm]$. Due to our \textbf{Claim I} in section \ref{subsec:sumsdominant}, we can express these polynomials in terms of Weyl characters $\chi_{(\lambda_1,\lambda_2)}(z_1,z_2)$, where $(\lambda_1,\lambda_2)$ is a dominant weight corresponding to an irreducible representation of $SU(3)$
\footnote{For instance, the weights $(1,0)$ and $(1,1)$ correspond to the fundamental representation $\bm{3}$ and the adjoint representation $\bm{8}$ respectively.}. 
In table \ref{tab:WeylSU3}, we provide the Weyl character expansions of polynomials $p_{n-1}(z_1,z_2)$ up to $n=4$. On the one hand, we find superextremal states at each excitation level
\begin{equation}
n(\lambda_1,\lambda_2) = \frac{1}{3}\left(\lambda_1^2 + \lambda_1 \lambda_2 + \lambda_2^2\right)\,, \quad (\lambda_1,\lambda_2) \in L_{\text{w}}(A_2) \cap L^\vee(A_2)\,,
\end{equation}
which corresponds to nearly tensionless strings of mass $M_n^2 = 4 T(n-1)$ and tension $T = \text{Vol}_{\omega}(F)$~\cite{Lee:2018spm}. 
On the other hand, as proved in section \ref{subsec:najac}, there is a contribution deriving from (\ref{eqn:thetaChambersDec}) that counts at every $(\lambda_1,\lambda_2) \in P_+(A_2) \cap L^\vee(A_2)$ superextremal resonances transforming in the $SU(3)$ irreducible representation $R$ with highest weight $(\lambda_1,\lambda_2)$. In table \ref{tab:WeylSU3}, we marked such contributions with red color.  With these two points together, the Weyl invariant sublattice $L^\vee(A_2) \subset L_{\text{w}}(A_2)$ satisfies the non-Abelian sLWGC.      

More generally, the non-Abelian sLWGC holds for all $SU(3)$ geometries with polarization lattice $\Lambda = U \oplus L^\vee(A_2)(-m_F)$. There, the Weyl invariant sublattice that fulfils the conjecture is $m_F L^\vee(A_2) \subset L_{\text{w}}(A_2)$. To illustrate this, we provide a complementary notebook~\cite{DataLink} in which we provide higher order Weyl characters expansions of the elliptic genera (\ref{eqn:ZFA2}), together with the terms in the first line of (\ref{eqn:thetaChambersDec}).

In the following we regard the computation of BPS invariants obtained by using the GW-NL correspondence theorem (\ref{eqn:GWNL}), as well as the KKP conjecture (\ref{eqn:KKP}). In our example of interest, i.e. those geometries that give rise to elliptic genera with numerator $\Phi^{(m_B,1)}$, the Noether-Lefschetz generator read 

{\small
\begin{equation}
\Phi^\pi = \sum_{\bm{\mu}\in\{\bm{0},\bm{\omega}_1,\bm{\omega}_2\}} \left[E_4E_6 h_{\bm{\mu}} (\phi_0)-12^{-2}\left((13+2m_B)E_4^3+(11-2m_B)E_6^2\right)h_{\bm{\mu}}(\phi_{2}) \right] \mathrm{e}_{\bm{\mu}}\,.\\
\end{equation}}
The Noether-Lefschetz discriminant $\Delta_{NL}$ is given by
\begin{equation}
\Delta_J = 2n\ell  - \frac23 (\lambda_1^2 +\lambda_1 \lambda_2 + \lambda_2^2) \,, \quad \Delta_{NL} = \frac{\Delta_J}{2} + 1 - h \,, 
\end{equation}
where $(n,\ell, \lambda_1,\lambda_2)$ are the degrees of the curves (\ref{eqn:curvesdeg}) correponding to the $TU\bm{V}$ parameters in the Heterotic language (\ref{eqn:HetTypII}). With this information we calculate BPS invariants associated to the $K3$ fiber. We provide a few GV invariants in the table \ref{tab:$K3$gvSU3} and their corresponding refinements in the table \ref{tab:SU3ref}.

\begin{table}[h!]
\centering\scalebox{.7}{
  \begin{tabular}{c|ccccccc}
  $g\backslash \frac{\Delta_J}{2}$&$-1$&$-\frac{1}{3}$&$0$&$ \frac{2}{3}$ &$1$&$\frac{5}{3}$ & $ 2$ \\ \hline
 $0$ & $-2 $ & $24+12m_B $ & $348-72 m_B $ & $26580 +168 m_B $ & $121176-648 m_B$ & $1747434 +1104 m_B$& $5687144-3888 m_B$ \\
$1$ & $0$ & $ 0$ & $4 $ & $-48-24m_B$ & $-684+144 m_B$ & $-53304 -408 m_B$ &$-244424+1728 m_B$ \\
$2$ & $ 0$ &$ 0$ &$ 0$ &$ 0$  & $-6$ & $72+36m_B$ &$1012-216 m_B$ \\
 $3$ & $0$ & $ 0$ & $0 $& $0 $& $ 0$& $0 $ &$8$ \\
       \end{tabular}
        \caption{$n^g_{[\Delta_J/2]}$ GV invariants of the $K3$ fiber with polarization lattice $\Lambda = U \oplus L^\vee(A_2)(-1)$. Here $m_B \in \{0,1,2\}$.}
	\label{tab:$K3$gvSU3}. }
	\end{table}

\begin{table}[h!]
	\centering
	{\scriptsize
	
	\begin{tabular}{|r|cc|}\hline
			$\mathsf{N}_{j_-,j_+}^{[-1]}$ &$2j_+=$0&1\\ \hline
			$2j_-=$0&& $1$\\ 
			\hline \end{tabular}
		\begin{tabular}{|r|c|}\hline
			$\mathsf{N}_{j_-,j_+}^{[-\frac{1}{3}]}$ &$2j_+=$0\\ \hline
			$2j_-=$0& $24+12m_B$\\ 
			\hline \end{tabular}
			\\
                        \begin{tabular}{|r|ccc|}\hline
			$\mathsf{N}_{j_-,j_+}^{[0]}$ &$2j_+=$0&1&2\\ \hline
			$2j_-=$0&$356-72m_B  $&&\\ 
			1 &1&&1\\ 
			\hline \end{tabular}
	 \begin{tabular}{|r|cc|}\hline
			$\mathsf{N}_{j_-,j_+}^{[\frac{2}{3}]}$ &$2j_+=$0&1\\ \hline
			$2j_-=$0&$26484+120m_B$&\\ 
			1 &&$24+12m_B$\\ 
			\hline \end{tabular}
         \\
		\begin{tabular}{|r|cccc|}\hline
			$\mathsf{N}_{j_-,j_+}^{[1]}$ &$2j_+=$0&1&2&3\\ \hline
			$2j_-=$0&$119780-360m_B$&1&&\\ 
			1 &1&$356-72m_B$&1&\\ 
			2 &&1&&1\\  
			\hline \end{tabular}
		\begin{tabular}{|r|ccc|}\hline
			$\mathsf{N}_{j_-,j_+}^{[\frac{5}{3}]}$ &$2j_+=$0&1&2\\ \hline
			$2j_-=0$& $1641186 +468 m_B$ & & \\
			1 &  &$26508 +132 m_B$&  \\
			2 & &  &$24+12m_B$\\
			\hline \end{tabular}
			\\
					\begin{tabular}{|r|ccccc|}\hline
			$\mathsf{N}_{j_-,j_+}^{[2]}$ &$2j_+=$0&1&2& 3 & 4\\ \hline
			$2j_-=0$ & $5203476 -1512 m_B $ & $2$ & & $1$ &\\ 
			1 & $2$ &$120136 -432 m_B$& $2$  & & \\ 
			2 & &  $2$ &$356 -72 m_B$ &  $1$ & \\  
			3&  &  & $1$ & & $1$  \\
			\hline \end{tabular}
						}
			\caption{Refined BPS numbers $\mathsf{N}_{j_-,j_+}^{[\Delta_J/2]}$ of the $K3$ fiber with polarization lattice $\Lambda = U \oplus L^\vee(A_2)(-1)$. Here $m_B\in \{0,1,2\}$. }
	\label{tab:SU3ref}
\end{table}

\subsection{$G= G_2$ geometry}
\label{subsec:G2}

Let us now discuss an example of the non-simply laced Lie algebra $G_2$. 
To this end we consider the geometry introduced in~\cite{Bouchard:2003bu} that corresponds to the toric data in Table~\ref{tab:g2points}.
\begin{table}[h!]
\begin{align*}
\begin{blockarray}{rrrrrrrrrl}
	&&&&C_1&C_2&C_3&C_4&C_5\\
\begin{block}{|rrrr|rrrrr|l}
	-1& 0& 0& 0& 0& 0& 0& 0& 1&\\
	 0&-1& 0& 0& 0& 0& 0& 1& 0&\\
	 2& 3& 0& 0& 1&-2& 0& 0& 0&\\
	 2& 3&-1& 0&-2& 1& 0& 0& 1&\leftarrow\text{fibral divisor }E_1\\
	 2& 3&-2& 0& 1& 0& 0& 1&-2&\leftarrow\text{fibral divisor }E_2\\
	 1& 1&-1& 0& 0& 0& 0&-2& 3&\\
	 2& 3& 2& 1& 0& 0& 1& 0& 0&\\
	 2& 3& 0&-1& 0& 0& 1& 0& 0&\leftarrow\text{vertical divisor }\pi^{-1}(F)\\
	 2& 3& 1& 0& 0& 1&-2& 0& 0&\leftarrow\text{vertical divisor }\pi^{-1}(B)\\
	 0& 0& 0& 0& 0& 0& 0& 0&-3&\\
\end{block}
\end{blockarray}
\end{align*}
	\caption{Toric data associated to an elliptic fibration that leads to a $G_2$ gauge symmetry in F-theory.}
\label{tab:g2points}
\end{table}
It is an elliptic fibration over the Hirzebruch surface $\mathbb{F}_2$ and we choose a basis of fibral divisors that leads to the polarization lattice 
\begin{equation}
\Lambda = U \oplus L^\vee(G_2)(-1)\,.
\end{equation}

As discussed in pervious sections, our main object to compute is the elliptic genus $Z_{F}(\tau,\lambda,\bm{z})$ resulting from D3 branes wrapping the Hirzebruch surface fiber curve $F$. For this, we consider the basis for the ring of $G_2$ Weyl-invariant Jacobi forms given by~\cite{MR1775220}
\begin{align}
\label{eqn:G2WeylJac}
\begin{split}
\varphi_6 (\tau,\bm{z})&= \frac{1}{2} \phi_3^2(\tau, \bm{x})\Big\vert_{x_*\rightarrow z_*}\,,\\
\varphi_2(\tau,\bm{z})&  = \phi_2(\tau, \bm{x})\Big\vert_{x_*\rightarrow z_*}\,,\\
\varphi_0(\tau,\bm{z})&  = \phi_0(\tau, \bm{x})\Big\vert_{x_*\rightarrow z_*}\,,
\end{split}
\end{align}
where $\{\phi_3,\phi_2,\phi_0\}$ is the basis of $A_2$ Weyl-invariant Jacobi forms as defined in (\ref{eqn:Bertola}). Moreover, we introduced in (\ref{eqn:G2WeylJac}) the reparametrization
\begin{equation}
x_1 = z_2-2z_1 \,, \quad x_2 = -z_1\,.
\end{equation} 
We observe that the elliptic genus that follows an expansion of the form
\begin{equation}
\label{eqn:WeylG2Exp}
Z_F(\tau,\lambda,\bm{z}) \Big\vert_{\lambda^{-2}}= -2 q^{-1} -2 \chi_{\bm{\lambda}_{\bm{14}}}(\bm{z}) + 2n_{\bm{7}} \chi_{\bm{\lambda}_{\bm{7}}}(\bm{z}) - \chi - 2n_{\bm{7}} +2\text{rk}(G_2)+  \mathcal{O}(q)\,.
\end{equation}
Here the Weyl characters are those associated with the Lie algebra $G_2$ and not $A_2$. Moreover, $\bm{14}$ is the adjoint representation of $G_2$ that has an associated highest weight $\bm{\lambda}_{\bm{14}} = \bm{\omega}_2$; $\bm{7}$ is the fundamental representation with associated highest weight $\bm{\lambda}_{\bm{7}} = \bm{\omega}_1$. See figure \ref{fig2}. In this particular geometry, we identify that $n_{\bm{7}} = 16$.  For further discussion on counting matter representations via Gromov-Witten invariants, see the references~\cite{Kashani-Poor:2019jyo,Paul-KonstantinOehlmann:2019jgr}. Thus, the explicit form of the elliptic genus reads
\begin{equation}
\label{eqn:G2Num}
Z_F(\tau,\lambda,\bm{z}) = \frac{\Phi(\tau,\bm{z} )}{\eta^{24}(\tau) \phi_{-2,1}(\tau,\lambda)}\,, \quad \Phi(\tau,\bm{z}) = E_4 E_6 \varphi_0 - \frac{1}{72}\left( 7 E_4^3 +5  E_6^2\right)\varphi_{2}\,.
\end{equation}
As in previous sections, we provide an additional notebook~\cite{DataLink} that contains higher order Weyl character expansions (\ref{eqn:WeylG2Exp}) of the elliptic genus (\ref{eqn:G2Num}). Moreover, we include the terms (\ref{eqn:thetaChambersDec}) required to satisfy the non-Abelian sLWGC. In this case, the Weyl invariant sublattice $L^\vee(G_2) \subset L_{\text{w}}(G_2)$ fulfils the conjecture. 

As exemplified in section \ref{subsec:sumsdominant}, we take as representatives of the discriminant group $G_{L^\vee(G_2)}$ the lattice points $\bm{\mu}_0 = (0,0), \bm{\mu}_1 =(4,-2) ,  \bm{\mu}_2 = (2,-1)$.  We remark that generally speaking the space of $G_2$ theta functions (\ref{eqn:spaceTheta}) differs from those of $A_2$. 
Despite this difference, we find out the following equivalences 
\begin{align}
\label{eqn:hthetaG2}
\begin{split}
h_{\bm{0}}(\varphi_2) & =h_{\bm{0}}(\phi_2)\,, \quad h_{\bm{0}}(\varphi_0)  =h_{\bm{0}}(\phi_0) \,,\\
h_{\bm{\mu}_1}(\varphi_2) & =h_{\bm{\mu}_2}(\varphi_2)=  h_{\tilde{\bm{\omega}}_1}(\phi_2) =   h_{\tilde{\bm{\omega}}_2}(\phi_2)\,, \\
h_{\bm{\mu}_1}(\varphi_0) & =h_{\bm{\mu}_2}(\varphi_0)=  h_{\tilde{\bm{\omega}}_1}(\phi_0) =   h_{\tilde{\bm{\omega}}_2}(\phi_0)\,.
\end{split}
\end{align}
Here we used the labels $\tilde{\bm{\omega}}_i$ to indicate the fundamental weights in $L_{\text{w}}(A_2)$. We remind that the explicit expressions for (\ref{eqn:hthetaG2}) are given in (\ref{eqn:ThetaDecBer}). Similarly as in the previous section, we calculate the Noether-Lefschetz generator via the $h$-map projection of  (\ref{eqn:G2Num}) by using the theta coefficients (\ref{eqn:hthetaG2}). Moreover, now the Noether-Lefschetz discriminant $\Delta_{NL}$ now reads
\begin{equation}
\Delta_J = 2n\ell  - 2 \left(\lambda_1^2 +\lambda_1 \lambda_2 +\frac13 \lambda_2^2\right) \,, \quad \Delta_{NL} = \frac{\Delta_J}{2} + 1 - h \,, 
\end{equation} 
Using the GW-NL correspondence theorem (\ref{eqn:GWNL}), we obtain the GV invariants shown in table \ref{tab:$K3$gvG2}. Note that our results are exactly those provided by the Heterotic 1-loop computation in table 5.4 of~\cite{Weiss:2007tk}. By using the KKP conjecture (\ref{eqn:KKP}),  we obtain the refinement for the latter invariants that appear in the table \ref{tab:XrefG2}.
\begin{table}[h!]
\centering\scalebox{.7}{
  \begin{tabular}{c|ccccccccccc }
  $g\backslash \frac{\Delta_J}{2}$&$-1$&$-\frac{1}{3}$&$0$&$ \frac{2}{3}$ &$1$&$\frac{5}{3}$ & $ 2$ & $\frac{8}{3}$ & $3$ & $\frac{11}{3}$ & $4$\\\hline
 $0$ & $-2 $ & $30 $ & $312 $ & $26664 $ & $120852$ & $1747986$& $5685200$ & $49588776$ & $135063180$ & $886842720$ & $2156305536$\\ 
$1$ & $0$ & $ 0$ & $4 $ & $-60$ & $-612$ & $-53508$ &$-243560$ &  $-3656196$ &$-12097980$& $-109879200$ &$-305208720$ \\
$2$ & $ 0$ &$ 0$ &$ 0$ &$ 0$  & $-6$ & $90$ &$904$ & $80472$ & $367458$ & $5671932$ &$19003080$ \\
$4$ &  $0$ & $0$ & $0$ & $0$ & $0$ & $0 $ & $0$ & $ 0$ & $-10$ & $150$ &$1464$ \\
$5$ & $0$ & $0$ & $0$ & $0$ & $0$ & $0$ & $0$ & $0$ & $0$ & $0$ & $12$
       \end{tabular}
        \caption{$n^g_{[\Delta_J/2]}$ GV invariants of the $K3$ fiber of $M_{G_2}$. }
	\label{tab:$K3$gvG2}. }
	\end{table}
	
	\begin{table}[h!]
	\centering
	\centering
	{\scriptsize
	
	\begin{tabular}{|r|cc|}\hline
			$\mathsf{N}_{j_-j_+}^{[-1]}$ &$2j_+=$0&1\\ \hline
			$2j_-=$0&& $1$\\ 
			\hline \end{tabular}
		\begin{tabular}{|r|c|}\hline
			$\mathsf{N}_{j_-j_+}^{[-\frac{1}{3}]}$ &$2j_+=$0\\ \hline
			$2j_-=$0& $30$\\ 
			\hline \end{tabular}
			\\
                        \begin{tabular}{|r|ccc|}\hline
			$\mathsf{N}_{j_-j_+}^{[0]}$ &$2j_+=$0&1&2\\ \hline
			$2j_-=$0&$320  $&&\\ 
			1 &1&&1\\ 
			\hline \end{tabular}
	 \begin{tabular}{|r|cc|}\hline
			$\mathsf{N}_{j_-j_+}^{[\frac{2}{3}]}$ &$2j_+=$0&1\\ \hline
			$2j_-=$0&$26544$&\\ 
			1 &&$30$\\ 
			\hline \end{tabular}
         \\
		\begin{tabular}{|r|cccc|}\hline
			$\mathsf{N}_{j_-j_+}^{[1]}$ &$2j_+=$0&1&2&3\\ \hline
			$2j_-=$0&$119600$&1&&\\ 
			1 &1&$320$&1&\\ 
			2 &&1&&1\\  
			\hline \end{tabular}
		\begin{tabular}{|r|ccc|}\hline
			$\mathsf{N}_{j_-j_+}^{[\frac{5}{3}]}$ &$2j_+=$0&1&2\\ \hline
			$2j_-=0$& $1641420 $ & & \\
			1 &  &$26574 $&  \\
			2 & &  &$30$\\
			\hline \end{tabular}
			\\
					\begin{tabular}{|r|ccccc|}\hline
			$\mathsf{N}_{j_-j_+}^{[2]}$ &$2j_+=$0&1&2& 3 & 4\\ \hline
			$2j_-=0$ & $5202720 $ & $2$ & & $1$ &\\ 
			1 & $2$ &$119920 $& $2$  & & \\ 
			2 & &  $2$ &$320$ &  $1$ & \\  
			3&  &  & $1$ & & $1$  \\
			\hline \end{tabular}
						}
			\caption{Refined BPS invariants $\mathsf{N}_{j_-,j_+}^{[\Delta_J/2]}$ of $M_{G_2}$. }
	\label{tab:XrefG2}
\end{table}
\section{Sublattice conjectures for M-theory on genus one fibrations}
\label{sec:genusone}
With our previous discussion we have focussed on F-theory compactifications on elliptically fibered Calabi-Yau threefolds that, in order to exhibit an appropriate weak coupling limit for the gauge theory, also had a $K3$ fibration.
In particular, the identification of the tower of states that satisfies the sublattice weak gravity conjecture and its non-Abelian extension was based on the duality with Heterotic strings, compactified on another $K3$, that arise from D3 branes wrapping the base of the $K3$ fiber.
More precisely, the elliptic genus of the Heterotic strings encode certain particle-like exitations and the properties of Jacobi forms imply that a superextremal subset populates the desired sublattice.

If we compactify on a circle to five dimensions, the excitations of the wrapped strings turn into BPS particles and the same argument about the sublattice holds.
However, we also have the freedom to perform an orbifolding on the Heterotic side of the duality and consider compactifications on $(K3\times S^1)/\mathbb{Z}_N$.
These are dual to M-theory compactified on genus one fibered Calabi-Yau threefolds that do not have a section but only $N$-sections~\cite{Kachru:1997bz,Cota:2019cjx,Banlaki:2019bxr}.
In this situation the elliptic genus of the Heterotic strings is replaced by the twisted elliptic genera and the arguments for the sublattice conjectures have to be modified.

\subsection{Genus one fibered $K3$ surfaces without section}
As a starting point, let us consider the 5d Heterotic strings on $(K3\times S^1)/\mathbb{Z}_N$ with no Wilson lines turned on.
The $\mathbb{Z}_N$ orbifold group acts on the circle via the shift
\begin{equation}
 \delta: x \mapsto x + \frac{2\pi}{N}R\,,
\end{equation}
and on the $K3$ via some non-trivial order $N$ automorphism.
We choose the momentum and winding numbers $(k,w)$ along the circle to be quantized in $\mathbb{Z}\oplus\mathbb{Z}$ such that the left- and right-moving momenta satisfy
\begin{equation}
	(p_L,p_R) = \frac{1}{\sqrt{2}} \left( \frac{k}{R} - \frac{wR}{N},\,\frac{k}{R} + \frac{wR}{N}\right)\,,
\end{equation}
With this in mind, the momentum lattice $\Gamma_N^{1,1}$ is equipped with the quadratic form
\begin{equation}
\label{eqn:CHLint}
	p_L^2 - p_R^2 =  \frac{2kw}{N}\,.
\end{equation}
Note that a state with momentum number $k$ will have eigenvalue $\exp(2\pi i k/N)$ under the action of $\delta$.
Moreover, with $\psi_{r,k}$ we denote a state in the $r$-twisted sector on $K3/\mathbb{Z}_N$ with eigenvalue $g=\exp(-2\pi i k/N)$, where $g$ acts as a $\mathbb{Z}_N$ automorphism on the CFT associated with the $K3$~\cite{Chattopadhyaya:2016xpa}.
Together the invariant states in the spectrum of the compactification on $(K3\times S^1)/\mathbb{Z}_N$ are then of the form~\cite{Persson:2015jka,Persson:2017lkn}
\begin{equation}
\label{eqn:CHLstate}
\ket{k,w, \psi_{r,k}}\,, \quad r \equiv w \text{ mod } N \,.
\end{equation}

\begin{figure}[h!]
	\centering
	\begin{minipage}[b]{.35\textwidth}
	\includegraphics[width=\linewidth]{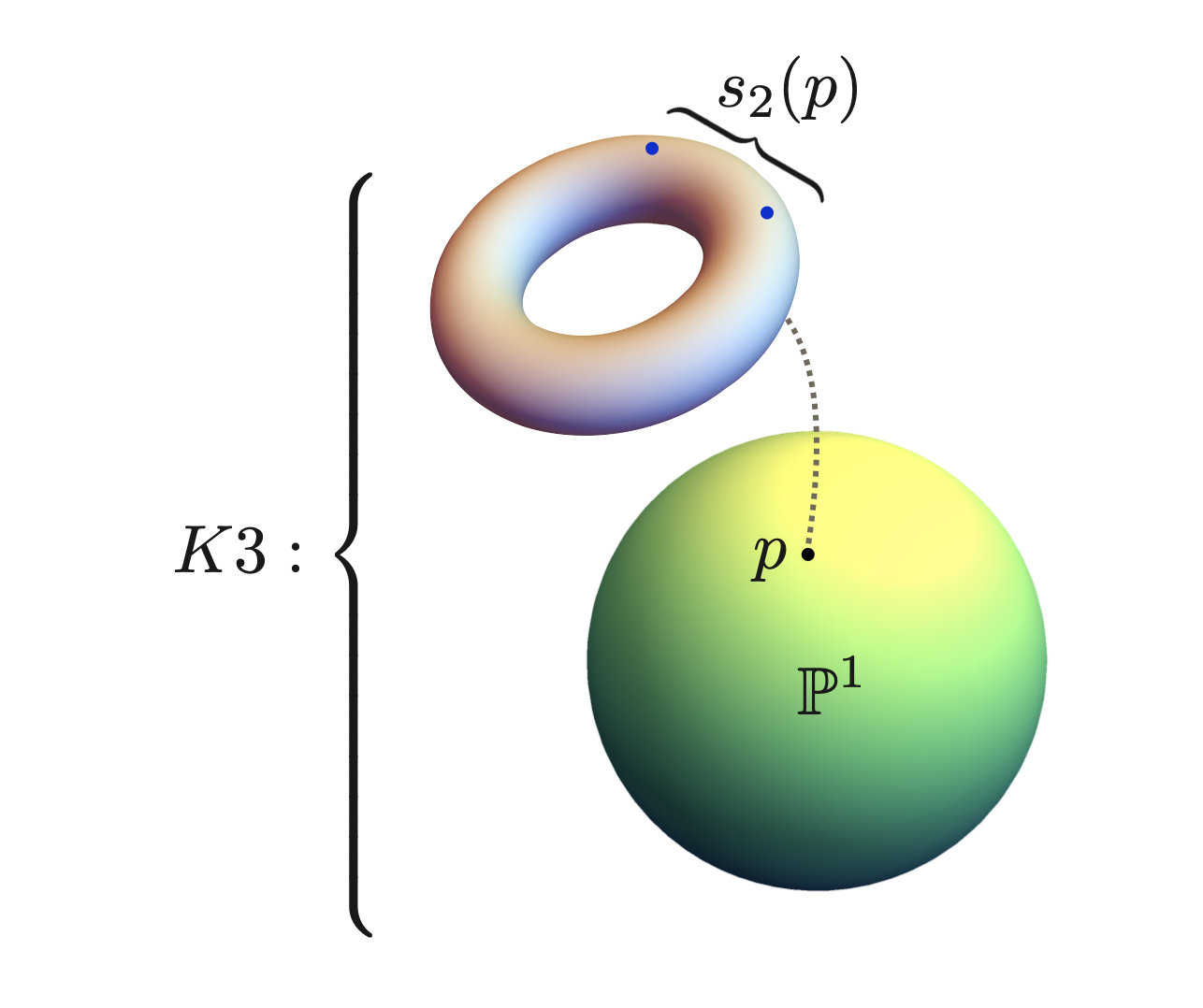}
	\end{minipage}
	\hspace{1cm}
	\caption{Schematic representation of a genus one fibered $K3$ surface with 2-section.} 
	\label{fig4}
\end{figure}

Now we turn back to the M-theory compactification on a genus one fibered Calabi-Yau threefold $\pi_N: M\rightarrow B$ with $N$-section.
By the latter, we mean a multivalued function $s_N : B\rightarrow M$ such that $|\pi^{-1}(p)|=N$ for almost all points $p\in B$. See figure \ref{fig4}.
For recent reviews on F-theory and M-theory that discuss genus one fibrations we refer the reader to the references~\cite{Weigand:2018rez,Cvetic:2018bni}.
We want the Calabi-Yau to also exhibit a $K3$ fibration $\pi : M \rightarrow \mathbb{P}_{\mathrm{b}}^1$ such that in the limit of large $\mathbb{P}_{\mathrm{b}}^1$ we have a dual description in terms of weakly coupled Heterotic strings.
Recall that a $K3$ fiber with polarization lattice $\Lambda$ of rank two has an intersection form
\begin{equation}
I_\Lambda\Big\vert_{\text{rk} (\Lambda) = 2} =\left(\begin{array}{rr} 2a\, &b\,\\ \,b \text{ }& \,2c\end{array}\right)\,, \quad a,b,c \in \mathbb{Z}\,, \quad 4ac -b^2<0\,.
\end{equation}
An algebraic $K3$ surface with $\Lambda$-polarized lattice of rank two admits a genus one fibration with $N$-section iff $b^2 -4ac = N^2$~\cite{MR0284440,Braun:2016sks,Kimura:2017rhk}.

For elliptic fibrations, i.e. when $N=1$, five-dimensional light BPS states arise in the limit $\text{Vol}_{\mathbb{C}}(\mathbb{P}^1_{\mathrm{b}}) \rightarrow \infty$, where the dual Heterotic string is weakly coupled, from M2 branes that wrap curves of the type~\cite{Lee:2019wij}
\begin{equation}
	C_{\text{M2}} = k T^2 + w (\mathbb{P}^1_f+T^2)\,,
\end{equation}
with $\mathbb{P}^1_f$ being the base of the $K3$ fiber and $T^2$ the generic fiber of the genus one fibration.
The shift of $\mathbb{P}^1_f$ by $T^2$ is fixed by demanding that the self-intersection of $C_{\text{M2}}$ inside the $K3$ fiber matches the quadratic form on the momentum lattice.
However, on a genus one fibration with $N$-section we find components of reducible fibers that intersect the $N$-section only once so that the analogous expansion reads
\begin{equation}
	C_{\text{M2}} = k \frac{T^2}{N} + w (\mathbb{P}^1_f+T^2)\,,
\end{equation}
with self-intersection given by~\eqref{eqn:CHLint}, i.e.
\begin{equation}
C_{\text{M2}}^2 = \frac{2kw}{N}\,.
\end{equation}
A crucial implication is, that the coefficients $Z_{n\cdot\mathbb{P}^1_f},\,n=1,\dots,N-1$ of the topological string partition function~\eqref{eq:baseexpansion} encode twisted sectors of the dual Heterotic strings,
while the contributions from untwisted sectors appear in $Z_{n\cdot\mathbb{P}^1_f},\,n=0\,\text{mod}\,N$.

As a next step we extend the construction and turn on Heterotic Wilson lines.
On the M-theory side this amounts to considering $K3$ fibrations with $\Lambda$-polarized lattices of the form
\begin{equation}
\label{eqn:CHLlattice}
\Lambda = U(N) \oplus L(-m)\,,
\end{equation}
where $m\in\mathbb{N}$ is some twist that signals a non-perturbative realization of the Heterotic gauge group.
So far the situation is completely analogous to that of elliptically fibered $K3$ surfaces which we described in section \ref{sec:EGandNL}.
The new feature here is, that the discriminant group $G_\Lambda$ associated with the lattice~\eqref{eqn:CHLlattice} decomposes into two factors $G_\Lambda \cong G_{U(N)} \oplus G_{L(-m)}$.
Consequently, the Noether-Lefschetz generators of these geometries will be vector-valued modular forms that transform under the dual of the Weil representation $\rho_{\Lambda}^*: \text{Mp}(2,\mathbb{Z}) \rightarrow \text{End}\left(\mathbb{C}[G_\Lambda]\right)$, where $\mathbb{C}[G_\Lambda] \cong \mathbb{C}[G_{U(N)}] \otimes_{\mathbb{C}} \mathbb{C}[G_{L(-m)}]$.

If we recall that for Heterotic strings on $K3\times T^2$ the elliptic genus could be obtained from the Noether-Lefschetz generator by inverting the map~\eqref{eqn:hmap}, this leads to the question
what happens when the latter transforms non-trivially with respect to
\begin{equation}
G_{U(N)} \cong \left( \mathbb{Z} / N \mathbb{Z}\right)^2\,.
\end{equation}
It turns out that the objects we obtain are the twisted twined elliptic genera
\begin{equation}
\label{eqn:ttgen}
	Z^{(r,s)}(\tau,\bm{z})= \text{Tr}_{\mathcal{H}_{r}}g^s(-1)^F F^2 q^{H_L} \bar{q}^{H_R} \prod_{a=1}^{n_V}(\zeta^a)^{J_a}\,, 
\end{equation}
that have been introduced by~\cite{Gaberdiel:2012gf} and further studied in~\cite{Datta:2015hza,Chattopadhyaya:2016xpa,Chattopadhyaya:2017zul,Banlaki:2019bxr,Chattopadhyaya:2020qqq}.
Here we use the same notation that we already defined for the ordinary elliptic genus in~\eqref{eqn:ellipticgenus1}.
The only difference here is that the trace is taken over the $r$-twisted Ramond-Ramond sector $\mathcal{H}_r$, whereas the insertion $g^s$ projects  multiples of the phase eigenvalue of states (\ref{eqn:CHLstate}). 
Moreover, the twisted twined elliptic genera are lattice Jacobi forms for $\Gamma_1(N)$, while the vector $\left(Z^{(r,s)}(\tau,\bm{z})\right)_{r,s \in \mathbb{Z}_N}$  transforms as vector-valued lattice Jacobi form under $\text{SL}(2,\mathbb{Z})$.
In particular, the relevant object to extend the discussion of the sublattice weak gravity conjecture to M-theory on genus one fibrations is the untwisted, untwined elliptic genus $Z^{(0,0)}(\tau,\bm{z})$.
Contrary to the elliptic case this receives contributions not only from base degree one invariants of the topological string partition function but from the first $N$ degrees with respect to the base of the $K3$ fiber.
However, before we flesh out the details of the sublattice conjecture, let us discuss how exactly the twisted twined elliptic genera can be obtained from the Noether-Lefschetz generator and also how the latter relates to the new supersymmetric index.

\subsection{Twisted elliptic genera and Noether-Lefschetz theory}
\label{sec:nssiandtwistedelliipticgenus}
We will now extend the discussion of the Noether-Lefschetz generator, the new supersymmetric index and the elliptic genera to Heterotic strings on $(K3\times S^1)/\mathbb{Z}_N$, where we in fact have to consider the twisted twined elliptic genera~\eqref{eqn:ttgen}.
For simplicity we again restrict to the case where the Heterotic gauge algebra consists of a single simple factor $\mathfrak{g}$.
The dual $K3$ fibered Calabi-Yau threefolds $S \hookrightarrow M \xrightarrow{\pi} \mathbb{P}_{\mathrm{b}}^1$ then exhibit a polarization lattice
\begin{equation}
 \Lambda = U(N) \oplus L(-m)\,, \quad \Lambda' = U(-1) \oplus \Lambda\,,
\end{equation}
where $\Lambda'$ is the extended lattice defined by $U(-1) \cong H^0(S,\mathbb{Z}) \oplus H^4(S,\mathbb{Z})$.
Moreover, $L$ is a rank $\text{rk}(\mathfrak{g})$ lattice associated with the Wilson lines parameters, whose intersection form is given by $-m(\cdot\,, \cdot)$ with $(\cdot\,,\cdot)$ the Killing form of $\mathfrak{g}$ for some twist $m\in\mathbb{N}$.
If the gauge group is realized perturbatively, the twist is $m=1$.

In~\cite{Datta:2015hza} the authors introduced the shifted Siegel theta function
\begin{align}
\label{eqn:shiftTheta}
\Gamma_{\bm{\mu}}^{(r,s)}  = \sum_{\bm{\lambda} \in L(m) + \bm{\mu}}\sum_{k_0,w_0, k \in \mathbb{Z}} \sum_{w \in \mathbb{Z} +\frac{r}{N}} \mathrm{e}^{-2\pi i\frac{ k s}{N}} q^{\frac{p_L^2}{2}} \bar{q}^{\frac{p_R^2}{2}} \,,
\end{align}
where they considered the case $L= L^\vee(A_1)$ and $\bm{\mu} \in \mathbb{Z}/2\mathbb{Z}$.
The left- and right moving momenta can be obtained from the relations
\begin{align}
	\begin{split}
\frac12 (p_L^2 - p_R^2) & = \frac{1}{2} m \left(\bm{\lambda},\bm{\lambda}\right) - k w + k_0 w_0 \,,\\
\frac{1}{2} p_R^2 & = -\frac{1}{2\langle\text{Im}(\bm{y}), \text{Im}(\bm{y}) \rangle } \Big\vert \bm{\lambda} \cdot \bm{V} +k U + w T + k_0 -w_0 \frac{\langle \bm{y},\bm{y} \rangle}{2}\Big\vert^2\,,
	\end{split}
\end{align}
where $\bm{y}$ denotes the $(T,U,\bm{V})$ Heterotic moduli with intersection pairing
\begin{equation}
\langle \bm{y} , \bm{y} \rangle \equiv -2 TU + m(\bm{V},\bm{V})\,.
\end{equation}
Note that here we choose the signs to match the conventions for the quadratic form of~\cite{Harvey:1995fq,Enoki:2019deb}.
Based on the results from~\cite{Chattopadhyaya:2016xpa,Chattopadhyaya:2017zul,Banlaki:2019bxr,Chattopadhyaya:2020qqq} we observe that the new supersymmetric index for Heterotic strings on $(K3\times T^2)/\mathbb{Z}_N$ factorizes as
\begin{equation}
\mathcal{Z}(\tau,\bar{\tau}) =  \frac{1}{N}\sum_{ \bm{\mu} \in G_L} \sum_{r,s \in \mathbb{Z}/ N \mathbb{Z}} \Gamma_{\bm{\mu}}^{(r,s)}(\tau,\bar{\tau}) Z_{\bm{\mu}}^{(r,s)}(\tau)\,,
\end{equation}
where we denote by $G_{L(m)}$ the discriminant group of $L(m)$ and $Z_{\bm{\mu}}^{(r,s)}$ are the components of a meromorphic vector-valued modular form
\begin{equation}
\label{eqn:ttgen}
Z(\tau) = \sum_{ \bm{\mu} \in G_{L(m)}} \sum_{r,s \in \mathbb{Z}/ N \mathbb{Z}}  Z_{\bm{\mu}}^{(r,s)} (\tau) \mathrm{e}_{\bm{\mu}} \otimes \mathrm{e}_{(r,s)}\,, \quad  Z_{\bm{\mu}}^{(r,s)}(\tau)  = \frac{h_{\bm{\mu}}^{(r,s)}(\tau)}{\eta^{24}(\tau)}  \,.
\end{equation}
We claim that $(Z^{(r,s)}_{\bm{\mu}})_{ \bm{\mu} \in G_{L(m)}}$ is nothing but the $h$\textit{-map projection} of the twisted twined elliptic genus (\ref{eqn:ttgen}).
In other words, the vector of twisted twined elliptic genera, which form a vector-valued Jacobi form under $\text{SL}(2,\mathbb{Z})$ transformations, can be obtained by replacing
\begin{equation}
\label{eqn:ttegrepl}
\mathrm{e}_{\bm{\mu}} \mapsto \vartheta_{L(m),\bm{\mu}}(\tau,\bm{z})\,.
\end{equation}

To make the connection with the Noether-Lefschetz generator, we expand the new supersymmetric index into
\begin{align}
\label{eqn:calculation}
\begin{split}
\mathcal{Z}(\tau,\bar{\tau}) & = \sum_{\bm{\mu} \in G_{L(m)}} \sum_{r,s\in \mathbb{Z}/N\mathbb{Z}} \sum_{k_0,w_0 \in \mathbb{Z} } \sum_{\substack{w \in \mathbb{Z} +\frac{r}{N} \\ k \in \mathbb{Z}}}  q^{\frac{p_L^2}{2}} \bar{q}^{\frac{p_R^2}{2}} \frac{\mathrm{e}^{-2\pi i\frac{ k s}{N}}}{N} Z_{\bm{\mu}}^{(r,s)}(\tau) \\
& = \sum_{\bm{\mu} \in G_{L(m)}} \sum_{r,\ell \in \mathbb{Z}/N \mathbb{Z}} \sum_{k_0,w_0 \in \mathbb{Z} } \sum_{\substack{w \in \mathbb{Z} +\frac{r}{N} \\ k \in N \mathbb{Z} + \ell } } q^{\frac{p_L^2}{2}} \bar{q}^{\frac{p_R^2}{2}} \widehat{Z}_{\bm{\mu}}^{(r,\ell)}(\tau) \\
& =  \sum_{\bm{\mu} \in G_{L(m)}}\sum_{(k_0,w_0) \in U} \sum_{(r,\ell) \in G_\delta} \sum_{(k,w) \in L_\delta^{\text{unt}} + (r,\ell)} q^{\frac{p_L^2}{2}} \bar{q}^{\frac{p_R^2}{2}} \widehat{Z}_{\bm{\mu}}^{(r,\ell)}(\tau) \\
& =  \sum_{\bm{\mu} \in G_{L(m)}} \sum_{(r,\ell) \in G_\delta} \theta_{\Lambda'(-1) + \bm{\mu}\oplus (r,\ell)  } (\tau,\bar{\tau})  \widehat{Z}_{\bm{\mu}}^{(r,\ell)}(\tau)\,,
\end{split}
\end{align}
where we used several new definitions.
First, we introduced the Fourier discrete transform of a $\mathbb{C}\left[ \mathbb{Z} / N \mathbb{Z} \times \mathbb{Z} / N \mathbb{Z} \right]$-valued function $F$ on $\mathbb{H}$, whose component read~\cite{MR2904095}
\begin{equation}
\widehat{F}_{(r,\ell)} \equiv  \frac{1}{N} \sum_{s\in \mathbb{Z}/N \mathbb{Z}} \mathrm{e}^{-2\pi i \frac{s \ell }{N}} F_{(r,s)} \,.
\end{equation}
Second, following the discussion of section 2.2 from~\cite{Persson:2017lkn}, we introduced the CHL lattice $L_\delta = N^{-1} \mathbb{Z} \oplus \mathbb{Z}$ spanned by the vectors $\delta = \mathrm{e}_1 / N = (1/N,0)$ and $\mathrm{e}_2 = (0,1)$, such that
\begin{equation}
\delta^2 = \mathrm{e}^2_2 = 0\,, \quad \delta \cdot \mathrm{e}_2  = \frac{1}{N}\,,
\end{equation}
whereas the untwisted $\delta$-invariant sublattice $L_{\delta}^{\text{unt}}  \subset U = \text{Span}_{\mathbb{Z}}\left\{\mathrm{e}_1,\mathrm{e}_2 \right\} \cong \mathbb{Z} \oplus \mathbb{Z}$ is defined as 
\begin{equation}
L_{\delta}^{\text{unt}} = \{ (w,k) \in U \mid  (w,k) \cdot \delta \in \mathbb{Z} \} \cong \mathbb{Z} \oplus N \mathbb{Z} \,.
\end{equation}
Note that by definition $(L_{\delta}^{\text{unt}})^* =L_{\delta}$ and the intersection form of $L_{\delta}^{\text{unt}}$ is that of $U(N)$. Moreover, the CHL discriminant group is given by 
\begin{equation}
G_\delta = L_\delta/L_\delta^{\text{unt}} \cong G_{U(N)} \cong \left(\mathbb{Z}/N\mathbb{Z}\right)^2\,.
\end{equation}
Third, we introduced the components of the \textit{Siegel theta function} $\Theta_{\Lambda'(-1)}(\tau,\bar{\tau})$ associated with the lattice $\Lambda'(-1)$, which we describe in more detail in (\ref{eqn:SiegelTheta}) in Appendix~\ref{app:VVMF}. 
With this in mind, 
let us now recall the claim of the authors~\cite{Enoki:2019deb}
\begin{equation}
\mathcal{Z}(\tau,\bar{\tau}) = \sum_{\gamma \in G_{\Lambda}}\theta_{\Lambda'(-1) +\gamma}(\tau,\bar{\tau}) \frac{\Phi^\pi_\gamma(\tau)}{\eta^{24}(\tau)} \,,
\end{equation}
where in our case $G_\Lambda = G_{L(m)} \oplus G_{U(N)}$.
By comparing (\ref{eqn:calculation}) and (\ref{eqn:ttgen}) we find that the Noether-Lefschetz generator is given by
\begin{equation}
\label{eqn:nlgenchl}
\Phi^\pi_{\gamma} (\tau)\Big\vert_{\gamma = \bm{\mu} \oplus (r,\ell)} = 
\widehat{h}_{\bm{\mu}}^{(r,\ell)}(\tau) \,.
\end{equation}

Together with the NL-GW correspondence theorem~\eqref{eqn:GWNL} we have thus found a straightforward recipe to obtain the twisted twined elliptic genera directly from the topological string partition function
of a $K3$ fibered Calabi-Yau threefold that also exhibits a compatible genus one fibration:
\begin{enumerate}
	\item Calculate the genus zero Gromov-Witten invariants, for example via mirror symmetry, and use~\eqref{eqn:GWNL} to determine the Noether-Lefschetz generator~\eqref{eqn:nlgenchl}.
	\item Invert the discrete Fourier transformation to recover~\eqref{eqn:ttgen}.
	\item Apply the replacement~\eqref{eqn:ttegrepl} to obtain the twisted twined elliptic genera.
\end{enumerate}
Below we will illustrate this procedure at the hand of genus one fibrations with $4$-sections.
However, let us first extend the previous discussion of the sublattice weak gravity conjecture to genus one fibrations.

\subsection{The sublattice weak gravity conjectures for genus one fibrations}
We will now discuss how the arguments for the sublattice and the non-Abelian sublattice weak gravity conjecture apply to M-theory on a genus one fibered Calabi-Yau threefold $M$ with $N$-sections.
The existence of a limit in which the gauge coupling goes to zero while the Planck mass remains finite still implies that the base is a Hirzebruch surface or a blowup thereof~\cite{Lee:2018urn}.
Moreover, in the same limit a dual Heterotic string, that is now compactified on $(K3\times S^1)/\mathbb{Z}_N$, becomes weakly coupled and, using the prescription that we worked out in the previous subsection,
one can calculate the twisted twined elliptic genera $Z^{(r,s)}(\tau,\bm{z})$~\eqref{eqn:ttgen} which are of the form
\begin{align}
	Z^{(r,s)}(\tau,\bm{z})= \text{Tr}_{\mathcal{H}_{r}}g^s(-1)^F F^2 q^{H_L} \bar{q}^{H_R} \prod_{a=1}^{n_V}(\zeta^a)^{J_a}=\frac{\Phi^{(r,s)}(\tau,\bm{z})}{\eta(\tau)^{24}} \,,
	\label{eqn:twittytwine}
\end{align}
with $\Phi^{(r,s)}(\tau,\bm{z})$ being holomorphic, Weyl invariant lattice Jacobi forms for $\Gamma_1(N)$.
We can therefore again perform an expansion~\eqref{eqn:elgenthetaexp}
\begin{align}
	\Phi^{(r,s)}(\tau,\bm{z})=\sum\limits_{\bm{\mu}\in L^*/L}h_{\bm{\mu}}^{(r,s)}(\tau)\vartheta_{\underline{L},\bm{\mu}}(\tau,\bm{z})\,,
	\label{eqn:ttelgenthetaexp}
\end{align}
where $\vartheta_{\underline{L},\bm{\mu}}(\tau,\bm{z})$ are Jacobi theta functions associated with the, in general twisted, coroot lattice $\underline{L}$ of the gauge group
and $(h_{\bm{\mu}}^{(r,s)})_{(r,s)\oplus \bm{\mu}}$ are vector-valued modular or, in the presence of NS5-branes, vector-valued quasi-modular forms that transform under the dual of the Weil representation  $\rho^*:\text{Mp}(2,\mathbb{Z})\rightarrow \mathbb{C}\left[\mathbb{Z}/N\mathbb{Z}\times\mathbb{Z}/N\mathbb{Z} \right] \otimes \mathbb{C} \left[ L^*/L\right]$.

The Tachyon is part of the untwisted sector and appears in $Z^{(0,0)}(\tau,\bm{z})$. We can therefore apply the same arguments that held for elliptic fibrations to conclude that there is a sublattice of superextremal states which satisfies
the sublattice weak gravity conjecture and, in the case of a single simple factor of the Heterotic gauge group, also the non-Abelian generalization.
In fact, the untwisted untwined elliptic genus $Z^{(0,0)}(\tau,\bm{z})$ is identical to the elliptic genus of the Heterotic string on $K3$.
The crucial difference to the elliptic case is now that it is not identical to the genus zero contribution to the topological string partition function $Z_{\beta=F}(\tau,\lambda)$ but receives contributions from $Z_{\beta=n\cdot F}(\tau,\lambda)$ for $n=1,\dots,N$.\\\\
\noindent\textbf{Example:} We consider the genus one fibered Calabi-Yau threefold
\begin{align*}
	M = (F_4 \rightarrow \mathbb{F}_1)\left[ SU(2) \times \mathbb{Z}_2\right]_4^{-144}
\end{align*}
that has been discussed in~\cite{Cota:2019cjx}.
Here the polarization lattice is of the form $\Lambda \simeq U(2) \oplus L^\vee (A_1)(-2)$.
The twisted elliptic genera~\eqref{eqn:twittytwine} can be fixed via the procedure outlined in Section~\ref{sec:nssiandtwistedelliipticgenus}
and we obtain the following result:

{\scriptsize
\begin{align}
\begin{split}
\Phi^{(0,0)}(\tau,z) & = -\frac{11}{12} E_6(\tau) E_{4,2}(\tau,z) -\frac{13}{12} E_4(\tau)E_{6,2}(\tau,z)\,, \\
\Phi^{(0,1)}(\tau,z) & =-\frac{1}{2592}\left(E_4(\tau)-4 \mathcal{E}_2^2(\tau)\right)^2 \Bigg[2 \phi_{-2,1}^2(\tau,z) \mathcal{E}_2(\tau) \left(7 \mathcal{E}_2^2(\tau)-5 E_4(\tau)\right)\\
&\qquad +\phi_{-2,1}(\tau,z) \phi_{0,1}(\tau,z) \left(E_4(\tau)-9
\mathcal{E}_2^2(\tau)\right)+4 \phi_{0,1}^2(\tau,z) \mathcal{E}_2(\tau)\Bigg]\,,\\
\Phi^{(1,0)}(\tau,z) & =-\frac{1}{2592}\left(E_4(\tau)- \mathcal{E}_2^2\left(\frac{\tau}{2}\right)\right)^2 \Bigg[ \phi_{-2,1}^2(\tau,z) \mathcal{E}_2\left(\frac{\tau}{2}\right) \left(5 E_4(\tau)-\frac{7}{4} \mathcal{E}_2^2\left(\frac{\tau}{2}\right)\right)\\
&\qquad +\phi_{-2,1}(\tau,z) \phi_{0,1}(\tau,z) \left(E_4(\tau)-\frac{9}{4}
\mathcal{E}_2^2\left(\frac{\tau}{2}\right)\right)- 2 \phi_{0,1}^2(\tau,z) \mathcal{E}_2\left(\frac{\tau}{2}\right)\Bigg]\,, \\
\Phi^{(1,1)}(\tau,z) & =-\frac{1}{2592}\left(E_4(\tau)- \mathcal{E}_2^2\left(\frac{\tau+1}{2}\right)\right)^2 \Bigg[ \phi_{-2,1}^2(\tau,z) \mathcal{E}_2\left(\frac{\tau+1}{2}\right) \left(5 E_4(\tau)-\frac{7}{4} \mathcal{E}_2^2\left(\frac{\tau+1}{2}\right)\right)\\
&\qquad +\phi_{-2,1}(\tau,z) \phi_{0,1}(\tau,z) \left(E_4(\tau)-\frac{9}{4}
\mathcal{E}_2^2\left(\frac{\tau+1}{2}\right)\right)- 2 \phi_{0,1}^2(\tau,z) \mathcal{E}_2\left(\frac{\tau+1}{2}\right)\Bigg] \,.
\end{split}
\end{align} }
Here $\mathcal{E}_2(\tau) \equiv E_2(2\tau)-E_2(\tau)$. 
Indeed, the untwined untwisted elliptic genus is identical to the one we obtained in~\eqref{eqn:NumSU2ab} for the Heterotic strings dual to elliptic fibrations.
Morever, we find out the following exchange transformation under $S,T \in \mathrm{SL}(2,\mathbb{Z})$:
\begin{equation}
T \circlearrowright \Phi^{(0,1)} \stackrel[]{S} \longleftrightarrow \Phi^{(1,0)}\stackrel[]{T} \longleftrightarrow \Phi^{(1,1)}\circlearrowleft S\,.
\end{equation}
This observation derives from the trivial relations $\Phi^{(1,0)}(\tau+1,z) = \Phi^{(1,1)}(\tau,z)$ and $\Phi^{(0,1)}(\tau+1,z) = \Phi^{(0,1)}(\tau,z)$, whereas the exchange under $S$ transformations follows from the identities
\begin{align}
\mathcal{E}_2\left(-\frac{1}{2\tau}\right) = -2 \tau^2 \mathcal{E}_2(\tau) \,, \quad \mathcal{E}_2\left( -\frac{1}{2\tau} + \frac{1}{2} \right) = \tau^2 \mathcal{E}_2\left( \frac{\tau +1}{2} \right)\,.
\end{align}
\subsection{New supersymmetric index of CHL strings on $(K3\times T^2)/\mathbb{Z}_4$}
\label{sec:4sec}
In order to construct Calabi-Yau threefolds that exhibit not only a $K3$ fibration structure but are at the same time genus one fibered with four sections, we have to go beyond hypersurfaces
in toric ambient spaces and consider complete intersections.
If we consider complete intersections in codimension two and, moreover, assume that the genus one fibration arises from a compatible fibration of the toric ambient space,
then there is exactly one choice for the toric fiber that leads to four sections, namely $\mathbb{P}^3$.
In order to ensure that the geometry also exhibits a compatible $K3$ fibration structure, we will first construct $K3$ varieties as complete intersections in $\mathbb{P}^3$ fibrations over $\mathbb{P}^1$.
We will then fiber the corresponding geometries over another $\mathbb{P}^1$.
For a general choice of the fibrations this will lead to Calabi-Yau threefolds with $h^{1,1}=3$.

To construct $K3$ varieties with Picard rank two that are genus one fibered with four sections, we consider toric ambient spaces that correspond to polytopes $\Delta^\circ(n_1,n_2)$ with vertices
\begin{align}
\begin{blockarray}{crrrr}
\begin{block}{c(rrrr)}
	\nu^1& 1& 0& 0& 0\\
	\nu^2& 0& 1& 0& 0\\
	\nu^3& 0& 0& 1& 0\\
	\nu^4&-1&-1&-1& 0\\
	\nu^5& n_1&n_2& 0& 1\\
	\nu^6& 0& 0& 0&-1\\
\end{block}
\end{blockarray}\,,
\label{eqn:$K3$points1}
\end{align}
where we can restrict to $n_1,n_2=0,...,2$, and determine all nef-partitions $\Delta^\circ(n_1,n_2)=\langle\nabla_1,\nabla_2\rangle$ such that two of the points $\nu^i,\,i=1,...,4$ are contained in each
component $\nabla_i,\,i\in\{1,2\}$.
It turns out that we can restrict the values $(n_1,n_2)$ even further and only need to consider the four pairs
\begin{align}
	I:\,(0,0)\,,\quad II:\,(1,0)\,,\quad III:\,(1,1)\,,\quad IV:\,(2,0)\,,
\end{align}
in order to obtain all inequivalent models that do not exhibit additional divisors that are torically realized.
Judging from the analogous construction of $K3$ hypersurfaces in toric ambient spaces with two- and three-sections we expect that one of the ambient spaces actually
leads to $K3$ varieties with Picard rank five which arises from a toric four section that is actually a union of four independent sections.
However, once we consider fibrations of the $K3$ over another $\mathbb{P}^1$, those sections can merge into a genuine four section and therefore it makes sense to consider those geometries as well.
One can check that, independently of the choice of nef-partition, the four choices for $(n_1,n_2)$ lead to the intersection forms
\begin{align}
	I_I=\left(\begin{array}{rr}4&4\\4&0\end{array}\right)\,,\quad I_{II}=\left(\begin{array}{rr}2&4\\4&0\end{array}\right)\,,\quad I_{III}=\left(\begin{array}{rr}0&4\\4&0\end{array}\right)\,,\quad I_{IV}=\left(\begin{array}{rr}0&4\\4&0\end{array}\right)\,.
\end{align}
Only the last two cases are compatible with T-duality on a dual Heterotic compactification on $(K3\times T^2)/\mathbb{Z}_4$.

Let us now take the $K3$ complete intersections and construct $K3$ fibered Calabi-Yau threefolds.
To this end we consider the toric ambient spaces that correspond to the polytopes $\tilde{\Delta}^\circ(n_1,n_2,\nu)$ with vertices $\tilde{\nu}^i=(\nu^i,0)$
and $\tilde{\nu}^7=(\nu, 0),\,\tilde{\nu}^8=(0^4,-1)$, where $\nu\in2\Delta^\circ(n_1,n_2)$, and nef partitions that extend those that we considered for the $K3$ fibers.
The condition on $\nu$ ensures that the polytope is reflexive.
We find that for geometries with fibers of type $I$, $III$ and $IV$, the corresponding Gopakumar-Vafa invariants of degree zero with respect to the base
of the $K3$ fibration only depend on the type of $(n_1,n_2)$ and the Euler characteristic $\chi$ of the complete intersection.
The invariants are listed in table~\ref{tab:gvIII}.
The inequivalent Euler characteristics that we obtain for each type of fiber are
\begin{align}
	\begin{split}
		I:& -104,\\
		II:& -92,\,-96,\,-100,\,-104,\,-108,\,-116,\\
		III:& -96,\,-100,\,-104,\,-108,\,-112,\\
		IV:& -96,\,-100,\,-104,\,-112,\,-116,\,-128\,.
	\end{split}
\end{align}
Let us now focus on the cases $III$ and $IV$. 

In the following, we want to find a set of modular forms in $M_{10}\left(\Gamma_1(N)\right)$ that fix the Noether-Lefschetz generator components (\ref{eqn:nlgenchl}). We can do this via the GW-NL correspondence theorem relation (\ref{eqn:GWNL}). By following the discussion of section \ref{sec:nssiandtwistedelliipticgenus}, the GV invariants along the $K3$ fiber with polarization lattice $\Lambda = U(N)$ results into the following expression
\begin{table}
\centering
\tiny
\begin{align}
\begin{array}{|c|cccccc|}
	\hline
	w\backslash k&0&1&2&3&4&5\\\hline
	0& 0 & 128 & \chi +224 & 128 & -\chi  & 128 \\
	1&128 & 800-2 \chi & 5120  & 5 \chi +21040  &  70656  &  217792-12 \chi \\
	2&\chi +224 & 5120 & 69632 - 8 \chi & 626688 & 4268800 + 40 \chi & 24164352 \\
	3& 128 & 21040 + 5 \chi & 626688 & 10345120 - 58 \chi & 119377920 & 1078170408 + \frac{819}{2} \chi\\
	4& -\chi & 70656 & 4268800 + 40 \chi & 119377920 & 2151677952 - 552 \chi & 28827891712 \\
	5& 128 & 217792 - 12 \chi & 24164352 & 1078170408 + \frac{819}{2} \chi & 28827891712 & 547018864352 - 6302 \chi \\ \hline
\end{array}
\end{align}
	\caption{Genus zero Gopakumar-Vafa invariants for fibrations of type $III$ and $IV$. 
	Here $(w,k)$ indicates the degrees of the $K3$ fiber.}
	\label{tab:gvIII}
\end{table}
\begin{equation}
\label{eqn:GWNLCHLrk2}
n_{(w,k)}^g = \sum_{h = 0}^\infty \sum_{s \in \mathbb{Z}/N\mathbb{Z}} \frac{\mathrm{e}^{-2\pi i \frac{s \ell}{N}}}{N} \text{Coeff}\left(\Phi^{(r,s)}(\tau),q^{\Delta_{NL}(h,w,k)}\right) \cdot r_h^g \,,
\end{equation}
where $\Phi^{(r,s)} \in M_{10}\left( \Gamma_1(N)\right)$,  $r \equiv w \text{ mod } N$ and $ \ell \equiv k \text{ mod } N$ denote the  entries in the discriminant group  $(r,\ell) \in G_{U(N)}$, whereas the Noether-Lefschetz discriminant form reads
\begin{equation}
\Delta_{NL}(h,w,k) = \frac{kw }{ N } +1 - h \,.
\end{equation}
Note that we kept the label $N$ for generality of the discussion, although our case of study here concerns the case $N=4$. \footnote{ These expressions can also be implemented for the geometries that were studied in~\cite{Cota:2019cjx,Banlaki:2019bxr}.} It turns out that the genus zero GV invariants obtained by mirror symmetry computations suffice to fix the modular forms $\Phi^{(r,s)}$. We provide a few genus zero invariants of this type in the table \ref{tab:gvIII}. We remark that the latter objects must fulfil  vector-valuedness, as described in the previous sections. To take into account this property in our 4-section example, we make an Ansatz for the various $\Phi^{(r,s)}$ such that they follow the same vector-valued transformations under $\mathrm{SL}(2,\mathbb{Z})$ as the twisted elliptic genus of class $4B$ in  \cite{Chattopadhyaya:2017zul}.
For this purpose, we introduce generators for the ring of modular forms that transform under the congruence subgroups $\Gamma_1(N)$
\begin{equation}
\mathcal{E}_{\tilde{N}} (\tau) \equiv - \frac{1}{\tilde{N}-1} \mathfrak{d}_\tau \log \left( \frac{\eta(\tau)}{\eta(\tilde{N}\tau)}  \right)\,, \quad \tilde{N} \in \mathbb{N}\,.
\end{equation}
 For the 4-section geometries of current interest, we only need the generators $\mathcal{E}_2$ and $\mathcal{E}_4$. With this in mind, we obtain the following result:
 
{\scriptsize
\begin{align}
\begin{split}
\Phi^{(0,0)} & = -2E_4(\tau) E_6(\tau)\,,\\
\Phi^{(0,1)} & =  \frac{1}{2^5} \Bigg[ 2\mathcal{E}_2^2(\tau)\mathcal{E}_4^3(\tau) ( 816+9 \chi)-\mathcal{E}_2^3(\tau)
  \mathcal{E}_4^2(\tau) (656+7 \chi )+\mathcal{E}_2^4(\tau)\mathcal{E}_4(\tau) (
   96+\chi)\\
   &\qquad \qquad-4\mathcal{E}_2(\tau)\mathcal{E}_4^4(\tau) (5 \chi +432)+8\mathcal{E}_4^5(\tau)
   (\chi +80)\Bigg]  \,,\\
   \Phi^{(0,2)} & =  -12\mathcal{E}_2^3(\tau)\mathcal{E}_4^2(\tau)+16\mathcal{E}_2^2(\tau)\mathcal{E}_4^3(\tau)+4\mathcal{E}_2^4(\tau)
  \mathcal{E}_4(\tau)-\frac{1}{2}\mathcal{E}_2^5(\tau)-8\mathcal{E}_2(\tau)\mathcal{E}_4^4(\tau)\,,\\
\Phi^{(0,3)} & =\Phi^{(0,1)}\,,\\
 \Phi^{(1,s)}& = \frac{1}{2^{12}} \Bigg[ -(192+2\chi )\mathcal{E}_2^4\Big(\frac{\tau +s }{2}\Big)\mathcal{E}_4\Big(\frac{\tau +s }{4}\Big)-(80+\chi )\mathcal{E}_4^5\Big(\frac{\tau +s }{4}\Big)\\ 
                  & \qquad \qquad +(816+9\chi)\mathcal{E}_2^2\Big(\frac{\tau +s }{2}\Big)\mathcal{E}_4^3\Big(\frac{\tau +s }{4}\Big) +(432+5\chi) \mathcal{E}_2\Big(\frac{\tau +s }{2}\Big)\mathcal{E}_4^4\Big(\frac{\tau +s }{4}\Big)\\
                   & \qquad \qquad +(656+7\chi)\mathcal{E}_2^3\Big(\frac{\tau +s }{2}\Big)\mathcal{E}_4^2\Big(\frac{\tau +s }{4}\Big) \Bigg] \,,\\
\Phi^{(3,3s)}& = \Phi^{(1,s)} \,, \\
\Phi^{(2,2s)} & =  \frac{1}{2^7}\Bigg[ -6\mathcal{E}_2^3\Big(\frac{\tau +s }{2}\Big)\mathcal{E}_4^2\Big(\frac{\tau +s }{4}\Big)+4\mathcal{E}_2^2\Big(\frac{\tau +s }{2}\Big)\mathcal{E}_4^3\Big(\frac{\tau +s }{4}\Big)+4\mathcal{E}_2^4\Big(\frac{\tau +s }{2}\Big) 
  \mathcal{E}_4\Big(\frac{\tau +s }{4}\Big)\,,\\
                     & \qquad \qquad -\mathcal{E}_2^5\Big(\frac{\tau +s }{2}\Big)-\mathcal{E}_2\Big(\frac{\tau +s }{2}\Big)\mathcal{E}_4^4\Big(\frac{\tau +s }{4}\Big) \Bigg]\,,\\ 
  \Phi^{(2,1)} & = \mathcal{E}_2^5(\tau) -\frac{1}{16}(32-\chi)\mathcal{E}_2^4(\tau) \mathcal{E}_4 (\tau) - \frac{1}{16} (272+7\chi)\mathcal{E}_2^3(\tau)\mathcal{E}_4^2(\tau) +\frac{1}{8} (560 +9\chi) \mathcal{E}_2^2(\tau) \mathcal{E}_4^4(\tau)\,, \\
  & \qquad \qquad - \frac{1}{4}(368 +5\chi)\mathcal{E}_2(\tau) \mathcal{E}_4^4(\tau) + \frac{1}{2} \mathcal{E}_4^5(\tau) \,,\\
   \Phi^{(2,3)} &  = \Phi^{(2,1)}   \,.
   \end{split}
\end{align}
  }
Note that the corresponding Noether-Lefschetz generator reads
\begin{equation}
\Phi^\pi = \sum_{r,\ell \in \mathbb{Z}/N\mathbb{Z}}\widehat{\Phi}^{(r,\ell)}(\tau)\mathrm{e}_{(r,\ell)}\,, \quad \widehat{\Phi}^{(r,\ell)} = \frac{1}{N} \sum_{s\in \mathbb{Z}/N \mathbb{Z}} \mathrm{e}^{-2\pi i \frac{s \ell}{N}} \Phi^{(r,s)}\,,
\end{equation}
as implicitely expressed in (\ref{eqn:GWNLCHLrk2}). With this information at hand, we apply the KKP conjecture and obtain the following refined BPS numbers $\mathsf{N}_{j_L,j_R}^{(w,k)}$:

\begin{table}[h!]
	\centering
	\centering
	{\small
	
                        \begin{tabular}{|r|c|}\hline
			$\mathsf{N}_{j_-,j_+}^{(1,0)}$ & $2j_+=$0\\ \hline
			$2j_-=$0& $128$\\
			\hline \end{tabular}
	                 \begin{tabular}{|r|c|}\hline
			$\mathsf{N}_{j_-,j_+}^{(1,1)}$ & $2j_+=$0\\ \hline
			$2j_-=$0& $800-2\chi$\\
			\hline \end{tabular}
		         \begin{tabular}{|r|c|}\hline
			$\mathsf{N}_{j_-,j_+}^{(1,2)}$ & $2j_+=$0\\ \hline
			$2j_-=$0& $5120$\\
			\hline \end{tabular}
	 \begin{tabular}{|r|cc|}\hline
			$\mathsf{N}_{j_-,j_+}^{(1,3)}$ &$2j_+=$0&1\\ \hline
			$2j_-=$0&$20816+3 \chi$&\\ 
			1 &&$56 + \frac{\chi}{2}$\\ 
			\hline \end{tabular}
						}
			\caption{Refined BPS invariants $\mathsf{N}_{j_-,j_+}^{(w,k)}$ with $w=1$, which corresponds to the twisted sector $r=1$. }
	\label{tab:Xref1}
\end{table}

\begin{table}[h!]
	\centering
	\centering
	{\small
	
                        \begin{tabular}{|r|c|}\hline
			$\mathsf{N}_{j_-,j_+}^{(2,0)}$ & $2j_+=$0\\ \hline
			$2j_-=$0& $224+\chi$\\
			\hline \end{tabular}
		         \begin{tabular}{|r|c|}\hline
			$\mathsf{N}_{j_-,j_+}^{(2,1)}$ & $2j_+=$0\\ \hline
			$2j_-=$0& $5120$\\
			\hline \end{tabular}\\
	 \begin{tabular}{|r|cc|}\hline
			$\mathsf{N}_{j_-,j_+}^{(2,2)}$ &$2j_+=$0&1\\ \hline
			$2j_-=$0&$69504-4 \chi$&\\ 
			1 &&$32 -\chi$\\ 
			\hline \end{tabular}
         	 \begin{tabular}{|r|cc|}\hline
			$\mathsf{N}_{j_-,j_+}^{(2,3)}$ &$2j_+=$0&1\\ \hline
			$2j_-=$0&$606208$&\\ 
			1 &&$5120$\\ 
			\hline \end{tabular}
						}
			\caption{Refined BPS invariants $\mathsf{N}_{j_-,j_+}^{(w,k)}$ with $w=2$, which corresponds to the twisted sector $r=2$. }
	\label{tab:Xref1}
\end{table}

\begin{table}[h!]
	\centering
	\centering
	{\small
	
                        \begin{tabular}{|r|c|}\hline
			$\mathsf{N}_{j_-,j_+}^{(3,0)}$ & $2j_+=$0\\ \hline
			$2j_-=$0& $128$\\
			\hline \end{tabular}
	 \begin{tabular}{|r|cc|}\hline
			$\mathsf{N}_{j_-,j_+}^{(3,1)}$ &$2j_+=$0&1\\ \hline
			$2j_-=$0&$20816+3 \chi$&\\ 
			1 &&$56 + \frac{\chi}{2}$\\ 
			\hline \end{tabular}
         \\
                  	 \begin{tabular}{|r|cc|}\hline
			$\mathsf{N}_{j_-,j_+}^{(3,2)}$ &$2j_+=$0&1\\ \hline
			$2j_-=$0&$606208$&\\ 
			1 &&$5120$\\ 
			\hline \end{tabular}
					\begin{tabular}{|r|ccc|}\hline
			$N_{j_-j_+}^{(3,3)}$ &$2j_+=$0&1&2\\ \hline
			$2j_-=$0&$9476352-16 \chi $&&\\ 
			1 &&$215392-6 \chi$&\\ 
			2 &&& $800-2\chi$\\  
			\hline \end{tabular}
						}
			\caption{Refined BPS invariants $\mathsf{N}_{j_-,j_+}^{(w,k)}$  with $w = 3$, which corresponds to the twisted sector $r=3$. }
	\label{tab:Xref1}
\end{table}

\begin{table}[h!]
	\centering
	\centering
	{\small
	
                        \begin{tabular}{|r|ccc|}\hline
			$\mathsf{N}_{j_-,j_+}^{(4,0)}$ &$2j_+=$0&1&2\\ \hline
			$2j_-=$0&$8-\chi  $&&\\ 
			1 &1&&1\\ 
			\hline \end{tabular}
	 \begin{tabular}{|r|cc|}\hline
			$\mathsf{N}_{j_-,j_+}^{(4,1)}$ &$2j_+=$0&1\\ \hline
			$2j_-=$0&$70144$&\\ 
			1 &&$128$\\ 
			\hline \end{tabular}
         \\
		\begin{tabular}{|r|ccc|}\hline
			$\mathsf{N}_{j_-,j_+}^{(4,2)}$ &$2j_+=$0&1&2\\ \hline
			$2j_-=$0&$3982752+11\chi $&&\\ 
			1 &&$71008+5 \chi$&\\ 
			2 &&& $224+\chi$\\  
			\hline \end{tabular}
		\begin{tabular}{|r|cccc|}\hline
			$\mathsf{N}_{j_-,j_+}^{(4,3)}$ &$2j_+=$0&1&2 & 3\\ \hline
			$2j_-=$0&$102530688 $& & $128$ & \\ 
			1 &&$4052992$& & \\ 
			2 &$128$&& $70272$&\\  
			3 & & & & $128$\\
			\hline \end{tabular}
			
						}
			\caption{Refined BPS invariants $\mathsf{N}_{j_-,j_+}^{(w,k)}$  with $w = 4$, which corresponds to the untwisted sector $r=0$. }
	\label{tab:Xref0}
\end{table}
A caveat to have in mind are the 4-section geometries of type $IV$ with $\chi = -116, -128$. In these cases, a special the Noether-Lefschetz numbers $N^\pi_{1,(1,3)}$ and $N^\pi_{1,(3,1)}$ result with negative value. We attribute these values to Noether-Lefschetz divisors of the type $T_{\iota}$ in (\ref{eqn:NLdivdiv}). Then, implementing the formula (\ref{eqn:RNLpos}), we obtain the refined BPS numbers shown in the table \ref{tab:Xref128116}. Moreover, in these cases the following holds $\mathsf{N}_{j_-,j_+}^{(1,3)}=\mathsf{N}_{j_-,j_+}^{(3,1)}$.
\begin{table}[h!]
	\centering
	\centering
	{\small
			${\chi=-116}:$
	 \begin{tabular}{|r|ccc|}\hline
			$\mathsf{N}_{j_-,j_+}^{(1,3)}$ &$2j_+=$0&1&2\\ \hline
			$2j_-=$0&$20468$& & \\ 
			1 & 1 &$$& 1\\ 
			\hline \end{tabular}
			\,, \quad
         ${\chi=-128:}$
         	 \begin{tabular}{|r|ccc|}\hline
			$\mathsf{N}_{j_-,j_+}^{(1,3)}$ &$2j_+=$0&1&2\\ \hline
			$2j_-=$0&$20432$& & \\ 
			1 & 4&$$& 4\\ 
			\hline \end{tabular}
						}
			\caption{Special cases for Refined BPS invariants of type $IV$ geometries. }
	\label{tab:Xref128116}
\end{table}

\section{Conclusion}
The enumerative geometry that is captured by the topological string partition function on elliptic and genus one fibered Calabi-Yau threefolds exhibits a deep relation to the theory of lattice Jacobi forms.
On the other hand, if the Calabi-Yau also exhibits a $K3$ fibration one can study the associated Noether-Lefschetz theory which expresses a subset of the invariants in terms of vector-valued modular forms.
In this paper we have explored the relation among these theories and then utilized them in various applications.

For elliptic fibrations we could identify an isomorphism between the space of lattice Jacobi forms and vector-valued modular forms that essentially maps the topological string partition function $Z_F$ with $F$ being the base of the $K3$, which is equal to the elliptic genus of a dual Heterotic string on $K3\times T^2$, to the generator of Noether-Lefschetz numbers.
In the case of genus one fibrations that do not have a section but only $N$-sections we found that the relation is more intricate and higher base degrees have to be taken into account.
The reason is that the dual Heterotic strings are compactified on $(K3\times T^2)/\mathbb{Z}_N$ and  invariants with base degree $\beta=n\cdot F,\,n=1,...,N$ do not correspond to states with multiple Heterotic strings but to twisted sectors of a single string.
Nevertheless, we worked out the relation among the twisted twined elliptic genera of the dual Heterotic string and the Noether-Lefschetz generator.
Together with the Gromov-Witten $\leftrightarrow$ Noether-Lefschetz correspondence theorem this allows to determine the former two objects in terms of the genus zero free energy of the topological string.

We then used the theory of lattice Jacobi forms and proved the non-Abelian sublattice weak gravity conjecture for F-theory on elliptically fibered Calabi-Yau threefolds.
To this end we built upon a previous derivation of the ordinary sublattice weak gravity conjecture and proved a novel relation that expresses lattice theta functions in terms of Weyl characters.
We also showed that the discussion extends straightforwardly to genus one fibrations.

Let us stress that the original motivation for the non-Abelian sublattice weak gravity conjecture was stability under compactifications as well as confirmation in simple perturbative string models.
The fact that it holds true in the vast landscape of six-dimensional F-theory compactification is highly non-trivial and begs for a deeper physical understanding of the conjecture.
It would also be very interesting to study the physical consequences that have been invertigated in~\cite{Heidenreich:2015nta}, in particular a potential unification of the strong coupling scales for the gauge theory and gravity, in terms of the dual Heterotic strings. Technically this mechanism seems close enough to the BPS saturated amplitudes calculated here, as they encode  at least in four 
dimensions  threshold corrections to the gauge as well as to gravitational couplings. However we leave a more detailed discussion of the interesting  estimates 
for the loop  corrections in~\cite{Heidenreich:2015nta}  in the light of the Jacobi-form structure for the BPS multiplicities for future work. Let us note however that the  
asymptotic contribution in the limit of large charge and large spin of exactly these BPS degeneracies in the $K3$ confirms detailed predictions for the microscopic 
entropy of  5d small  black holes~\cite{Huang:2007sb}.

As another application, we have checked a conjecture on the refinement of the Noether-Lefschetz invariants for elliptic and genus one fibrations.
The calculation of refined enumerative invariants for Calabi-Yau threefolds is a very active area of mathematical research but few examples have been explicitly worked out.
Also from the physical side there are different proposals to which amplitudes the refined invariants actually contribute.
Our results therefore not only provide non-trivial evidence for the conjecture~\cite{MR3524167} but provide important test cases 
against which each of those proposals can be checked.  

As recalled in the introduction one the most stringent consistency condition for a description of quantum gravity is the 
absence of global symmetries~\cite{Banks:2010zn}\cite{Harlow:2018tng}, an argument that has been extended to higher form 
global symmetries~\cite{Gaiotto:2014kfa}\cite{Heidenreich:2020pkc}. In the geometrical context of string compactificatios 
it has direct bearings on the question when it is possible to embed  a  member  of the infinite class of non-gravitational  quantum theories 
based on a local or non compact Calabi-Yau geometry geometries into the presumably finite number of topological classes of compact  
Calabi-Yau manifolds, where finiteness is proven for 3-folds with elliptic fiberation structure~\cite{MR1272978}. 
If it is possible, the global symmetry must be either broken or gauged. This explains on the one hand why the definition of 
refined  invariant is so hard~\cite{Huang:2020dbh} on compact Calabi-Yau manifolds. It is precisely because the global $U(1)_R$ symmetry 
used to define their index in the local cases \cite{Choi:2012jz} is broken upon embedding them into a gobal Calabi-Yau manifold. 
On the other hand the local elliptic Calabi-Yau spaces, for which the refined invariants have been calculated by now in almost all
situations~\cite{Gu:2018gmy,Gu:2019dan,Gu:2019pqj,Gu:2020fem} have rich flavour-- and in the case of $E$-strings 
higher form global symmetries. They hence are very interesting class to study how this algebraic censorship on global symmetries works in 
their embeddings into compact Calabi-Yau fibrations.           

\appendix 
\section{Summary of Lie algebras and representation theory}
\label{app:LART}
To guide the reader with notions of Lie algebras and representation theory, we include a summary thereof. We stick mainly to the conventions and notation of~\cite{Fuchs:1997jv}. 
\begin{itemize}
\item A Lie algebra $\mathfrak{g}$ is a vector space equipped with a commutator operation $\left[ \cdot\,, \cdot \right]:\mathfrak{g} \times \mathfrak{g} \rightarrow \mathfrak{g}$, that is bilinear, antisymmetric, and satisfies the Jacobi identity
\begin{equation}
\left[ X, \left[ Y, Z \right] \right] + \left[ Z, \left[ X, Y \right] \right]  +\left[ Y, \left[ Z, X \right] \right] = 0 \quad \forall \text{ } X,Y,Z \in \mathfrak{g}\,.
\end{equation}
\item We say that a Lie algebra $\mathfrak{g}$ is \textit{abelian} if $[ \mathfrak{g} , \mathfrak{g} ] = \{0\}$. We say that a Lie algebra is \textit{simple} if it is not abelian and it contain no proper ideals, i.e., there is no non-trivial subset $\mathfrak{k} \subset \mathfrak{g}$ such that $[\mathfrak{k},\mathfrak{g}] \subset \mathfrak{k}$. A direct sum of simple Lie algebras is called \textit{semisimple}.
\item For a complex simple Lie algebra $\mathfrak{g}$, the Cartan-Weyl basis consists of the decomposition
\begin{equation}
\mathfrak{g} = \mathfrak{h} \oplus \bigoplus_{\bm{\alpha} \in \Phi(\mathfrak{g})}  \mathfrak{g}_{\bm{\alpha}}\,.
\end{equation}
Here $\mathfrak{h}$ denotes the Cartan subalgebra of $\mathfrak{g}$, which is spanned my a maximal set of commuting Hermitian generators $\{ H^i\}_{i = 1, \ldots, \text{rk}(\mathfrak{g})}$, where $\text{rk}(\mathfrak{g})\equiv\text{dim}(\mathfrak{h})$ . Moreover, $\Phi(\mathfrak{g})$ is the root system, which is a set of vectors $\bm{\alpha} = ( \alpha^1, \ldots, \alpha^{\text{rk}(\mathfrak{g})})\in \mathbb{R}^{\text{rk}(\mathfrak{g})}$ called \textit{roots}.
Furthermore, Each $\mathfrak{g}_{\bm{\alpha}}$ is the linear span over $\mathbb{C}$ of a generator $E^{\bm{\alpha}}$ in $\mathfrak{g}$. Putting all together, the Cartan-Weyl basis obeys the following commutator relations
\begin{equation}
[ H^i , H^j ] = 0 \,,  \quad [H^i ,E^{\bm{\alpha}}] = \alpha^i E^{\bm{\alpha}}\,, \quad [E^{\bm{\alpha}},E^{\bm{\beta}}] =
{\scriptsize\begin{cases}
\text{const} \cdot E^{\bm{\alpha +\beta}} &\text{ if }\bm{\alpha + \beta} \in \Phi(\mathfrak{g}) \\ 
\frac{2}{(\bm{\alpha},\bm{\alpha})} \bm{\alpha} \cdot \bm{H} &\text{ if }  \bm{\alpha + \beta} = 0 \\
0 & \text{ otherwise}
\end{cases}}\,.
\end{equation}
\item The Killing form on $\mathfrak{g}$ is the symmetric bilinear map $\left( \cdot\,, \cdot \right) : \mathfrak{g} \times \mathfrak{g} \rightarrow \mathbb{C}$ given by 
\begin{equation}
\left( X , Y\right) = \frac{1}{2g}\text{tr}\left( \text{ad}_X \circ \text{ad}_Y\right)\,,
\end{equation}
where $\text{ad}_X : \mathfrak{g} \rightarrow \mathfrak{g}$ is the linear map defined by $\text{ad}_X (Y) = \left[ X, Y \right]$, and $g$ is a normalization constant that will not be relevant for us. A finite dimensional Lie algebra $\mathfrak{g}$ is semisimple if and only if $\left(\cdot\, , \cdot\right)$ is non-degenerate. 
\item Through the linear map $\bm{\alpha}(H^i) = \alpha^i$, we associate roots $\bm{\alpha} \in \Phi(\mathfrak{g})$ with elements in the dual vector space $\mathfrak{h}^*$. In fact $\text{Span}_{\mathbb{C}}(\Phi(\mathfrak{g})) = \mathfrak{h}^*$. Thus, we can transfer the Killing form for elements in the Cartan subalgebra $\mathfrak{h}$ as follows
\begin{equation}
\left( \bm{ \alpha} , \bm{\beta}\right) = \left( H^{\bm{\alpha}} , H^{\bm{\beta}}\right), \quad H^{\bm{\gamma}} \equiv \sum_{i=1}^{\text{rk}(\mathfrak{g})} \gamma^i  H^i \text{ for } \bm{\gamma}\in \Phi(\mathfrak{g})\,.
\end{equation}
\item  For each root $\bm{\alpha} \in \Phi(\mathfrak{g})$, then $-\bm{\alpha} \in \Phi(\mathfrak{g})$. It is possible find an hyperplane $\mathcal{H}$ that contains no roots, such that it divides the root system into two disjoint half-spaces $\mathcal{H}_\pm$. We can declare the set of \textit{positive roots} $\Phi^+(\mathfrak{g}) \subset \Phi(\mathfrak{g})$ to be those roots lying on $\mathcal{H}_+$. A \textit{simple root} $\bm{\alpha}_i$ is a root that cannot be written as a sum of positive roots, where $i = 1,\ldots, \text{rk}(\mathfrak{g})$. There are 
 $\text{rk}(\mathfrak{g})$ simple roots that provide a basis for $\mathfrak{h}^*$.
\item To each root $\bm{\alpha} \in \Phi(\mathfrak{g})$ there is an associated coroot $\bm{\alpha}^\vee$ defined by the normalization
\begin{equation}
\bm{\alpha}^\vee = \frac{2 \bm{\alpha}}{\left(\bm{\alpha},\bm{\alpha}\right)}\,.
\end{equation}
\item The set of simple roots $\{\bm{\alpha}_i\}_{i=1,\dots, \text{rk}(\mathfrak{g})}$ with associated simple coroots $\{\bm{\alpha}^\vee_i\}_{i=1,\dots, \text{rk}(\mathfrak{g})}$ define the Cartan matrix via the intersections
\begin{equation}
C_{ij} = \left(\bm{\alpha}_i,\bm{\alpha}_j^\vee \right)\,.
\end{equation}
\item Given a complex semisimple Lie algebra $\mathfrak{g}$, every $\mathfrak{g}$-module $V$ has a basis in which the Cartan subalgebra $\mathfrak{h}$ acts diagonally. Thus, there is a decomposition
\begin{equation}
V = \bigoplus_{\bm{\lambda}} V_{\bm{\lambda}}\,, 
\end{equation}
into weight spaces $V_{\bm{\lambda}}$, such that  $R(H^i) \ket{ \bm{\lambda} } = \lambda^i \ket{ \bm{\lambda} }$ for all $\ket{ \bm{\lambda} } \in V_{\bm{\lambda}}$.  Here $R : \mathfrak{g} \rightarrow \text{End}(V)$ is a representation of the Lie algebra $\mathfrak{g}$. The vectors $\bm{\lambda} = ( \lambda^1, \ldots , \lambda^{\text{rk}(\mathfrak{g})})$ are called \textit{weights} of the module $V$.
\item We associate the weights with elements in $\mathfrak{h}^*$ via the linear maps $\bm{\lambda}(H^i) = \lambda^i$. It is convenient to introduce the basis of \textit{fundamental weights} $\{\bm{\omega}_i\}_{i=1,\ldots, \text{rk}(\mathfrak{g})}$ defined by
\begin{equation}
\left( \bm{\omega}_i,\bm{\alpha}_j^\vee\right) = \delta_{ij}\,.
\end{equation}
In this basis, each weight $\bm{\lambda}$ has an expansion
\begin{equation}
\bm{\lambda} = \sum_{i = 1}^{\text{rk}(\mathfrak{g})} \lambda^i \bm{\omega}_i \,, \quad  \Longleftrightarrow \quad
 \lambda^i = \left( \bm{\lambda}, \bm{\alpha}^\vee_i \right)\,,
\end{equation}
where each $\lambda^i$ is the eigenvalue assoiated to the representation $R(H^i)$, also called \textit{Dynkin label}. Moreover, $\bm{\lambda}$ is a weight of a finite dimensional module if and only if its Dynkin labels $\lambda^i \in \mathbb{Z}$.
\item Given two weights $\bm{\lambda} = \sum_{i = 1}^{\text{rk}(\mathfrak{g})} \lambda^i \bm{\omega}_i$ and $\bm{\mu}  = \sum_{i = 1}^{\text{rk}(\mathfrak{g})} \mu^i \bm{\omega}_i$, their intersection pairing reads
\begin{equation}
\left( \bm{\lambda}, \bm{\mu} \right) = \sum_{i,j} Q_{ij} \lambda^i \mu^j\,, \quad Q_{ij} \equiv  \left(\bm{\omega}_i,\bm{\omega}_j\right)\,,
\end{equation}
where we identify $Q_{ij}$ with the weight space metric.
\item There are three lattices relevant for Lie algebras. These are the weight lattice, the root lattice and the coroot lattice
\begin{equation}
L_{\text{w}}(\mathfrak{g}) \equiv \text{Span}_{\mathbb{Z}}\left\{\bm{\omega}_i \right\}\,, \quad
L(\mathfrak{g}) \equiv \text{Span}_{\mathbb{Z}}\left\{\bm{\alpha}_i \right\} \,, \quad
L^\vee(\mathfrak{g}) \equiv \text{Span}_{\mathbb{Z}}\left\{\bm{\alpha}_i^\vee \right\}\,.
\end{equation}
Note that the weight lattice is the dual lattice over $\mathbb{Z}$ to the coroot lattice, i.e.,
\begin{equation}
L_{\text{w}}(\mathfrak{g}) = \left( L^\vee(\mathfrak{g}) \right)^* = \left\{ \bm{\lambda} \mid \left( \bm{\lambda},\bm{\alpha}^\vee\right) \in \mathbb{Z} \text{ } \forall \text{ }\bm{\alpha^\vee} \in L^\vee(\mathfrak{g}) \right\}  \,.
\end{equation}
Throughout this paper we use the upperscript $\vee$ to denote coroots, whereas we reserve the notation $(\,\cdot\,)^*$ for dual vector spaces.
\item Another integral space of interest is the set of \textit{dominant weights} defined by
\begin{equation}
P_+(\mathfrak{g}) \equiv \mathbb{Z}_{\geq0} \bm{\omega}_1 + \cdots + \mathbb{Z}_{\geq0} \bm{\omega}_{\text{rk}(\mathfrak{g})}\,.
\end{equation}
\item For every finite-dimensional $\mathfrak{g}$-module $V$ there exists a maximal weight $\bm{\lambda}_R \in P_+(\mathfrak{g})$, such that
\begin{equation}
\label{eqn:hw}
R(E^{\bm{\alpha}}) \ket{ \bm{\lambda}_R } = 0 \,, \text{ } \forall \text{ } \bm{\alpha} \in \Phi^+(\mathfrak{g})\,, 
\end{equation}
If the $\mathfrak{g}$-module $V$ is irreducible, then there is exactly one weight with the property (\ref{eqn:hw}). In this case, we call $\bm{\lambda}_R$ the \textit{highest weight} of the \textit{irreducible highest weight module} $V_{\bm{\lambda}_R}$.

\item The highest weight theorem states that for any dominant weight $\bm{\lambda} \in P_+(\mathfrak{g})$ there exists a unique, irreducible, finite-dimensional representation $R_{\bm{\lambda}}$ of $\mathfrak{g}$ with highest weight $\bm{\lambda}$. Here $R_{\bm{\lambda}}$ acts on an irreducible highest weight module $V_{\bm{\lambda}}$. A proof of this theorem can be found in~\cite{MR1153249}.

\item For each root $\bm{\alpha} \in \Phi(\mathfrak{g})$, there is a map $w_{\bm{\alpha}}: \Phi(\mathfrak{g}) \rightarrow \Phi(\mathfrak{g})$  whose action reads
\begin{equation}
w_{\bm{\alpha}} : \bm{\beta} \mapsto \bm{\beta} - \left(\bm{\alpha}^\vee,\bm{\beta}\right) \bm{\alpha}\,.
\end{equation} 
The Weyl group $W(\mathfrak{g})$ is the group generated by all the $w_{\bm{\alpha}}$'s with $\bm{\alpha}\in \Phi(\mathfrak{g})$. The action of the Weyl group extends to the weight space of $\mathfrak{g}$.

\item We say that a connected component of the complement of the union of the hyperplanes
\begin{equation}
\left\{ \bm{\lambda} \in \mathbb{R}^{\text{rk}(\mathfrak{g})} \mid \left( \bm{\lambda}, \bm{\alpha}\right)=0 \quad \forall\,\, \bm{\alpha} \in \Phi(\mathfrak{g})\right\}\,,
\end{equation}
is an \textit{open Weyl chamber}. We call \textit{Weyl chamber} the closure of an open Weyl chamber.  The Weyl group acts simply transitively on the set of Weyl chambers, i.e. for every pair of Weyl chambers $\mathcal{C}, \mathcal{C}'$ there is exactly one $w \in W(\mathfrak{g})$ such that $w \mathcal{C} = \mathcal{C}'$. We define the fundamental Weyl chamber $\mathcal{W}(\mathfrak{g})$ by the set
\begin{equation}
\mathcal{W}(\mathfrak{g}) = \left\{ \bm{\lambda} \in \mathbb{R}^{\text{rk}(\mathfrak{g})} \mid \left(  \bm{\lambda} ,  \bm{\alpha}_i \right)\geq 0 \,, \quad i = 1, \ldots, \text{rk}(\mathfrak{g}) \right\}\,.
\end{equation} 
\end{itemize}

\section{Modular Appendix}
\subsection{Vector-valued modular forms}
\label{app:VVMF}
In this section, we review the theory of vector-valued modular forms based on the literature \cite{MR1773561,MR2512363,MR3675870}. For more details on the subject, we encourage the reader to consult the latter references.

\begin{itemize}
\item The metaplectic group $\text{Mp}(2,\mathbb{Z})$ is a double cover of $\mathrm{SL}(2,\mathbb{Z})$. Its elements are pairs of the form
       \begin{equation}
           \left(\gamma = \left(\begin{array}{rr} a\,\, &\,\, b \\  c\,\,  &\,\, d \end{array}\right)\,, \pm \sqrt{c \tau +d}\right)\,, 
        \end{equation}   
where $\gamma \in \mathrm{SL}(2,\mathbb{Z})$ and $w_\gamma(\tau) \equiv \sqrt{c \tau + d}$ is a holomorphic function on $\mathbb{H}$ whose square is $c\tau + d$. The multiplication in $\text{Mp}(2,\mathbb{Z})$ follows the composition law $\left( \gamma_1, w_{\gamma_1 }(\tau)\right) \cdot \left( \gamma_2, w_{\gamma_2}(\tau)\right) = \left( \gamma_1 \gamma_2, w_{\gamma_1} (\gamma_2 \tau) w_{\gamma_2}(\tau) \right) $.

 \item Let $V$ be an $r$-dimensional vector space and let $\rho: \text{Mp}(2,\mathbb{Z}) \rightarrow \text{End}(V)$ be a representation of $\text{Mp}(2,\mathbb{Z})$. A vector-valued modular form of weight 
$k$ and type $\rho$, where $k\in \frac{1}{2}\mathbb{Z}$, is a real analytical function $ \Phi : \mathbb{H} \rightarrow V$ that transforms as follows
        \begin{equation}
             \Phi(\gamma \tau)  = w_\gamma(\tau)^{2k} 
               \rho\left(\gamma, w_\gamma(\tau) \right) \Phi(\tau)\,,  \text{ for all } \left(\gamma, w_\gamma(\tau)\right) \in \text{Mp} _2(\mathbb{Z})\,. 
           \end{equation}
We denote the space of such vector-valued modular forms by $\text{Mod}( \text{Mp}(2,\mathbb{Z}), k, \rho)$.  An element $\Phi \in \text{Mod}( \text{Mp}(2,\mathbb{Z}),k,\rho)$ has a Fourier expansion at the cusp at infinity of the form
     \begin{equation}
        \Phi(\tau) =  \sum_{\ell=1}^r \sum_{n \in \mathbb{Q}} c_{\ell}(n) q^n \mathrm{e}_\ell \,,
     \end{equation}
     where $\{\mathrm{e}_{\ell}\}$ is a basis of $V$.
     
\item  Let $G = L^*/ L$ be the discriminant group endowed with a quadratic form $P: L^*/L \rightarrow \mathbb{Q}/\mathbb{Z}$ defined by $\bm{\lambda} +L \mapsto \frac{1}{2}(\bm{\lambda}\,,\bm{\lambda}) \text{ mod } \mathbb{Z}$, where  $L$ is an even lattice with quadratic form $(\cdot\,, \cdot)$. A canonical choice for a representation $\rho: \text{Mp}(2,\mathbb{Z}) \rightarrow \text{End} \left(\mathbb{C}[G]\right)$ is the \textit{Weil representation}. The latter is defined by the following action on the standard basis $\{\mathrm{e}_{\bm{\lambda}}\}_{\bm{\lambda} \in G}$ of $\mathbb{C}[G]$:
     \begin{align}
      \begin{split}
       \rho (T) (\mathrm{e}_{\bm{\lambda}})& = \exp\left( 2\pi i P(\bm{\lambda}) \right) \mathrm{e}_{\bm{\lambda}}\,,\\
        \rho(S) (\mathrm{e}_{\bm{\lambda}})  & = \frac{ \left(-i \right)^{\frac{\text{sign}(L)}{2}}}{\sqrt{\vert L^* / L\vert}} \sum_{\bm{\mu} \in L^*/L} 
             \mathrm{e}^{ 2\pi i \Big(P(\bm{\lambda})  + P(\bm{\mu}) - P(\bm{\lambda} +\bm{\mu})\Big)}\mathrm{e}_{\bm{\mu}}\,,
      \end{split}
     \end{align}
 where $\text{sign}(L)$ denotes the signature of the lattice $L$ and $T,S \in \text{Mp}(2,\mathbb{Z})$ read
   \begin{equation}
      T = \left( \left(\begin{array}{rr}\, 1 \,\,&\,\, 1\, \\ \,0\,\, &\,\, 1\, \end{array}\right),\, 1\right)\,, \quad S = \left( \left(\begin{array}{rr}\, 0 \,\,& -1\, \\ \,1\,\, & 0\, \end{array}\right),\, \sqrt{\tau}\right)\,.
   \end{equation}    
   Moreover, $\rho^*$ denotes the inverse transpose of $\rho$.
   
\item   A relevant normal subgroup of $\text{Mp}(2,\mathbb{Z})$ is the following one 
   \begin{align}
      \Gamma(4N)^* = \left\{ \left(\gamma\,, \frac{\theta(\gamma \tau)}{\theta (\tau)} \right) \in  \text{Mp}(2,\mathbb{Z}) \, \Big\vert \, \gamma \in \Gamma(4N) \right\}  \,,
    \end{align}
      where $\theta(\tau) = \sum_{n\in\mathbb Z} q^{n^2}$ and $\Gamma(4N)$ is a principal congruence subgroup of $\mathrm{SL}(2,\mathbb{Z})$ defined by
       \begin{equation}
       \Gamma(4N) = \left\{ {\scriptsize\left(\begin{array}{rr} a\,\, &\,\, b \\  c\,\,  &\,\, d \end{array}\right)} \in \mathrm{SL}(2,\mathbb{Z}) \, \Big\vert  \, b \equiv c \equiv 0 \text{ (mod $N$) and } a \equiv d \equiv 1 \text{ (mod $N$)} \right\}\,. 
             \end{equation}
  \item We denote by $M_{k}(4N)$, where $k\in \frac{1}{2} \mathbb{Z}$, the space of holomorphic functions $f: \mathbb{H}\rightarrow \mathbb{C}$ such that $f(\gamma \tau) = w_\gamma(\tau)^{2k} f(\tau)$ for all $\left( \gamma, w_\gamma(\tau)\right) \in \Gamma(4N)^*$.  To span the latter kind of modular forms, we use as basis the space $M_{\frac{1}{2}} (4N)$. This can be written as \cite{MR806354,gritsenko2019theta}
     \begin{equation}
     M_{\frac{1}{2}} \left( 4N \right) = \bigoplus_{\substack{ d \mid N \\ \frac{N}{d} \text{ squarefree }}} \Theta^{\text{null}} \left(\underline{\mathbb{Z}}(2d)\right)\,,
     \label{eqn:VVMFbasis}
     \end{equation}      
     where $\Theta^{\text{null}}\left(\underline{\mathbb{Z}}(2m)\right)$ is the space of \textit{``Thetanullwerte''} defined by
     \begin{equation}
    \Theta^{\text{null}}\left(\underline{\mathbb{Z}}(2m)\right) = \text{Span}_{\mathbb{C}}\left\{ \vartheta_{m,\ell} (\tau,0) \mid \ell\in \mathbb{Z}/2m \mathbb{Z} \right\} \,,
     \end{equation}
     and
     \begin{equation}
       \vartheta_{m,\ell} (\tau,z) = \sum_{\substack{r \in \mathbb{Z}\\ r \equiv \ell \text{ mod } 2m }} q^{\frac{r^2}{2m}} \zeta^r\,,  \quad \zeta \equiv \mathrm{e}^{2\pi i z}\,.
     \end{equation}
     
     \item Let $\Gamma$ be an even lattice with signature $(b^+,b^-)$, dual lattice $\Gamma^*$, and an isometry $P: \Gamma \otimes \mathbb{R} \rightarrow \mathbb{R}^{b^+,b^-}$.  The isometry  $P$ defines projections on $\mathbb{R}^{b^+,0}, \mathbb{R}^{0,b^-}$ written as $P_+(p) = p_R,$  $P_-(p)= p_L$ respectively. The Siegel Theta function $\Theta_\Gamma$ of $\Gamma$ is defined by
     \begin{equation}
     \label{eqn:SiegelTheta}
 \Theta_\Gamma(\tau,\bar{\tau}) = \sum_{\gamma \in \Gamma^*/ \Gamma } \theta_{\Gamma+\gamma} (\tau,\bar{\tau}) \mathrm{e}_\gamma \,, \quad \theta_{\Gamma +\gamma} (\tau,\bar{\tau}) = \sum_{p  \in \Gamma +\gamma } q^{\frac{p_L^2}{2}}\bar{q}^{\frac{p_R^2}{2}}\,, 
\end{equation}
where $\tau \in \mathbb{H}$. Further generalizations of Siegel theta functions can be found in~\cite{MR1625724}, such as the one introduced in (\ref{eqn:SiegelNarain}).
\end{itemize}

\subsection{Modular operators}
\label{app:ModOp}
\begin{itemize}
\item The Eisenstein series $E_{2k}(\tau),\,k>1$ are modular forms for $\mathrm{SL}(2,\mathbb{Z})$ of weight $2k$ and can be written as
\begin{align}
	E_{2k}(\tau)=1+\frac{2}{\zeta(1-2k)}\sum\limits_{n=1}^\infty\frac{n^{2k-1}q^n}{1-q^n}\,.
\end{align}
For $k=1$ one obtains the quasi modular Eisenstein series $E_2(\tau)$ of weight $2$.
\item The Dedekind $\eta$-function
\begin{align}
	\eta(\tau)=q^{\frac{1}{24}}\prod\limits_{i=1}^\infty (1-q^i)\,,
\end{align}
is also not quite modular but satisfies
\begin{align}
	\eta(\tau+1)=e^{\frac{\pi i}{12}}\eta(\tau)\,,\quad\eta(-1/\tau)=\sqrt{-i\tau}\eta(\tau)\,.
\end{align}
However, $\Delta_{12}(\tau)=\eta^{24}(\tau)$ is a modular form of weight 12 for $\mathrm{SL}(2,\mathbb{Z})$ that vanishes as $\tau\rightarrow i\infty$.
\item Analogous to the ordinary Eisenstein series, the Jacobi Eisenstein series can be defined as orbit sums~\cite{MR781735}
\begin{align}
	E_{k,m}(\tau,z)=\frac12\sum\limits_{\small\begin{array}{c}c,d\in\mathbb{Z}\\(c,d)=1\end{array}}\sum\limits_{\lambda\in\mathbb{Z}}(c\tau+d)^{-k}\exp\left(m\lambda^2\frac{a\tau+b}{c\tau+d}+2m\lambda\frac{z}{c\tau+d}-\frac{mcz^2}{c\tau+d}\right)\,,
\end{align}
with $a,b$ such that the M\"obius transformation acting on $\tau$ is in $\Gamma_1$.
They admit a Fourier expansion
\begin{align}
	E_{k,m}(\tau,z)=\sum\limits_{\small\begin{array}{c}n,r\in\mathbb{Z}\\4nm\ge r^2\end{array}}e_{k,m}(n,r)q^n\zeta^r\,,
\end{align}
and for $E_{k,1}$ the coefficients are
\begin{align}
	e_{k,1}(n,r)=\frac{H(k-1,4n-r^2)}{\zeta(3-2k)}\,.
\end{align}
Here $H(r,N)$ are Cohen's functions~\cite{MR0389784} and for our purposes it will be sufficient to know the generating functions
\begin{align}
	\mathcal{H}_r(q)=\sum\limits_{N\ge 0}H(r,N)q^N\,,
\end{align}
for $r=3$ and $r=5$.
They can be expressed in terms of $\Gamma_0(4)$ modular forms~\cite{MR0389784}
\begin{align}
	F_2(\tau)=\frac{\eta(4\tau)^8}{\eta(2\tau)^4}\,,\quad \theta(\tau)=1+2\sum\limits_{n\ge 1}q^{n^2}\,,
\end{align}
and read
\begin{align}
	\begin{split}
		\mathcal{H}_3=&-\frac{1}{252}\left(\theta^7-14\theta^3F_2\right)\,,\\
		\mathcal{H}_5=&-\frac{1}{132}\left(\theta^{11}-22\theta^7F_2+88\theta^3F_2^2\right)\,.
	\end{split}
\end{align}
\item The Hecke operator $V_l:\,J_{k,m}\rightarrow J_{k,lm}$ acts on a Jacobi form $\phi\in J_{k,m}$
\begin{align}
	\phi=\sum\limits_{n,r}c(n,r)q^n\zeta^r\,,
\end{align}
as
\begin{align}
	V_l\phi=\sum\limits_{n,r}\left(\sum\limits_{a|(n,r,l)}a^{k-1}c\left(\frac{nl}{a^2},\frac{r}{a}\right)\right)q^n\zeta^r\,,
\end{align}
and, for $m$ square free, it can be shown that~\cite{MR781735}
\begin{align}
	E_{k,m}=\sigma_{k-1}(m)^{-1}V_mE_{k,1}\,.
\end{align}
\item We introduce the normalized Eisenstein series $\tilde{E}_{4,4}$ and $\tilde{E}_{6,4}$
\begin{align}
	\begin{split}
	\tilde{E}_{4,4}=\sum\limits_{n,r}e_{4,1}(n,r/2)q^n\zeta^r\,,\\
	\tilde{E}_{6,4}=\sum\limits_{n,r}e_{6,1}(n,r/2)q^n\zeta^r\,.
	\end{split}
\end{align}

\item
 The ring of weak Jacobi forms of a single variable and integer index is freely generated over the ring of modular froms by the two generators
\begin{align}
\label{eqn:weakJac}
	\begin{split}
		\phi_{-2,1}(\tau,z)=&-\frac{\theta_1(\tau,z)^2}{\eta(\tau)^6}\\
		\phi_{0,1}(\tau,z)=&4\left[\frac{\theta_2(\tau,z)^2}{\theta_2(\tau,0)^2}+\frac{\theta_3(\tau,z)^2}{\theta_3(\tau,0)^2}+\frac{\theta_4(\tau,z)^2}{\theta_4(\tau,0)^2}\right] \,,
	\end{split}
\end{align}
of index one and respective weight $-2$ and $0$.
Here the expressions are written in terms of Jacobi theta functions, which are defined as
\begin{align}
	\begin{split}
		\theta_1(\tau,z)=&\vartheta_{\frac{1}{2}\frac{1}{2}}(\tau,z)\,,\quad \theta_2(\tau,z)=\vartheta_{\frac{1}{2}0}(\tau,z)\,,\\
		\theta_3(\tau,z)=&\vartheta_{00}(\tau,z)\,,\quad \theta_4(\tau,z)=\vartheta_{0\frac{1}{2}}(\tau,z)\,,\\
		\vartheta_{ab}(\tau,z)=&\sum\limits_{n=-\infty}^\infty e^{\pi i(n+a)^2\tau+2\pi i z(n+a)+2\pi i b(n+a)}\,.
	\end{split}
	\label{eqn:jacobitheta}
\end{align}

\item When we discuss the refinement of the topological string partition function we will also need the Jacobi form
\begin{align}
	\begin{split}
		\phi_{-1,\frac12}(\tau,z)=&i\frac{\theta_1(\tau,z)}{\eta(\tau)^3}\,,
	\end{split}
\end{align}
of weight $-1$ and index $1/2$.
\end{itemize}

\subsection{Modular expressions}

\noindent\textbf{Weyl invariant theta combinations:} Here we provide explicit combinations of theta functions that obey the formula (\ref{eqn:Claim}). These theta functions are defined by the Weyl invariant subsets in (\ref{eqn:KsWeyl})

{\scriptsize
\begin{align}
\label{eqn:KsThetas}
\begin{split}
\vartheta_{[\textcolor{harvardcrimson}{\bullet}]}(\tau,\bm{z})& =\left(\frac{1}{\zeta_1^2} +\frac{\zeta_1^2}{\zeta_2^2}+ \zeta_2^2 \right)q^{\frac{2}{3}} + \left(\zeta_1^4 + \frac{\zeta_2^4}{\zeta_1^4} + \frac{1}{\zeta_2^4} \right)q^{\frac{8}{3}} + \mathcal{O}\left(q^{\frac{14}{3}}\right)\\
& = \left( \chi_{(0,2)}(\bm{z}) - \chi_{(1,0)}(\bm{z})\right)q^{\frac{2}{3}} + \left( \chi_{(4,0)}(\bm{z}) - \chi_{(2,1)}(\bm{z}) + \chi_{(1,0)}(\bm{z})\right) q^{\frac{8}{3}}+ \mathcal{O}\left(q^{\frac{14}{3}}\right)\,, \\
\vartheta_{[\textcolor{yellow(ncs)}{\bullet}]}(\tau,\bm{z})& = \left(\frac{1}{\zeta_2^2} +\frac{\zeta_2^2}{\zeta_1^2}+ \zeta_1^2 \right)q^{\frac{2}{3}} + \left(\zeta_2^4 + \frac{\zeta_1^4}{\zeta_2^4} + \frac{1}{\zeta_1^4} \right)q^{\frac{8}{3}}+ \mathcal{O}\left(q^{\frac{14}{3}}\right)\\
& =\left( \chi_{(2,0)}(\bm{z}) - \chi_{(0,1)}(\bm{z})\right)q^{\frac{2}{3}} + \left( \chi_{(0,4)}(\bm{z}) - \chi_{(1,2)}(\bm{z}) + \chi_{(0,1)}(\bm{z})\right) q^{\frac{8}{3}}+ \mathcal{O}\left(q^{\frac{14}{3}}\right)  \,,\\
  \vartheta_{[\textcolor{lasallegreen}{\bullet}] \oplus[\textcolor{teal}{\bullet}]\oplus[\textcolor{limegreen}{\bullet}] }(\tau,\bm{z}) & =  \left(\zeta_\pm^{(2,1)} + \zeta_\pm^{(1,1)} + \zeta_\pm^{(1,2)}\right)q^{\frac{1}{2}} +  \left(\zeta_\pm^{(3,0)} + \zeta_\pm^{(3,3)} + \zeta_\pm^{(0,3)}\right)q^{\frac{3}{2}}  + \mathcal{O}\left(q^{\frac{7}{2}}\right) \\
  & = \left( \chi_{(1, 1)}(\bm{z}) -2 \chi_{(0, 0)} \right)q^{\frac{1}{2}} +  \left( 2\chi_{(0, 0)} + \chi_{(3,0)}(\bm{z})- 2\chi_{(1, 1)}(\bm{z}) + \chi_{(0,3)}(\bm{z})  \right)q^{\frac{3}{2}} +\mathcal{O}\left(q^{\frac{7}{2}}\right)\,, \\
  \vartheta_{[\textcolor{indigo(web)}{\bullet}] \oplus[\textcolor{jazzberryjam}{\bullet}]\oplus[\textcolor{darkscarlet}{\bullet}]}(\tau,\bm{z}) & =  \left( \zeta_2 + \frac{1}{\zeta_1} + \frac{\zeta_1}{\zeta_2}\right) q^{\frac{1}{6}} + \left( \frac{\zeta_2}{\zeta_1^3} + \frac{1}{\zeta_2 \zeta_1^2} + \frac{\zeta_2^3}{\zeta_1} + \zeta_2^2 \zeta_1 + \frac{\zeta_2^2}{\zeta_1^3} + \frac{\zeta_2^3}{\zeta_1^2} \right)q^{\frac{7}{6}} + \mathcal{O}\left(q^{\frac{13}{6}}\right)\\
& = \chi_{(0,1)}(\bm{z}) q^{\frac{1}{6}} + \left(\chi_{(1,2)}(\bm{z}) -\chi_{(2,0)}(\bm{z}) - \chi_{(0,1)}(\bm{z}) \right)q^{\frac{7}{6}} + \mathcal{O}\left(q^{\frac{13}{6}}\right)\,,\\
\vartheta_{[\textcolor{scarlet}{\bullet}] \oplus[\textcolor{outrageousorange}{\bullet}]\oplus[\textcolor{princetonorange}{\bullet}] }(\tau,\bm{z}) &=  \left( \zeta_1 + \frac{1}{\zeta_2} + \frac{\zeta_2}{\zeta_1}\right) q^{\frac{1}{6}} + \left( \frac{\zeta_1}{\zeta_2^3} + \frac{1}{\zeta_1 \zeta_2^2} + \frac{\zeta_1^3}{\zeta_2} + \zeta_1^2 \zeta_2 + \frac{\zeta_1^2}{\zeta_2^3} + \frac{\zeta_1^3}{\zeta_2^2} \right)q^{\frac{7}{6}} + \mathcal{O}\left(q^{\frac{13}{6}}\right)\\
& = \chi_{(1,0)}(\bm{z}) q^{\frac{1}{6}} + \left(\chi_{(2,1)}(\bm{z}) -\chi_{(0,2)}(\bm{z}) - \chi_{(1,0)}(\bm{z}) \right)q^{\frac{7}{6}} + \mathcal{O}\left(q^{\frac{13}{6}}\right)\,.\\
\end{split}
\end{align}}


\addcontentsline{toc}{section}{References}
\bibliographystyle{utphys}
\bibliography{References}







\end{document}